\documentclass[prd,aps,showpacs,nofootinbib]{revtex4}
\usepackage{amssymb,amsmath,epsfig}
\usepackage{mathrsfs}
\usepackage[usenames,dvipsnames]{color}
\usepackage[bookmarks]{hyperref}
\usepackage[svgnames]{xcolor}
\usepackage{subfigure}
\usepackage{times}

\newcommand{\Real}{\mathbb{R}}

\newcommand{\dvol}{\mbox{dvol}}

\newcommand{\Torus}{\mathbb{T}}

\begin{document}

\title{Axisymmetric, stationary collisionless gas configurations surrounding black holes}

\author{Carlos Gabarrete and Olivier Sarbach}
\address{Instituto de F\'isica y Matem\'aticas,
Universidad Michoacana de San Nicol\'as de Hidalgo,
Edificio C-3, Ciudad Universitaria, 58040 Morelia, Michoac\'an, M\'exico.}

\begin{abstract}
The properties of a stationary gas cloud surrounding a black hole are discussed, assuming that the gas consists of collisionless, identical massive particles that follow spatially bound geodesic orbits in the Schwarzschild spacetime. Several models for the one-particle distribution function are considered, and the essential formulae that describe the relevant macroscopic observables, like the current density four-vector and the stress-energy-momentum tensor are derived. This is achieved by rewriting these observables as integrals over the constants of motion and by a careful analysis of the range of integration. In particular, we provide configurations with finite total mass and angular momentum. Differences between these configurations and their nonrelativistic counterparts in a Newtonian potential are analyzed. Finally, our configurations are compared to their hydrodynamic analogues, the ``polish doughnuts".
\end{abstract}

\date{\today}


\maketitle

\section{Introduction}

The goal of this work is to extend the non-relativistic steady-state collisionless kinetic gas tori around compact objects discussed in our accompanying article~\cite{cGoS2022b} (referred to as paper I in the following) to the relativistic case, assuming that the central object is a black hole. Such tori have many potential astrophysical applications, including the modeling of low-density hot accretion disks and studying the behavior of dark matter or a distribution of stars surrounding supermassive black holes. In particular, the recent observations by the Event Horizon Telescope Collaboration (EHTC)~\cite{EHTC,EHTCI,EHTCXII} revealing the shadows of the supermassive black holes M87$^*$ and Sgr A$^*$ call for a profound understanding for the behavior of a hot plasma in the vicinity of a strong gravitational field beyond the hydrodynamic approximation.

The configurations discussed in this work are based on the same assumptions as the nonrelativistic model in paper I, except that the kinetic gas is treated in a fully relativistic description and the central object is assumed to be a black hole. For the sake of completeness and clarity, in the following we list all the assumptions made in this article. First, we assume the gas to be collisionless,\footnote{Note that for hot and diffuse plasmas accreting into black holes the mean free path is expected to be comparable or even larger than the macroscopic length scale, which justifies a kinetic description~\cite{mKjSeQ2016}.} neglecting the effects of collisions between the gas particles. Second, we consider a gas consisting of identical, massive and uncharged particles, and hence we do not take into account electromagnetic effects (which is justified for dark matter or star distributions but nor for plasmas). Third, we assume that the self-gravity of the gas can be neglected, the gravitational field being dominated by the one generated by the central black hole. This implies that we may regard the black hole as being described by an asymptotically flat, stationary solution of the Einstein vacuum equations with an event horizon. Modulo technical assumptions, the uniqueness theorems (see, for instance Refs.~\cite{Heusler-Book,lrr-2012-7}) imply that the central black hole belongs to the Kerr family. Fourth, in this article we further neglect the rotation of the black hole, implying that the exterior region is described by the Schwarzschild metric. Finally, similar to paper I and motivated by phase space mixing~\cite{pRoS2020,pRoS18} and dispersion~\cite{pRoS17a} on a fixed Schwarzschild background, we restrict ourselves to steady-states in which each gas particle follows a spatially bound geodesic trajectory in the Schwarzschild geometry. Moreover, for simplicity, we restrict our analysis to axisymmetric configurations in which the one-particle distribution function (DF) only depends on the energy and azimuthal component of the angular momentum of the gas particles.

Our models are based on the ans\"atze for the DF that have been used in previous works to construct self-gravitating axisymmetric configurations in the absence of a central black hole, see for example~\cite{sSsT85a,sSsT85b,sSsT1993a,sSsT1993b,hAmKgR11,hAmKgR14}. In particular, we consider the generalized polytropic ansatz used more recently in Refs.~\cite{eAhAaL16,eAhAaL19} to numerically construct configurations with toroidal, disk-like, spindle-like and composite structures. In these ans\"atze, the one-particle DF is the product of a function of the energy times a function of the azimuthal component of the angular momentum. Contrary to the works in~\cite{sSsT85a,sSsT85b,sSsT1993a,sSsT1993b,eAhAaL16,eAhAaL19}, the gravitational field in our configurations is dominated by the central black hole, such that the self-gravity can be neglected. One of the main achievements of this work is to derive explicit expressions for the spacetime observables (namely, the current density and energy-momentum-stress tensor) which have the form of a single integral of an analytic (albeit complicated) function of the energy. In the Newtonian limit these integrals simplify and reduce to the corresponding expressions in paper I. By analyzing the properties of these spacetime observables, we show that our configurations have the morphology of a thick torus extending all the way to spatial infinity, having a sharp inner boundary surface. Next, we show that although they have infinite extend, for suitable values of the free parameters our configurations have finite total particle number, energy and angular momentum. Finally, we provide a detailed analysis for the radial profile of the spacetime observables, including the particle density, the kinetic temperature and the principle pressures and compare them with an analogous fluid model.

Relevant to the self-consistency of our models is the recent work by Jabiri~\cite{fJ2022,fJ2021} who proves the mathematical existence of solutions to the stationary, axisymmetric Einstein-Vlasov system of equations describing self-gravitating tori of Vlasov matter which, in the limit of vanishing amplitude, reduce to configurations which are similar\footnote{The self-gravitating configurations constructed~\cite{fJ2022} have finite support, whereas in our case the support is infinite, although we could also easily obtain finite support configurations with our ansatz as discussed further below.} to the ones analyzed in the present article. In this sense, the work in~\cite{fJ2022} suggests that (for small enough amplitude) our configurations can likely be generalized to include their self-gravity. We also mention related recent work on the accretion of a Vlasov gas by a central black hole~\cite{pDeJmAeMdN17,pRoS17a,pRoS17b,aCpM2020,pMoA2021a,aGetal2021,pMoA2021b,pMaO2022,aCpMaO22} in which unbounded (instead of bounded) timelike geodesics are relevant.

This work is organized as follows. In section~\ref{Sec:Schwarzschild} we provide a brief review of the collisionless Boltzmann equation on a Schwarzschild spacetime. We focus our attention on the subset of phase space which corresponds to spatially bound future-directed timelike geodesics. Assuming that the one-particle DF is supported on this set, we establish general expressions for the current density vector and energy-momentum-stress tensor which allow one to compute these spacetime observables when the one-particle DF is a function only of the integrals of motion. Next, in section~\ref{Sec:StatAxiModel} we consider an ansatz in which the one-particle DF depends only on the energy and azimuthal component of the angular momentum of the gas particles. We concentrate on two models; the first one (rotating) describes a gas configuration that has nonvanishing total angular momentum while in the second one (even) the DF is an even function of the angular momentum such that the total angular momentum vanishes although the individual gas particles rotate. For these models, we reduce the expressions for the spacetime observables to single integrals over the energy variable. Furthermore, we compute the total particle number, energy and angular momentum of our configurations and compare them with the analogous Newtonian models in paper I. The properties of some spacetime observables are analyzed in section~\ref{Sec:Examples} where we also compare our kinetic configurations with the corresponding fluid models. Conclusions are drawn in section~\ref{Sec:Conclusions} and technical --yet important-- aspects of our calculations are described in appendices~\ref{App:Parametrization}--\ref{App:PolishDoughnuts}.

Throughout this work, we use the signature convention $(-,+,+,+)$ for the spacetime metric and work in geometrized units, such that $G_N = c = 1$. For recent reviews on mathematical results regarding the Einstein-Vlasov system and the geometric structure in the relativistic kinetic theory on curved spacetimes we refer the reader to Refs.~\cite{hA11,oStZ13,oStZ14b,rAcGoS2022}.

\section{Collisionless gas configurations trapped in a Schwarzschild potential}
\label{Sec:Schwarzschild}

In this section we review the basic formalism necessary to describe a collisionless kinetic gas propagating in the exterior of a non-rotating black hole spacetime. Since the self-gravity of the gas is neglected, we may assume that the spacetime is described by the Schwarzschild metric with fixed mass parameter $M > 0$, as explained in the introduction. We focus our attention on the case in which the individual gas particles follow future-directed spatially bound timelike geodesic. In the next subsection we provide a brief recapitulation of the Hamiltonian description of such geodesics, its integrals of motion and the properties of the effective potential describing the radial motion. In subsection~\ref{SubSec:VlasovEquation} we summarize a few well-known results concerning the formulation of general relativistic kinetic theory which are essential to this work, and in subsection~\ref{SubSec:ExplicitObservables} we express the particle current density vector and the energy-momentum-stress tensor in terms of an integral over the conserved quantities.

The case of unbound trajectories has been analyzed in detail in~\cite{pRoS17a} and applied to accretion problems~\cite{pRoS17a,pRoS17b,pMoA2021a,pMoA2021b,aGetal2021,pMaO2022}, see also Ref.~\cite{aCpM2020} for the analogous problem on the Reissner-Nordstr\"om background and Ref.~\cite{aCpMaO22} for a thin accretion disk confined to the equatorial plane of a Kerr black hole.

\subsection{Free-particle Hamiltonian, integrals of motion and effective potential}
\label{SubSec:SchwarzschildSpacetime}

Since we are only interested in the exterior region, it is sufficient to work in standard Schwarzschild coordinates $(x^\mu) = (t,r,\vartheta,\varphi)$, for which the spacetime metric is
\begin{equation}
g := -N(r) dt^2 + \frac{dr^2}{N(r)} + r^2 d\vartheta^2 + r^2 \sin^2\vartheta d\varphi^2, \qquad N(r) := 1-\frac{2M}{r} > 0.
\end{equation}
As this spacetime is static and spherically symmetric, the geodesic motion possesses several integrals of motion. First, we have the conserved energy $E$, which is associated with the timelike Killing vector field $\partial_t$. Next, we have the three components of the angular momentum $(L_x,L_y,L_z)$ which correspond to the generators of the rotations. Finally, the  particles' rest mass $m$ is conserved. The integral of motion corresponding to $-m^2/2$ is the free-particle Hamiltonian of the theory, given by
\begin{equation}
\mathcal{H}(x, p) := \frac{1}{2} g_x^{-1}(p,p) 
 = \frac{1}{2}\left( -\frac{p_t^2}{N(r)} + N(r)p_r^2 + \frac{p_\vartheta^2}{r^2} + \frac{p_\varphi^2}{r^2\sin^2\vartheta} \right),
\end{equation}
in adapted local coordinates $(x^\mu,p_\mu) = (t,r,\vartheta,\varphi,p_t,p_r,p_\vartheta,p_\varphi)$ on the cotangent bundle $T^*\mathcal{M}$ associated with the spacetime manifold $(\mathcal{M},g)$. In this article, we focus our attention on the conserved quantities
\begin{equation}
m = \sqrt{-2\mathcal{H}}, \quad E = -p_t, \quad L_z = p_\varphi, \quad \hbox{and} \quad 
L^2 := L_x^2 + L_y^2 + L_z^2 = p^2_\vartheta + \frac{L_z^2}{\sin^2\vartheta}, 
\label{Eq:ConservedQuantities}
\end{equation}
which Poisson commute with each other.\footnote{In terms of adapted local coordinates the Poisson bracket $\left\{ \mathcal{F}, \mathcal{G} \right\}$ between two functions $F$ and $G$ on $T^*\mathcal{M}$ is defined as
\begin{equation}
	\left\{\mathcal{F}, \mathcal{G} \right\} = \frac{\partial \mathcal{F}}{\partial p_\mu} \frac{\partial \mathcal{G} }{\partial x^\mu} - \frac{\partial \mathcal{G}}{\partial p_\mu} \frac{\partial \mathcal{F}}{\partial x^\mu},\nonumber
\end{equation}
see~\cite{rAcGoS2022} and references therein for more details and a coordinate-free definition.} For the following, it is useful to introduce the orthonormal basis of vector fields
\begin{equation}
e_{\hat{0}} := \frac{1}{\sqrt{N(r)}}\frac{\partial}{\partial t},\quad
e_{\hat{1}} := \sqrt{N(r)}\frac{\partial}{\partial r},\quad
e_{\hat{2}} := \frac{1}{r}\frac{\partial}{\partial \vartheta},\quad
e_{\hat{3}} := \frac{1}{r\sin\vartheta}\frac{\partial}{\partial \varphi}.
\label{Eq:OrthoBasis}
\end{equation}
In terms of the conserved quantities, the orthonormal components of $p$ have the form
\begin{equation}
(p_{\hat{\mu}})(\epsilon_r,\epsilon_\vartheta)
  = \left( -\frac{E}{\sqrt{N(r)}}, \frac{\epsilon_r}{\sqrt{N(r)}} \sqrt{E^2 - V_{m,L}(r)}, 
  \frac{\epsilon_\vartheta}{r}\sqrt{L^2 - \frac{L_z^2}{\sin^2\vartheta}}, 
  \frac{L_z}{r\sin\vartheta} \right), 
\label{Eq:Orthonormalbasis}
\end{equation}
where the signs $\epsilon_r = \pm 1$ and $\epsilon_\vartheta = \pm 1$ determine the corresponding signs of $p_r$ and $p_\vartheta$, and $V_{m,L}(r)$ is the effective potential for radial geodesic motion in the Schwarzschild spacetime, which is given by
\begin{equation}
V_{m,L}(r) := N(r) \left(m^2 + \frac{L^2}{r^2} \right).
\label{Eq:SchEffectivePotential}
\end{equation}
The properties of this potential are well-know, see for example Appendix~A in~\cite{pRoS17a} for a summary. The most relevant features for the purpose of the present article are the following: (i) there exists a potential well with associated stable bound orbits only if $L > L_{\textrm{ms}} := \sqrt{12} M m$. In the limit $L = L_{\textrm{ms}}$ the potential has an inflection point at $r = r_{\textrm{ms}} = 6M$ describing the marginally stable circular orbits which have energy $E_{\textrm{ms}} = \sqrt{8/9} m$. (ii) Bound orbits exist only when $E\in (E_{\textrm{ms}}, m)$ and the total angular momentum is confined to the interval $L\in (L_{\textrm{c}}(E),L_{\textrm{ub}}(E))$, with $L_{\textrm{c}}(E)$ ($L_{\textrm{ub}}(E)$) the critical angular momentum corresponding to the case in which the effective potential's maximum (minimum) is equal to $E^2$. These extrema of $V_{m,L}$ are located at
\begin{equation}
r_{\textrm{max}} = \frac{6M}{1 + \sqrt{1 - 12M^2m^2/L^2}},\qquad
r_{\textrm{min}} = \frac{6M}{1 - \sqrt{1 - 12M^2m^2/L^2}},
\end{equation}
with $r_{\textrm{max}}$ decreasing monotonously from $6M$ to $3M$ and $r_{\textrm{min}}$ increasing monotonously from $6M$ to $\infty$ as $L$ increases from $L_{\textrm{ms}}$ to $\infty$. The unstable circular orbit at $r = r_{\textrm{max}}$ separate those orbits that are reflected at the potential from those that plunge into the black hole, and they have energies smaller than $m$ as long as $L_{\textrm{ms}} < L < 4Mm$. Therefore, for fixed values of $L$ in this interval, the bound orbits with minimum radius are those whose energies and left turning points approach $E_{\textrm{max}} = \sqrt{V_{m,L}(r_{\textrm{max}})}$ and $r_{\textrm{max}}$, respectively. For these reasons, the limiting orbits with $E = E_{\textrm{max}}$  are called innermost stable orbits (ISO). In the limit $L = L_{\textrm{mb}} = 4Mm$ the ISOs are marginally bound, and they have energy $E = m$ and  minimum radius $r = r_{\textrm{mb}} = 4M$.

\subsection{Collisionless Boltzmann equation and spacetime observables}
\label{SubSec:VlasovEquation}

The collisionless Boltzmann (or Vlasov) equation on the curved spacetime manifold $(\mathcal{M}, g)$ can be written in a compact way as
\begin{equation}
\left\{\mathcal{H}, f \right\} = 0,
\label{Eq:Vlasov}
\end{equation}
where $f: T^*\mathcal{M}\to \Real$ is the one-particle DF, and it expresses the fact that $f$ is constant along the Liouville flow. When restricted to the subset $\Gamma_{\textrm{bound}}$ of $T^*\mathcal{M}$ corresponding to bound orbits, one can transform $(x^\mu,p_\mu)$ to generalized action-angle variables $(\mathcal{Q}^\mu, \mathcal{J}_\mu)$ which allows one to provide an explicit solution representation for the DF~\cite{pRoS2020}. The explicit form of these variables will not be used in this article, although the existence of the symplectic transformation $(x^\mu,p_\mu)\mapsto (\mathcal{Q}^\mu, \mathcal{J}_\mu)$ will turn out to be useful when computing the total particle number, energy and angular momentum in the next section. 

For the following, we restrict ourselves to gas configurations consisting of identical particles of positive rest mass $m > 0$. In this case, the momentum is confined to the future mass hyperboloid,
\begin{equation}
P_x^+(m) := \left\{p\in T_x^* \mathcal{M} \: : \: g_x^{-1}(p,p) = -m^2, \hbox{ the vector dual to $p$ is future directed} \right\}
\end{equation}
and $f$ is a function on the future mass shell
\begin{equation}
\Gamma_m^+ := \{ (x,p) : x\in \mathcal{M}, p\in P_x^+(m) \}.
\end{equation}
 
Due to phase space mixing (see~\cite{pRoS18,pRoS2020} and references therein), it is expected that a gas configuration consisting of purely bound particles (such that it is described by a DF supported in $\Gamma_{\textrm{bound}}$) relaxes in time to a stationary configuration which can be described by a one-particle DF $f$ depending only on the integrals of motion, that is, only on $(E,L_x,L_y,L_z)$. Notice that such a function $f$ automatically satisfies the collisionless Boltzmann equation~(\ref{Eq:Vlasov}) since $E,L_x,L_y,L_z$ Poisson commute with $\mathcal{H}$. In this article, we will assume, in addition, that $f$ depends only on $E$ and \emph{one} components of the angular momentum, say $L_z$, such that
\begin{equation}
f(x,p) = F(E,L_z),
\label{Eq:OneParticleDistributionFunction}
\end{equation}
for some suitable function $F$ that we specify further below. In particular, the ansatz~(\ref{Eq:OneParticleDistributionFunction}) implies that the gas configuration is stationary and axisymmetric. Provided that $f$ vanishes outside $\Gamma_{\textrm{bound}}$ and other suitable hypotheses hold, it has been shown recently~\cite{fJ2021,fJ2022} that such configurations can be made self-gravitating, provided their amplitude is sufficiently small.

To extract the physical content of our gas configurations, we compute the most relevant spacetime observables, that is, the particle current density $J$ and the energy-momentum-stress tensor $T$. In particular, these quantities contain the information about the particle density $n$, mean particle four-velocity, the energy density $\varepsilon$, heat flow, pressure tensor $P_{ij}$ and kinetic temperature. They are defined as follows (see, for instance, Refs.~\cite{CercignaniKremer-Book,rAcGoS2022})
\begin{equation}
J_\mu (x) := \int\limits_{P_x^+(m)} f(x,p) p_\mu \dvol_x(p), \qquad
T_{\mu \nu} (x) := \int\limits_{P_x^+(m)} f(x,p) p_\mu p_\nu \dvol_x(p), 
\label{Eq:CurrentDensityEnergyMomentumStress}
\end{equation}
where $\dvol_x(p)$ is the Lorentz-invariant volume form, which, in terms of the orthonormal basis~(\ref{Eq:OrthoBasis}) is defined as
\begin{equation}
\dvol_x(p) := \frac{dp_{\hat{1}}\wedge dp_{\hat{2}}\wedge dp_{\hat{3}} }{\sqrt{m^2 + p^2_{\hat{1}} + p^2_{\hat{2}} + p^2_{\hat{3}}}}
 = \frac{dp_{\hat{1}}\wedge dp_{\hat{2}}\wedge dp_{\hat{3}} }{|p_{\hat{0}}|}.
\label{Eq:VolumeForm}
\end{equation}
Any DF $f$ that satisfies the collisionless Boltzmann equation~(\ref{Eq:Vlasov}) and is positive implies that $J$ is future-directed timelike, that $T$ satisfies the standard energy conditions (strong, dominated, weak, null) and that $J$ and $T$ are divergence-free:
\begin{equation}
\nabla^\mu J_\mu = 0, \qquad \nabla^\mu T_{\mu\nu} = 0.
\end{equation}
In the next subsection, we rewrite $J_\mu$ and $T_{\mu\nu}$ in terms of integrals over the conserved quantities $E$, $L$ and $L_z$, such that these quantities can be evaluated more easily for any DF of the form~(\ref{Eq:OneParticleDistributionFunction}).

\subsection{Explicit representation of the observables in terms of integrals over the conserved quantities}
\label{SubSec:ExplicitObservables}

Re-expressing the fibre integrals in terms of the conserved quantities $E$, $L$ and $L_z$ involves two steps. The first one is to rewrite the volume form~(\ref{Eq:VolumeForm}) in terms of these quantities. The second is to determine the correct integration limits for them.

The first step is straightforward. Using Eqs.~(\ref{Eq:ConservedQuantities},\ref{Eq:Orthonormalbasis}) one obtains, fixing $m$, $r$ and $\vartheta$,
\begin{equation}
\dvol_x(p) = \frac{1}{r^2\sin\vartheta } \frac{dE\; LdL \; dL_z}{\sqrt{E^2 - V_{m,L}(r)} \sqrt{L^2 - L_z^2 /\sin^2\vartheta}}.
\end{equation}
The second step is more involved and has to be analyzed carefully. In order to determine the correct intervals over which $E$, $L$ and $L_z$ need to be integrated over, we make the following observations:
\begin{enumerate}
\item[(i)] As mentioned above, we restrict ourselves to DFs whose support lies in $\Gamma_{\textrm{bound}}$, that is the subset of relativistic phase space corresponding to bound orbits, which implies that $E_{\textrm{ms}} < E < m$ and $L_{\textrm{c}}(E) < L < L_{\textrm{ub}}(E)$.

\item[(ii)] Given the position $r = r_{\textrm{obs}}$ of the observer, one needs that $V_{m,L}(r) \leq E^2$ for $r$ to lie in the classically allowed region. This implies that the total angular momentum must satisfy the additional bound
\begin{equation}
L \leq L_{\textrm{max}}(E,r) := r\sqrt{\frac{E^2}{N(r)} - m^2}.
\end{equation}

\item[(iii)] A further restriction comes from the requirement that the radius of the observer $r = r_{\textrm{obs}}$ must lie between the turning points $r_1 < r_2$ of the bound orbit with given $E$ and $L$, that is, $r_1\leq r_{\textrm{obs}}\leq r_2$.

\item[(iv)] When $r_{\textrm{obs}} > r_{\textrm{ms}} = 6M$ the minimum possible value for the energy $E$, such that a bound trajectory satisfying $r_1 < r_{\textrm{obs}} < r_2$ exists, occurs when $r_{\textrm{obs}}$ coincides with the right turning point $r_2$ of the ISO, see figure~\ref{Fig:EffectivePotential}. As shown in Appendix~\ref{App:LimitsIntegration} (cf. Eq.~\ref{Eq:MinimumEnergy} and the right panel of figure~\ref{Fig:ParametrizationeP}) this implies that
\begin{equation}
E > E_{\textrm{c}}(r) := m\frac{r + 2M}{\sqrt{r\left(r + 6M\right)}},\qquad r > 6M.
\end{equation}
For $E\in (E_{\textrm{c}}(r),m)$ the lower bound $L = L_{\textrm{c}}(E)$ for the total angular momentum corresponds to the situation for which orbits with the same value of $E$ but slightly smaller value for $L$ plunge into the black hole.

\item[(v)] When $4M < r_{\textrm{obs}} < r_{\textrm{ms}}$ the minimum value for $E$ such that a bound trajectory satisfying $r_1 < r_{\textrm{obs}} < r_2$ exists occurs when $r_{\textrm{obs}}$ coincides with the position of the local maximum of the effective potential (see figure~\ref{Fig:EffectivePotential}), such that
\begin{equation}
E > E_{\textrm{c}}(r) := m\frac{r - 2M}{\sqrt{r\left(r - 3M\right)}},\qquad 4M < r < 6M.
\end{equation}
For $E\in (E_{\textrm{c}}(r),m)$ the lower bound for $L$ is again given by $L = L_{\textrm{c}}(E)$.

\item[(vi)] For $r = r_{\textrm{obs}} < r_{\textrm{mb}} = 4M$ there are not bound orbits.

\item[(vii)] Note that in both cases (iv) and (v), one has $L_{\textrm{max}}(E,r_{\textrm{obs}}) \leq L_{\textrm{ub}}(E)$. To prove this, consider a trajectory with $L = L_{\textrm{max}}(E,r_{obs})$, such that $r_{\textrm{obs}}$ is a turning point, and consider the effective potential $V_{m,L'}$ with $L' = L_{\textrm{ub}}(E)$, such that its local minimum is equal to $E^2$. Then, clearly, $V_{m,L'}(r_{\textrm{obs}}) > E^2$ which implies $L' > L_{\textrm{max}}(E,r_{\textrm{obs}})$. The limit $L' = L_{\textrm{max}}(E,r_{\textrm{obs}})$ occurs when $r_{\textrm{obs}}$ coincides with the position of the minimum of $V_{m,L'}$.

\item[(viii)] Finally, the range of $L_z$ is restricted by the requirement that $L^2\geq 0$, which yields $|L_z| \leq L\sin\vartheta$.
\end{enumerate}

\begin{figure}
\centerline{
\includegraphics[scale=0.5]{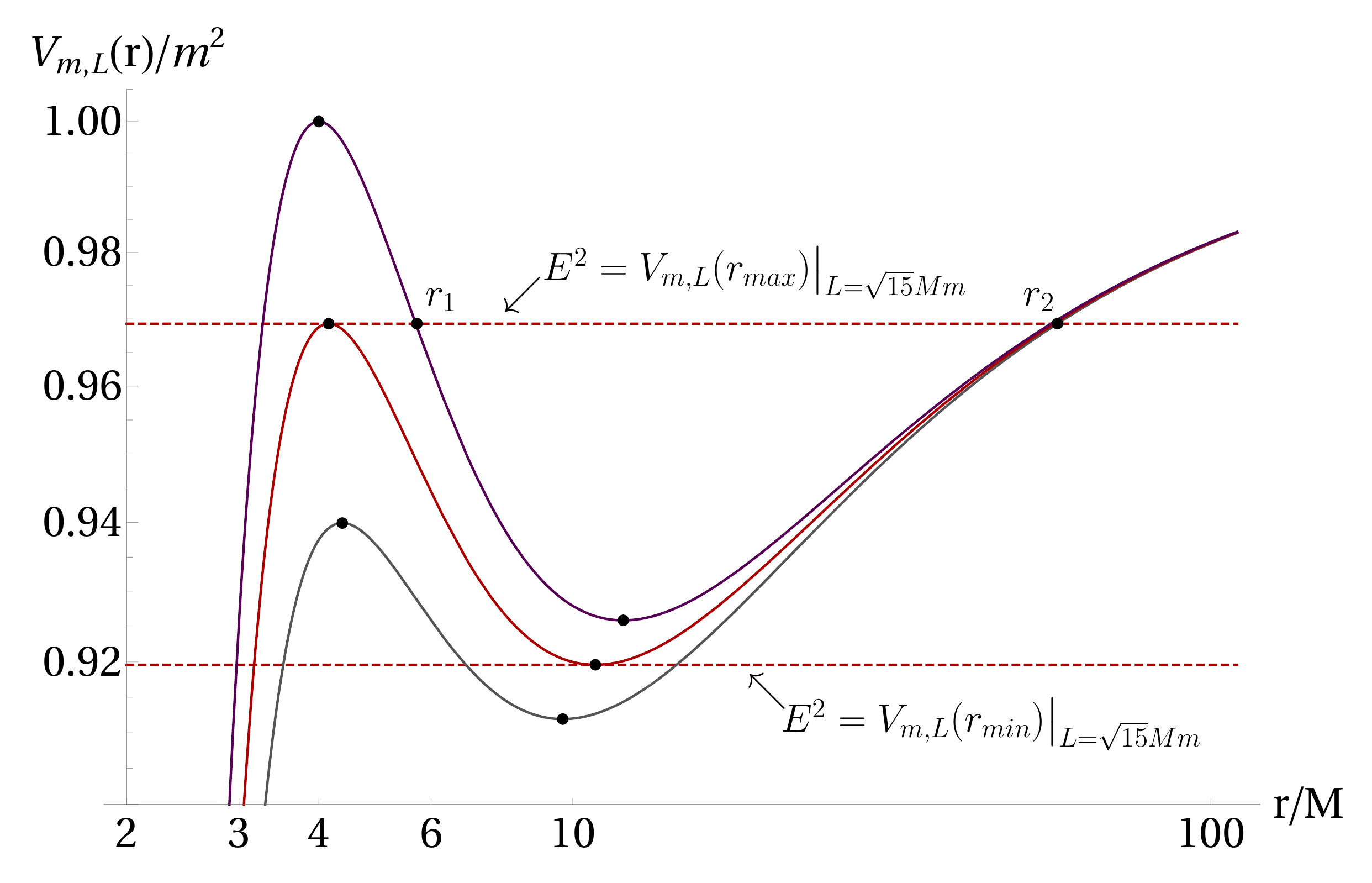}}
\caption{Behavior of effective potential vs radius for $L=\sqrt{14}Mm$ (gray), $L=\sqrt{15}Mm$ (red), and $L=4Mm$ (purple). $(r_1,r_2)$ are the turning points and the dashed (red) lines represent the energy level corresponding to the maximum and minimum values of the potential with $L = \sqrt{15} M m$.}
\label{Fig:EffectivePotential}
\end{figure}

\begin{figure}[h!]
\centerline{
\includegraphics[scale=0.28]{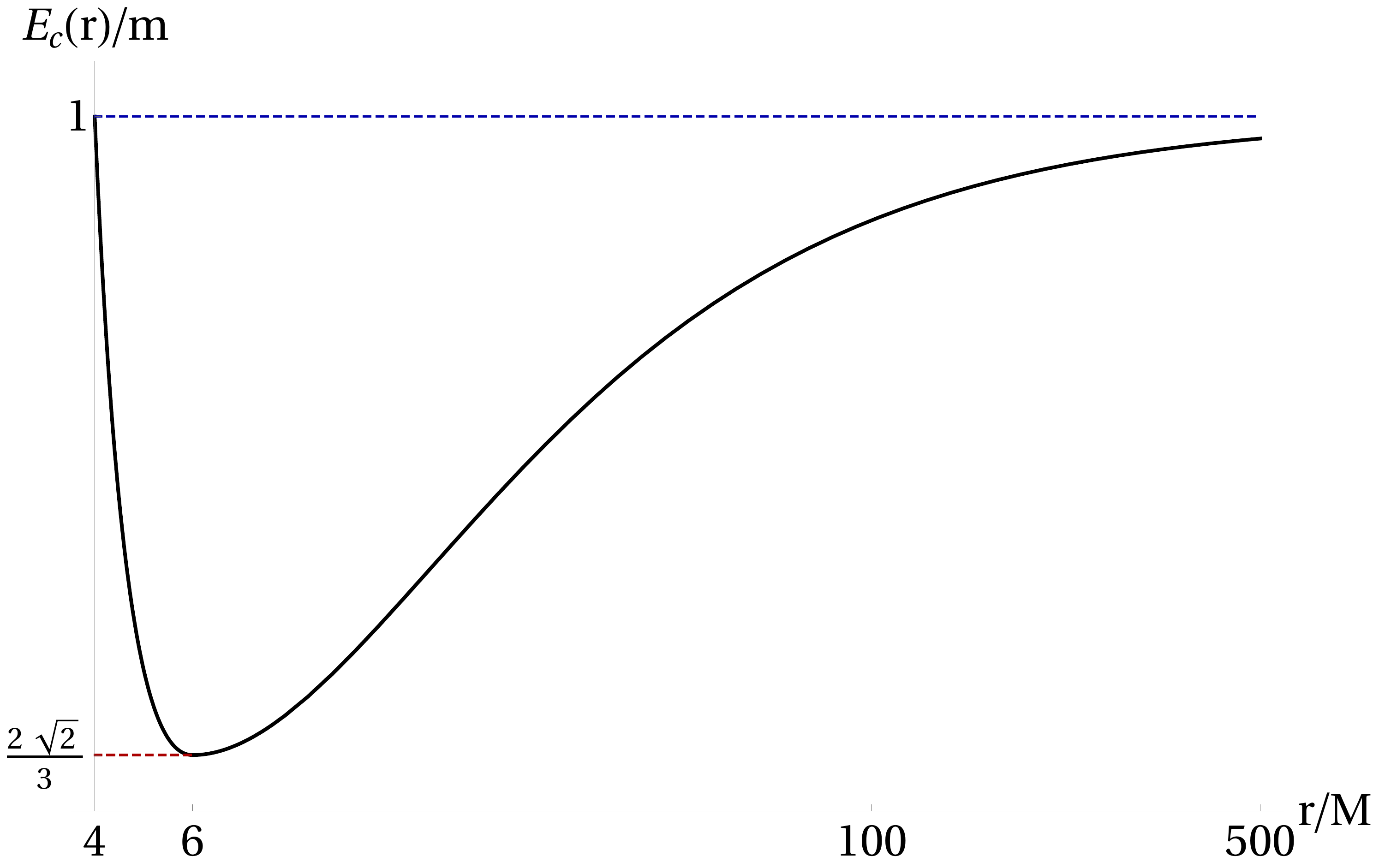}
\includegraphics[scale=0.28]{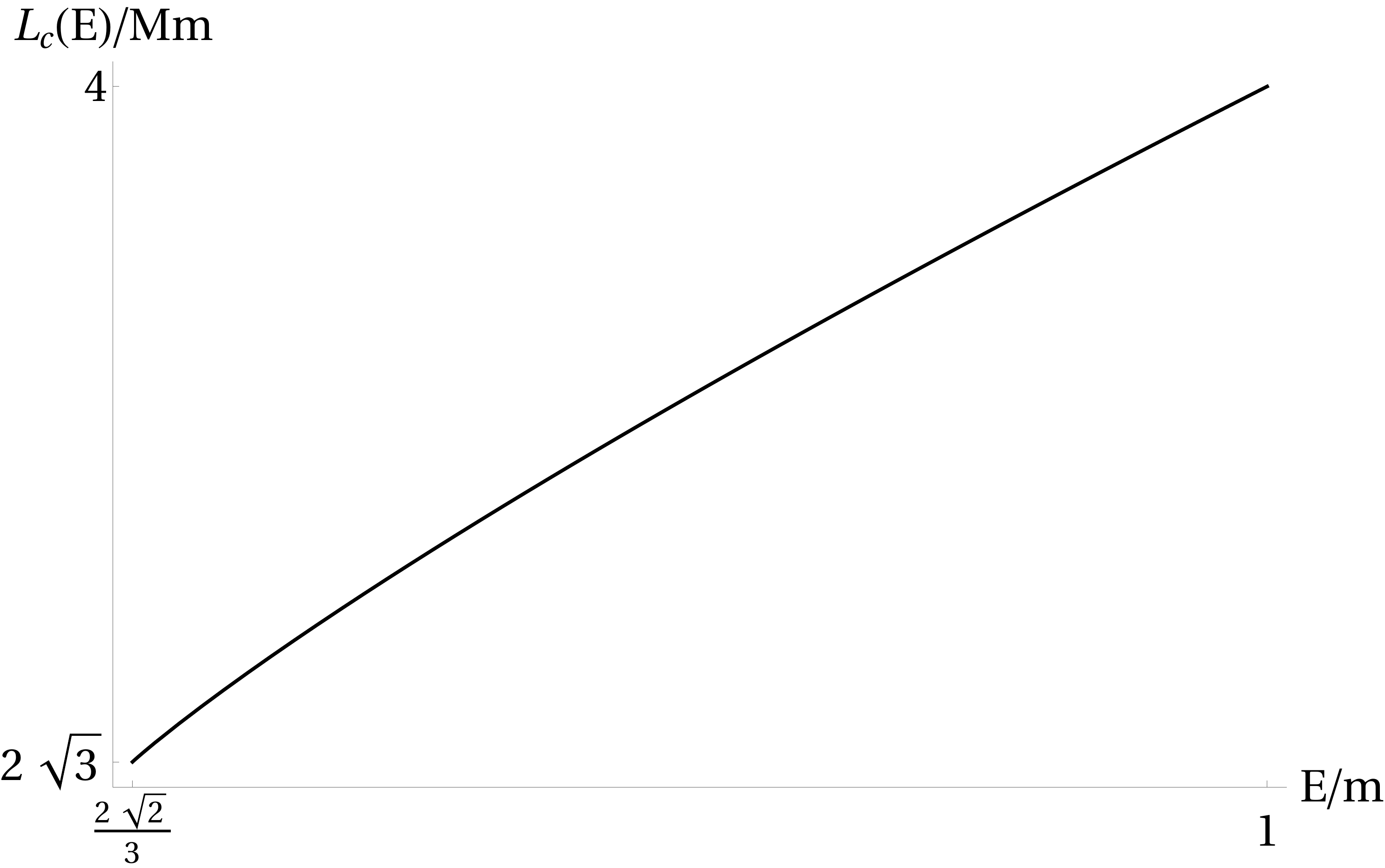}}
\caption{Behavior of the critical minimum energy $E_{\textrm{c}}$ vs radius (left) and the critical angular momentum $L_{\textrm{c}}$ vs energy (right).}
\label{Fig:MinimumEnergy}
\end{figure}

Based on these observations, we conclude that the relevant range of integration for the conserved quantities $(E,L,L_z)$ is given by
\begin{equation}
E_{\textrm{c}}(r) < E < m, \quad L_{\textrm{c}}(E) < L < L_{\textrm{max}}(E,r), \quad 
\hbox{and} \quad |L_z| < L\sin\vartheta,
\label{Eq:RangeIntegration}
\end{equation}
where
\begin{equation}
E_{\textrm{c}}(r) = \left\{
\begin{array}{lcl}
\displaystyle m\frac{r-2M}{\sqrt{r(r-3M)}}, && 4M \leq r \leq 6M, \\
 & & \\
\displaystyle m\frac{r + 2M}{\sqrt{r\left(r + 6M\right)}}, && r \geq 6M. 
\end{array}
\right.
\label{Eq:MinimumEnergy2}
\end{equation}
and
\begin{eqnarray}
L_{\textrm{c}}(E) &=& \frac{4\sqrt{2}Mm^3}{\sqrt{36m^2 E^2 - 8m^4 - 27E^4 + E\left( 9E^2 - 8m^2 \right)^{3/2} }}.
\label{Eq:AngularMomentumC}\\
L_{\textrm{max}}(E,r) &=& r\sqrt{\frac{E^2}{N(r)} - m^2}.
\label{Eq:MaximumAngularMomentum}
\end{eqnarray}
For an alternative derivation of these limits and further details, see Appendix~\ref{App:LimitsIntegration}. The behavior of the critical minimum energy $E_c(r)$ and the critical angular momentum $L_{\text{c}}(E)$ are shown in figure~\ref{Fig:MinimumEnergy}. Note that $E_c$ decreases from $m$ to $E_{ms}$ in the interval $[4M,6M]$ and increases from $E_{ms}$ to $m$ in the interval $[6M,\infty)$. Furthermore, this function is continuously differentiable at $r = 6M$. The function $L_{\text{c}}(E)$ increases monotonously from $L_{\text{ms}}$ to $L_{\text{mb}}$.

Now that the integration limits are understood, the explicit form for the particle current density and the energy-momentum-stress tensor in terms of conserved quantities can be given:
\begin{eqnarray}
J_{\hat{\mu}}(x) &=& \sum_{\epsilon_r,\epsilon_\vartheta = \pm1}\int\limits_{E_{\text{c}}(r)}^m \int\limits_{L_{\text{c}}(E)}^{L_{\text{max}}(E,r)} \int\limits_{-L\sin\vartheta}^{+L\sin\vartheta} f(x,p)\frac{p_{\hat{\mu}}(\epsilon_r,\epsilon_\vartheta)}{r^2 \sin\vartheta} \frac{dL_z \; LdL \; dE}{\sqrt{E^2 - V_{m,L}(r)} \sqrt{L^2 - L_z^2/\sin^2\vartheta}}, 
\label{Eq:Jhatmu}\\
T_{\hat{\mu}\hat{\nu}}(x)  &=& \sum_{\epsilon_r,\epsilon_\vartheta = \pm1}\int\limits_{E_{\text{c}}(r)}^m \int\limits_{L_{\text{c}}(E)}^{L_{\text{max}}(E,r)} \int\limits_{-L\sin\vartheta}^{+L\sin\vartheta} f(x,p)\frac{p_{\hat{\mu}}(\epsilon_r,\epsilon_\vartheta) p_{\hat{\nu}}(\epsilon_r,\epsilon_\vartheta)}{r^2 \sin\vartheta} \frac{dL_z \; LdL \; dE}{\sqrt{E^2 - V_{m,L}(r)} \sqrt{L^2 - L_z^2/\sin^2\vartheta}},
\label{Eq:Thatmunu}
\end{eqnarray}
where the functions $p_{\hat{\mu}}(\epsilon_r,\epsilon_\vartheta)$ are given by Eq.~(\ref{Eq:Orthonormalbasis}) and the effective potential $V_{m,L}$ is defined in Eq.~(\ref{Eq:SchEffectivePotential}). The expressions~(\ref{Eq:Jhatmu},\ref{Eq:Thatmunu}) are valid for any DF $f$ which decays sufficiently fast in the momentum space such that these integrals converge. In the following section, we further reduce these integrals under the assumption that $f(x,p)$ has the form~(\ref{Eq:OneParticleDistributionFunction}).

\section{Stationary, axisymmetric models}
\label{Sec:StatAxiModel}

In this section, we further reduce the expressions~(\ref{Eq:Jhatmu},\ref{Eq:Thatmunu}) for the current density and energy-momentum-stress tensor for the case that the DF has the form~(\ref{Eq:OneParticleDistributionFunction}), where we assume that $F(E,L_z)$ is given by a generalized polytropic ansatz as in Refs.~\cite{eAhAaL16,eAhAaL19}. Furthermore, we compute the total particle number, energy and angular momentum for these configurations and compare them with the analogous Newtonian models in paper I. For a related model in which the DF depends only on $E$ and $L_z/L$, see Ref.~\cite{cGoS2022a}.

\subsection{The relativistic $(E,L_z)$-models}
\label{SubSec:RelELzModel}

For the following, we assume that the function $F$ in Eq.~(\ref{Eq:OneParticleDistributionFunction}) has the product form
\begin{equation}
F(E,L_z) := F_0(E) \times I(L_z),
\label{Eq:DistributionFunctionProduct}
\end{equation}
with $F_0$ given by the polytropic ansatz
\begin{equation}
F_0(E) := \alpha\left(1- \frac{E}{E_0} \right)^{k-\frac{3}{2}}_+.
\label{Eq:PolytropeEnergy}
\end{equation}
Here, $\alpha$, $E_0$ are positive parameters, $k > 1/2$, and the notation $F_+$ refers to the positive part of the quantity $F$, that is $F_+ = F$ if $F > 0$ and $F_+ = 0$ otherwise. The cut-off parameter $E_0$ provides an upper bound for the energy. For $E_0 < m$, the resulting configurations have finite extend. However, in the results shown below we shall choose the limiting case $E_0 = m$ in which the configurations extend to infinity, similar to the Newtonian configurations constructed in paper I. In turn, the function $I(L_z) = I^{(\text{even},\text{rot})}_{\text{poly}}(L_z)$ is given by one of the following two functions:
\begin{equation}
I^{(\text{even})}(L_z) := \left(\frac{|L_z|}{L_0} - 1\right)^l_+
\label{Eq:PolytropeLzEven}
\end{equation}
or
\begin{equation}
I^{(\text{rot})}(L_z) := 2\left(\frac{L_z}{L_0} - 1\right)^l_+,
\label{Eq:PolytropeLzRot}
\end{equation}
with parameters $L_0$ and $l=0,1,2,\ldots$. Here, the superscript ``even" refers to the fact that $I(L_z)$ is an even function of $L_z$, meaning that for any particle orbiting in the positive sense around the black hole, there is a corresponding particle moving in the opposite direction with the same absolute value of $L_z$. As a consequence, the corresponding configurations are static and have vanishing total angular momentum. In contrast, the configurations denoted by the superscript ``rot" consist of particles with positive values of $L_z$, and hence they rotate. The cut-off parameter $L_0$ provides a lower bound for the absolute value of the azimuthal angular momentum; hence when $L_0 > 0$ the gas configuration vanishes in the vicinity of the $z$-axis. As follows from the analysis in paper I, the total particle number, energy and angular momentum for the infinitely extended configurations $E_0 = m$ are finite provided $2k > l + 7$. 

\subsection{Spacetime observables}
\label{SubSec:Observables}

Since the DF is independent of $L$, it is convenient to interchange the order of integration of $L$ and $L_z$ in Eq.~(\ref{Eq:Jhatmu}), which yields
\begin{equation}
J_{\hat{\mu}}(x) 
 = \sum_{\epsilon_r,\epsilon_\vartheta = \pm1}\int\limits_{E_{\text{c}}(r)}^m dE F_0(E)
\int\limits_{-L_{\text{max}}(E,r)\sin\vartheta}^{L_{\text{max}}(E,r)\sin\vartheta} dL_z I(L_z)
\int\limits_{\max\{ L_{\text{c}}(E), \frac{|L_z|}{\sin\vartheta} \}}^{L_{\text{max}}(E,r)}
\frac{dL L}{\sqrt{E^2 - V_{m,L}(r)} \sqrt{L^2 - L_z^2/\sin^2\vartheta}}
\frac{p_{\hat{\mu}}(\epsilon_r,\epsilon_\vartheta)}{r^2 \sin\vartheta}.
\end{equation}
Using the identity $E^2 - V_{m,L}(r) = N(r)[L_{\text{max}}(E,r)^2 - L^2]/r^2$, Eq.~(\ref{Eq:Orthonormalbasis}) and the formulae in Appendix~\ref{App:Integrals}, the integral over $L$ can be performed analytically, obtaining
\begin{eqnarray}
\left( \begin{array}{c} J_{\hat{0}} \\ \\ J_{\hat{3}} \end{array} \right)(x)
 &=& \frac{2\pi}{\sqrt{N}r\sin\vartheta}\int\limits_{E_{\text{c}}(r)}^m dE F_0(E)
\int\limits_{-L_{\text{max}}(E,r)\sin\vartheta}^{+L_{\text{max}}(E,r)\sin\vartheta} 
dL_z I(L_z)
\left( \begin{array}{r} -\frac{E}{\sqrt{N}} \\ \\ \frac{L_z}{r\sin\vartheta} \end{array} \right)
\nonumber\\
 &-& \frac{4}{\sqrt{N} r\sin\vartheta}\int\limits_{E_{\text{c}}(r)}^m dE F_0(E)
 \int\limits_{-L_{\text{c}}(E)\sin\vartheta}^{+L_{\text{c}}(E)\sin\vartheta} 
 dL_z I(L_z)\arctan\sqrt{\frac{L_{\text{c}}(E)^2 - L_z^2/\sin^2\vartheta}{L_{\text{max}}(E,r)^2 - L_{\text{c}}(E)^2}}
 \left( \begin{array}{r} -\frac{E}{\sqrt{N}} \\ \\ \frac{L_z}{r\sin\vartheta} \end{array} \right),
\label{Eq:Jhat03}
\end{eqnarray}
while the remaining components are zero: $J_{\hat{1}} = J_{\hat{2}} = 0$. Note that the contribution to $J_{\hat{0}}$ and $J_{\hat{3}}$ corresponding to the integrals in the first line are independent of $L_{\text{c}}(E)$ and very similar to the corresponding expression for the particle density $n$ in the Newtonian case (see Eq.~(38) in paper I). In contrast, the integrals on the second line depend on the critical angular momentum $L_{\text{c}}(E)$ which is related to the maximum of the potential, and thus they are due to relativistic effects. Similar expressions for $T_{\hat{\mu}\hat{\nu}}(x)$ are obtained following the same steps, starting from Eq.~(\ref{Eq:Thatmunu}).

The next step consists in computing the integrals over $L_z$ for the specific models~(\ref{Eq:PolytropeLzEven},\ref{Eq:PolytropeLzRot}). The integrals on the right-hand side of the first line of Eq.~(\ref{Eq:Jhat03}) are easily evaluated. For the corresponding integrals on the second line, it is convenient to introduce the shorthand notation
\begin{equation}
a = a(E,\vartheta) := \frac{L_{\text{c}}(E)\sin\vartheta}{L_0},\qquad
b = b(E,r) := \frac{L_{\text{c}}(E)}{L_{\text{max}}(E,r)}, \qquad L_z := a L_0 \lambda_z.
\label{Eq:abDef}
\end{equation}
Using integration by parts one finds, for instance,
\begin{equation}
\int\limits_{-L_{\text{c}}(E)\sin\vartheta}^{+L_{\text{c}}(E)\sin\vartheta} 
 dL_z I(L_z)\arctan\sqrt{\frac{L_{\text{c}}(E)^2 - L_z^2/\sin^2\vartheta}{L_{\text{max}}(E,r)^2 - L_{\text{c}}(E)^2}} = \frac{2L_0}{l+1} b\sqrt{1-b^2}\int\limits_{1/a}^1
  \frac{d\lambda_z \lambda_z}{\sqrt{1-\lambda_z^2}} 
 \frac{(a\lambda_z-1)_+^{l+1}}{1 - b^2\lambda_z^2},
\end{equation}
for both models~(\ref{Eq:PolytropeLzEven},\ref{Eq:PolytropeLzRot}). To write down the final result, it is convenient to introduce the functions
\begin{equation}
\tilde{\text{K}}_l(a,b) := \frac{2}{\pi}\sqrt{1 - b^2}\int\limits_{1/a}^1 \frac{d\lambda_z \lambda_z}{\sqrt{1-\lambda_z^2}} \frac{(a\lambda_z-1)_+^{l+1}}{1 - b^2\lambda_z^2},\qquad
a > 0,\quad 0 < b < 1,
\label{Eq:KtildeDef}
\end{equation}
and
\begin{equation}
\text{K}_l(a,b)
 := \frac{1}{l+1}\left[ \left( \frac{a}{b} - 1 \right)_+^{l+1} - b \tilde{\text{K}}_l(a,b) \right],\qquad
a > 0,\quad 0 < b < 1.
\label{Eq:KlDef}
\end{equation}
As shown in Appendix~\ref{App:IntegralKernel}, these functions are continuous in $(a,b)$ with $\tilde{\text{K}}_l(a,b)$ satisfying the bound $0\leq \tilde{\text{K}}_l(a,b)\leq (a-1)_+^{l+1}$ for all $a > 0$ and $0 < b < 1$, which implies $\text{K}_l(a,b)\geq 0$. In particular, $\tilde{\text{K}}_l(a,b)$ vanishes if $0 < a\leq 1$, that is, if $L_{\text{c}}(E)\sin\vartheta\leq L_0$, which is always the case if $L_0 \geq L_{\text{mb}} = 4M m$, see figure~\ref{Fig:MinimumEnergy}. Consequently, $\text{K}_l(a,b) = 0$ if $a \leq b\leq 1$, that is, if $L_{\text{max}}(E,r)\sin\vartheta\leq L_0$. Although we have not found a closed-form expression for the functions $\text{K}_l(a,b)$ for generic values of $l$, it is possible to obtain explicit expressions at least for $l=0,1,2$ using a symbolic integration package such as MAPLE or Wolfram Mathematica, see Appendix~\ref{App:IntegralKernel} for more details.

Gathering the results, one obtains
\begin{eqnarray}
&& J_{\hat{0}}^{(\textrm{even})}(x) = J_{\hat{0}}^{(\textrm{rot})}(x) = -\frac{4\pi L_0}{N(r) R}
\int\limits_{E_{\textrm{c}}(r)}^m dE E F_0(E) \text{K}_l(a,b),
\label{Eq:J0}\\
&& J_{\hat{3}}^{(\textrm{rot})}(x) = \frac{4\pi L_0^2}{\sqrt{N(r)} R^2}
\int\limits_{E_{\textrm{c}}(r)}^m dE F_0(E) \left[ \text{K}_{l+1}(a,b) + \text{K}_l(a,b) \right],
\label{Eq:J3}
\end{eqnarray}
while $J_{\hat{3}}^{(\textrm{even})}(x) = 0$, and the remaining orthonormal components of $J_{\hat{\mu}}$ are identically zero. Here, and in the following, $R := r\sin\vartheta$ refers to the cylindrical radius, and we recall the definitions of the quantities $a$ and $b$ in Eq.~(\ref{Eq:abDef}). The orthonormal components of the particle current density in Eqs.~(\ref{Eq:J0},\ref{Eq:J3}) determine the invariant particle density and four-velocity of the gas:
\begin{equation}
n := \sqrt{-J^{\hat{\mu}} J_{\hat{\mu}}} = \sqrt{J_{\hat{0}}^2 - J_{\hat{3}}^2},\qquad
u^{\hat{\mu}} = \frac{1}{n} J^{\hat{\mu}}.
\label{Eq:ParticleDensity}
\end{equation}

Repeating the calculations for the energy-momentum-stress tensor (see Appendix~\ref{App:Integrals} for further details) one obtains
\begin{eqnarray}
\left( \begin{array}{c} T_{\hat{0}\hat{0}} \\ \\ T_{\hat{0}\hat{3}} \\  \\ T_{\hat{3}\hat{3}} \end{array} \right)(x) &=& \frac{4}{\sqrt{N(r)} R}\left[ \frac{\pi}{2}\int\limits_{E_{\text{c}}(r)}^m dE F_0(E) \int\limits_{-L_{\text{max}}(E,r)\sin\vartheta}^{+L_{\text{max}}(E,r)\sin\vartheta} dL_z I(L_z) \left( \begin{array}{l} \frac{E^2}{N(r)} \\ \\ \frac{-E L_z}{\sqrt{N(r)} R} \\ \\ \frac{L_z^2}{R^2} \end{array} \right)\right.\nonumber\\
&-& \left.\int\limits_{E_{\text{c}}(r)}^m dE F_0(E) \int\limits_{-L_{\text{c}}(E)\sin\vartheta}^{+L_{\text{c}}(E)\sin\vartheta} dL_z I(L_z)\arctan\sqrt{\frac{L_{\text{c}}(E)^2 - L_z^2/\sin^2\vartheta}{L_{\text{max}}(E,r)^2 - L_{\text{c}}(E)^2}} \left( \begin{array}{l}\frac{E^2}{N(r)} \\ \\ \frac{-E L_z}{\sqrt{N(r)} R} \\ \\ 
\frac{L_z^2}{R^2} \end{array} \right)\right],
\label{Eq:T00T03T33}
\end{eqnarray}
and $T_{\hat{1}\hat{1}} = A - B$, $T_{\hat{2}\hat{2}} = A + B$ with
\begin{eqnarray}
A  &=& \frac{2}{\sqrt{N}r^2 R}\left\{ \frac{\pi}{2}\int\limits_{E_{\text{c}}(r)}^m dE F_0(E) \int\limits_{-L_{\text{max}}(E,r)\sin\vartheta}^{+L_{\text{max}}(E,r)\sin\vartheta} dL_z I(L_z) \left[ L_{\text{max}}(E,r)^2 - \frac{L_z^2}{\sin^2\vartheta} \right]  \right. \nonumber\\
&-& \left.\int\limits_{E_{\text{c}}(r)}^m dE F_0(E) \int\limits_{-L_{\text{c}}(E)\sin\vartheta}^{+L_{\text{c}}(E)\sin\vartheta} dL_z I(L_z)\left[ L_{\text{max}}(E,r)^2 - \frac{L_z^2}{\sin^2\vartheta} \right]\arctan\sqrt{\frac{L_{\text{c}}(E)^2 - L_z^2/\sin^2\vartheta}{L_{\text{max}}(E,r)^2 - L_{\text{c}}(E)^2}} \right\},
\label{Eq:A}\\
B &=& \frac{2}{\sqrt{N}r^2 R}\int\limits_{E_{\text{c}}(r)}^m dE F_0(E) \int\limits_{-L_{\text{c}}(E)\sin\vartheta}^{+L_{\text{c}}(E)\sin\vartheta} dL_z I(L_z) \sqrt{L_{\text{max}}(E,r)^2 - L_{\text{c}}(E)^2} \sqrt{  L_{\text{c}}(E)^2 -\frac{L_z^2}{\sin^2\vartheta}}.
\label{Eq:B}
\end{eqnarray}

For the specific model~(\ref{Eq:PolytropeLzRot}) one obtains, recalling the definitions of $a$, $b$, $\text{K}_l(a,b)$ and $\tilde{\text{K}}_l(a,b)$ in Eqs.~(\ref{Eq:abDef},\ref{Eq:KtildeDef},\ref{Eq:KlDef}):
\begin{eqnarray}
\label{Eq:T00}
T^{(\textrm{rot})}_{\hat{0}\hat{0}}(x) &=& \frac{4 \pi L_0}{\sqrt{N(r)^{3}} R} \int\limits_{E_{\text{c}}(r)}^m dE F_0(E) E^2 \text{K}_l (a,b), \\ 
\label{Eq:T03}
T^{(\textrm{rot})}_{\hat{0}\hat{3}}(x) &=& -\frac{4 \pi L_0^2}{N(r) R^2} \int\limits_{E_{\text{c}}(r)}^m dE F_0(E) E  \left[ \text{K}_l (a,b) + \text{K}_{l+1} (a,b) \right], \\
\label{Eq:T33}
T^{(\textrm{rot})}_{\hat{3}\hat{3}}(x) &=& \frac{4 \pi L_0^3}{\sqrt{N(r)} R^3} \int\limits_{E_{\text{c}}(r)}^m dE F_0(E) \left[ \text{K}_l(a,b) +2 \text{K}_{l+1}(a,b) + \text{K}_{l+2}(a,b)\right],\\
\label{Eq:T11}
T^{(\textrm{rot})}_{\hat{1}\hat{1}}(x) &=& \frac{2\pi L_0^3}{\sqrt{N(r)} R^3} \int\limits_{E_{\text{c}}(r)}^m dE F_0(E) \left[ \left(\frac{a^2}{b^2}-1\right) \text{K}_l(a,b) - 2 \text{K}_{l+1}(a,b) - \text{K}_{l+2}(a,b) - \frac{a^2}{b} \frac{\sqrt{1-b^2}}{l+1} \tilde{\text{K}}_l(a,0) \right],\\
\label{Eq:T22}
T^{(\textrm{rot})}_{\hat{2}\hat{2}}(x) &=& \frac{2\pi L_0^3}{\sqrt{N(r)} R^3} \int\limits_{E_{\text{c}}(r)}^m dE F_0(E) \left[ \left(\frac{a^2}{b^2}-1\right) \text{K}_l(a,b) - 2 \text{K}_{l+1}(a,b) - \text{K}_{l+2}(a,b) + \frac{a^2}{b} \frac{\sqrt{1-b^2}}{l+1} \tilde{\text{K}}_l(a,0) \right].
\end{eqnarray}
Here, the function $\tilde{\text{K}}_l(a,0)$ can be expressed in terms of hypergeometric functions ${}_{2}F_1(a,b;c; z)$, see Appendix~\ref{App:IntegralKernel} for the explicit form. For the even model~(\ref{Eq:PolytropeLzEven}) one finds $T^{(\textrm{even})}_{\hat{\mu}\hat{\nu}} = T^{(\textrm{rot})}_{\hat{\mu}\hat{\nu}}$ for the diagonal elements $(\hat{\mu}\hat{\nu}) = (\hat{0}\hat{0},\hat{1}\hat{1},\hat{2}\hat{2},\hat{3}\hat{3})$, whereas all non-diagonal components (including $(\hat{\mu}\hat{\nu}) = (03)$) are zero. The orthonormal components of the energy-momentum-stress tensor allow one to construct the energy density $\varepsilon$ and principal pressures $P_{\hat{r}}$, $P_{\hat{\vartheta}}$, $P_{\hat{\varphi}}$, which can be determined by diagonalizing $T^{\hat{\mu}}{}_{\hat{\nu}}$ \cite{Synge2-Book,oStZ13}. More specifically, $-\varepsilon$ is the eigenvalue of $T^{\hat{\mu}}{}_{\hat{\nu}}$ corresponding to the timelike eigenvector, and $P_{\hat{r}}$, $P_{\hat{\vartheta}}$, $P_{\hat{\varphi}}$ are the eigenvalues belonging to the spacelike ones. Explicitly, this yields
\begin{eqnarray}
\varepsilon &=& \frac{1}{2}\left[T_{\hat{0}\hat{0}} - T_{\hat{3}\hat{3}} + \sqrt{ \left(T_{\hat{3}\hat{3}} + T_{\hat{0}\hat{0}}\right)^2 - 4 T_{\hat{0}\hat{3}}^2 }\right], \\ 
P_{\hat{r}} &=& T_{\hat{1}\hat{1}}, \\ 
P_{\hat{\vartheta}} &=& T_{\hat{2}\hat{2}}, \\
P_{\hat{\varphi}} &=& \frac{1}{2}\left[ -T_{\hat{0}\hat{0}} + T_{\hat{3}\hat{3}} + \sqrt{ \left(T_{\hat{3}\hat{3}} + T_{\hat{0}\hat{0}}\right)^2 - 4 T_{\hat{0}\hat{3}}^2 }\right].
\end{eqnarray}
For the even model~(\ref{Eq:PolytropeLzEven}) one has $T^{(\text{even})}_{\hat{0}\hat{3}} = 0$ and the expressions above simplify to $\varepsilon = T_{\hat{0}\hat{0}}$ and $P_{\hat{\varphi}} = T_{\hat{3}\hat{3}}$, as expected since in this case the energy-momentum-stress tensor is diagonal.

We end this subsection by observing that the support of the configuration is determined by those values of $(r,\vartheta)$ for which $L_{\text{max}}(E,r)\sin\vartheta\geq L_0$ for some $E\in [E_{\text{c}}(r),m]$ in the allowed range (see the comments below Eq.~(\ref{Eq:KlDef})). Since for fixed $r$, $L_{\text{max}}(E,r)$ is an increasing function of $E$, we conclude that the support of the configuration is delimited by
\begin{equation}
L_{\text{max}}(m,r)\sin\vartheta\geq L_0,\qquad r\geq 4M,
\end{equation}
or
\begin{equation}
\frac{1}{N(r)} - 1 - \frac{L_0^2}{m^2 r^2 \sin^2\vartheta}\geq 0,\qquad r\geq 4M,
\label{Eq:SupportRel}
\end{equation}
which, in the Newtonian limit, reduces to the expression in Eq.~(51) of paper I with the dimensionless Kepler potential $\psi(r) = M m/(E_0 r)$. However, in stark contrast to the nonrelativistic case, the inner boundary of the configuration cannot lie arbitrarily close to the central object even if $L_0$ is small. The minimal value $r = 4M$ is due to the presence of the ISOs which are absent in the Newtonian case.

\subsection{Total particle number, energy and angular momentum of the kinetic gas cloud}
\label{SubSubSec:MassEnergyAngularMomKineticGas}

In this subsection we derive expressions for the total particle number, energy and angular momentum of the gas configuration. These are the conserved quantities associated with the divergence-free currents $J^\mu$, $J_{E}^\mu := -T^\mu{}_\nu k^\nu$ and $J_{L}^\mu := T^\mu{}_\nu v^\nu$, respectively, with $k = \partial_t$ and $v = \partial_\varphi$ the timelike and azimuthal Killing vector fields of the Schwarzschild metric. Denoting by $S$ a spacelike Cauchy surface and by $\hat{n}$ the associated future-directed unit normal, the corresponding conserved quantities are
\begin{equation}
N_{\textrm{gas}} = -\int\limits_S J^\mu\hat{n}_\mu \eta_S,\qquad
E_{\textrm{gas}} = -\int\limits_S J_E^\mu\hat{n}_\mu \eta_S,\qquad
J_{\textrm{gas}} = -\int\limits_S J_L^\mu\hat{n}_\mu \eta_S,
\label{Eq:NgasEgasJgas}
\end{equation}
with $\eta_S$ the induced volume form on $S$. Choosing without loss of generality $S$ to be a hypersurface of constant (Schwarzschild) time $t$ with spatial coordinates $(x^1,x^2,x^3)$, the first integral can be rewritten as~\cite{rAcGoS2022}
\begin{equation}
N_{\textrm{gas}} = \int\limits_{\Real^6} f(x,p_\mu dx^\mu) dx^1 dx^2 dx^3 dp_1 dp_2 dp_3,
\label{Eq:NumberOfParticles2}
\end{equation}
where $x$ is the manifold point with local coordinates $(t,x^1,x^2,x^3)$ and $p_\mu dx^\mu\in P_x^+(m)$ is the momentum covector on the future mass shell with spatial coordinates $(p_1,p_2,p_3)$. To make further progress, it is very useful to transform the spatial phase space coordinates $(x^i,p_i)$ to action-angle variables $(\mathcal{Q}^i,\mathcal{J}_i)$, see for example~\cite{pRoS2020}. Since the the transformation $(x^i,p_i)\mapsto (\mathcal{Q}^i,\mathcal{J}_i)$ is symplectic, the volume form transforms trivially, and one obtains
\begin{equation}
N_{\textrm{gas}} 
 = \int\limits_{\Omega_J}\int\limits_{\Torus^3} f(x,p) 
d\mathcal{Q}^1 d\mathcal{Q}^2 d\mathcal{Q}^3
d\mathcal{J}_1 d\mathcal{J}_2 d\mathcal{J}_3,
\label{Eq:NumberOfParticles3}
\end{equation}
with $\Omega_J\subset\Real^3$ the range of the action variables $\mathcal{J}_i$. In the models considered in this article the DF $f$ only depends on the integrals of motion and hence only on $\mathcal{J}_i$. Therefore, the integral over the angle variables $\mathcal{Q}^i$ is trivial and yields the simple factor $(2\pi)^3$ corresponding to the volume of the three-torus $\Torus^3$. On the other hand, the action variables can be expressed in terms of the conserved quantities $(E,L,L_z)$, and one can show that
\begin{equation}
d\mathcal{J}_1 d\mathcal{J}_2 d\mathcal{J}_3 = \frac{1}{2\pi} T_r(E,L) dE dL dL_z,
\end{equation}
where $T_r(E,L)$ denotes the period function for the radial motion (see~\cite{pRoS2020} for details). This yields
\begin{equation}
N_{\textrm{gas}} = 4\pi^2 \int\limits_{\Omega} F(E,L_z) T_r(E,L) dE dL dL_z,
\label{Eq:NumberOfParticles4}
\end{equation}
where $\Omega$ is the range for $(E,L,L_z)$ corresponding to $\Omega_J$. Similarly, one obtains the following expressions for the total energy and angular momentum:
\begin{equation}
E_{\textrm{gas}} = 4\pi^2 \int\limits_{\Omega} E F(E,L_z) T_r(E,L) dE dL dL_z,\qquad
J_{\textrm{gas}} = 4\pi^2 \int\limits_{\Omega} L_z F(E,L_z) T_r(E,L) dE dL dL_z.
\label{Eq:EJgas}
\end{equation}

The period function can be expressed analytically in terms of Legendre's elliptic integrals and the roots $r_0 < r_1 < r_2$ of the cubic equation $r^3(E^2 - V_{m,L}(r)) = 0$, with $r_1$ and $r_2$ the turning points (see Appendix~\ref{App:Parametrization} for more details). In terms of the dimensionless quantities
\begin{equation}
\varepsilon := \frac{E}{m}, \quad \varepsilon_0 := \frac{E_0}{m}, \quad \lambda := \frac{L}{M m}, \quad \lambda_0 := \frac{L_0}{M m}, \quad \xi := \frac{r}{M},
\label{Eq:DimensionlessQuantities}
\end{equation}
the period function has the form (cf. Appendix in~\cite{pRoS18})
\begin{equation}
T_r(\varepsilon,\lambda) = 2 M \varepsilon \left[ \mathbb{H}_2 - \mathbb{H}_0 \right],
\label{Eq:PeriodFunction2}
\end{equation}
where 
\begin{eqnarray}
\mathbb{H}_0 &:=& -\sqrt{\frac{\xi_{012} }{2\xi_1(\xi_2 -\xi_0)}} \left[ (\xi_0 \xi_{012} - \xi_1 \xi_2) \mathbb{F}(\kappa) + \xi_1(\xi_2 - \xi_0 )\mathbb{E}(\kappa) + \xi_{012} (\xi_1 - \xi_0)\Pi\left( b^2, \kappa \right) \right], \\
\mathbb{H}_2 &:=& \sqrt{\frac{8\xi_{012} }{\xi_1(\xi_2 -\xi_0)}} \left[ \frac{\xi_0^2}{\xi_0-2} \mathbb{F}(\kappa) + (\xi_1 -\xi_0)\Pi\left(b^2, \kappa \right) -\frac{4(\xi_1 -\xi_0)}{(\xi_1-2)(\xi_0-2)}\Pi\left(\beta^2, \kappa\right) \right].
\end{eqnarray}
Here, $\mathbb{F}(\kappa)$, $\mathbb{E}(\kappa)$, and $\Pi(b^2,\kappa)$ are Legendre's complete elliptic integrals of the first, second and third kind, respectively, as defined in~\cite{DLMF}. Further, we have abbreviated $\xi_{012} := \xi_0 + \xi_1 + \xi_2$ and have defined
\begin{equation}
b := \sqrt{\frac{\xi_2 -\xi_1}{\xi_2 -\xi_0}},\quad 
\kappa := \sqrt{\frac{\xi_0}{\xi_1}}b, \quad
\beta := \sqrt{\frac{\xi_0 - 2}{\xi_1 - 2}} b.
\label{Eq:bkappabeta}
\end{equation}
Using the DF~(\ref{Eq:DistributionFunctionProduct}) for the models~(\ref{Eq:PolytropeEnergy}-\ref{Eq:PolytropeLzRot}) and taking into account that $T_r(E,L)$ is independent of $L_z$, the integral over $L_z$ in Eq.~(\ref{Eq:NumberOfParticles4}) can be computed explicitly, and one obtains for both the (rot) and (even) models
\begin{equation}
\frac{N_{\textrm{gas}}}{M^3 m^3\alpha} = 4(2\pi)^2 \frac{\lambda_0}{l+1} \int\limits_{\varepsilon_{\text{min}}}^1 d\varepsilon \; \varepsilon\left(1-\frac{\varepsilon}{\varepsilon_0} \right)^{k-\frac{3}{2}}_+ \int\limits_{\lambda_{\text{c}}(\varepsilon)}^{\lambda_{\text{ub}}(\varepsilon)} d\lambda\; (\mathbb{H}_2-\mathbb{H}_0) \left(\frac{\lambda}{\lambda_0} -1 \right)^{l+1}_+.
\label{Eq:NumberParticles}
\end{equation}
The remaining two integrals on the right-hand side of Eq.~(\ref{Eq:NumberParticles}) are computed numerically. For this, it is very useful to perform a further change of variables from $(\varepsilon,\lambda)$ to $(p,e)$ which generalize the "semi-latus rectum" and eccentricity to the Schwarzschild case (see~\cite{wS02,jBmGtH15}). These new variables are related to the turning points through
\begin{equation}
\xi_1 = \frac{p}{1+e}, \quad \xi_2 = \frac{p}{1-e}.
\end{equation}
The main advantage of this transformation is the fact that it maps the region of integration to the simpler region $0 < e < 1$ and $p > 6 + 2e$. Furthermore, the third root $\xi_0$ and the dimensionless energy and angular momentum can be expressed explicitly in terms of $(p,e)$ as
\begin{equation}
\xi_0 = \frac{2p}{p-4},\quad
\varepsilon = \sqrt{\frac{(p-2)^2 - 4e^2 }{p\left( p - e^2 - 3\right)}},\quad
\lambda = \frac{p}{\sqrt{p - e^2 - 3}},
\qquad p > 6 + 2e,
\end{equation}
see Eqs.~(\ref{Eq:ParametrizationeP},\ref{Eq:EnergyAngularMomentum}) in Appendix~\ref{App:Parametrization}. Further, using Eqs.~(\ref{Eq:EnergyAngularMomentum},\ref{Eq:Elements}) one obtains
\begin{equation} 
\varepsilon d\varepsilon d\lambda = \frac{e \left[(p - 6)^2 - 4 e^2 \right]}{2p \sqrt{(p - e^2 - 3 )^5 } } de dp.
\end{equation}
Gathering these results yields the final expression for the total particle number:
\begin{equation}
\frac{N_{\textrm{gas}}}{M^3 m^3\alpha} = \frac{8\pi^2\lambda_0}{l+1}
 \int\limits_0^1 de e 
  \int\limits_{6+2e}^{\infty} dp \frac{(p - 6)^2 - 4 e^2}{p\sqrt{(p - e^2 - 3 )^5 } }
  \left(1-\frac{\varepsilon}{\varepsilon_0} \right)^{k-\frac{3}{2}}_+ \left(\frac{\lambda}{\lambda_0} -1 \right)^{l+1}_+ (\mathbb{H}_2-\mathbb{H}_0).
\label{Eq:NumberParticleseP}
\end{equation}
The corresponding expression for the total energy $E_{\textrm{gas}}$ is obtained from this by adding the factor $E = m\varepsilon$ in the integrand on the right-hand side of Eq.~(\ref{Eq:NumberParticleseP}). The total angular momentum is obtained from a similar calculation, starting from Eq.~(\ref{Eq:EJgas}), and yields
\begin{equation}
\frac{J^{\textrm{(rot)}}_{\textrm{gas}}}{M^4 m^4\alpha} = \frac{8\pi^2 \lambda_0^2}{l+1} 
 \int\limits_0^1 de e 
  \int\limits_{6+2e}^{\infty} dp \frac{(p - 6)^2 - 4 e^2}{p\sqrt{(p - e^2 - 3 )^5 } }
\left(1-\frac{\varepsilon}{\varepsilon_0} \right)^{k-\frac{3}{2}}_+ 
 \left[\left(\frac{\lambda}{\lambda_0} -1 \right)^{l+1}_+ + \frac{l+1}{l+2}\left(\frac{\lambda}{\lambda_0} -1 \right)^{l+2}_+\right](\mathbb{H}_2-\mathbb{H}_0),
\label{Eq:TotalAzimuthalAngularMomentum}
\end{equation}
whereas $J^{\textrm{(even)}}_{\textrm{gas}} = 0$.

These expressions simplify considerably in the Newtonian limit, in which $\lambda_0 \gg 1$. Since $\lambda \sim \sqrt{p}$ for large values of $\lambda$, this implies that $p\gg 1$, and using the asymptotic expressions
\begin{eqnarray}
&& \xi_0 = 2 + \frac{8}{p} + {\cal O}(p^{-2}),\\
&& b =  \sqrt{\frac{2e}{1+e}}\left[ 1 + {\cal O}(p^{-1}) \right], \quad
\kappa = 2\sqrt{\frac{e}{p}}\left[ 1 + {\cal O}(p^{-1}) \right],\quad
\beta = \frac{4\sqrt{e}}{p}\left[ 1 + {\cal O}(p^{-1}) \right],\\
&& \varepsilon = 1 - \frac{1-e^2}{2p} + {\cal O}(p^{-2}),\quad
\mathbb{H}_2 - \mathbb{H}_0 
 = \pi\left( \frac{p}{1-e^2} \right)^{3/2}\left[ 1 + {\cal O}(p^{-1}) \right],
\end{eqnarray}
one finds, choosing $\varepsilon_0 = E_0/m = 1$,
\begin{eqnarray}
\frac{N_{\textrm{gas}}}{M^3 m^3\alpha} &\simeq & \frac{\pi^3}{2^{k-\frac{9}{2}}} \frac{1}{(l+1)\lambda_0^l} \int\limits_0^1 de \; e (1-e^2)^{k-3} 
\int\limits_{\lambda_0^2}^{\infty} \frac{dp}{p^{k-\frac{3}{2}}} \left(\sqrt{p} - \lambda_0 \right)^{l+1}_+ \nonumber\\ 
&=& \frac{\pi^3}{2^{k-\frac{11}{2}}}\frac{\Gamma(l+1)\Gamma(2k-l-6)}{\Gamma(2k-3)}\frac{1}{\lambda_0^{2k-6}}.
\label{Eq:NumberParticlesLargeLambda} 
\end{eqnarray}
This agrees with the corresponding expression for the total rest mass of the Newtonian model, see Eq.~(C7) in paper I with $\chi=0$, $E_0 = m c^2$ and taking into account that in natural units $M$ should be replaced with the gravitational radius $r_g := GM/c^2$ and $m$ with $m c$ in the left-hand side of Eq.~(\ref{Eq:NumberParticlesLargeLambda}). Likewise, the internal energy, $E_{\textrm{gas}} - m N_{\textrm{gas}}$, reduces to the corresponding expression in the Newtonian limit for large values of $\lambda_0$,
\begin{equation}
\frac{E_{\textrm{gas}} - m N_{\textrm{gas}}}{M^3 m^4\alpha} 
 \simeq -\frac{\pi^3}{2^{k-\frac{9}{2}}}\frac{\Gamma(l+1)\Gamma(2k-l-4)}{\Gamma(2k-1)}\frac{1}{\lambda_0^{2k-4}},
\label{Eq:EnergyLargeLambda} 
\end{equation} 
see Eq.~(C8) in paper I.

\begin{figure}[h!]
\centerline{
\includegraphics[scale=0.28]{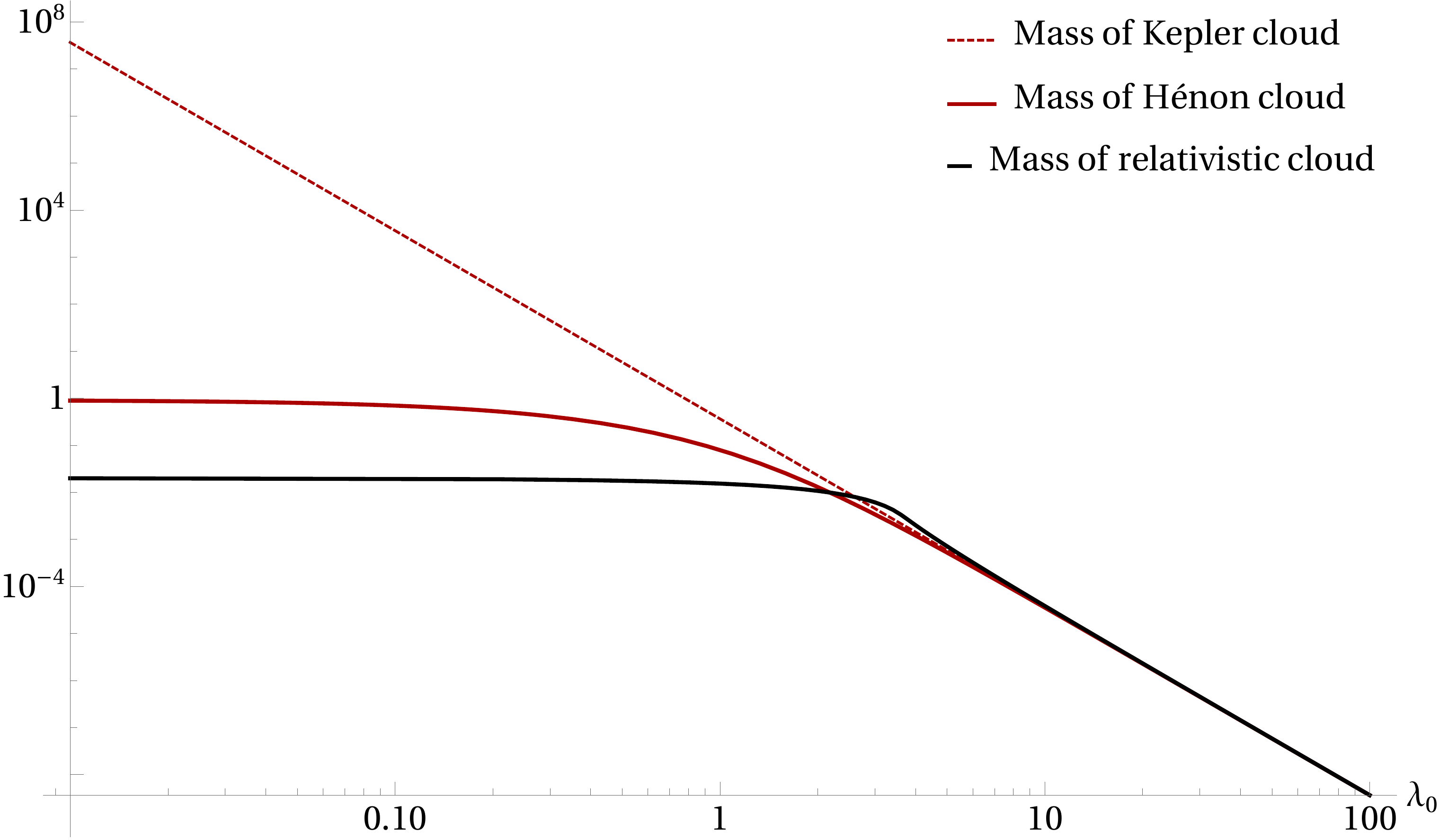}
\includegraphics[scale=0.28]{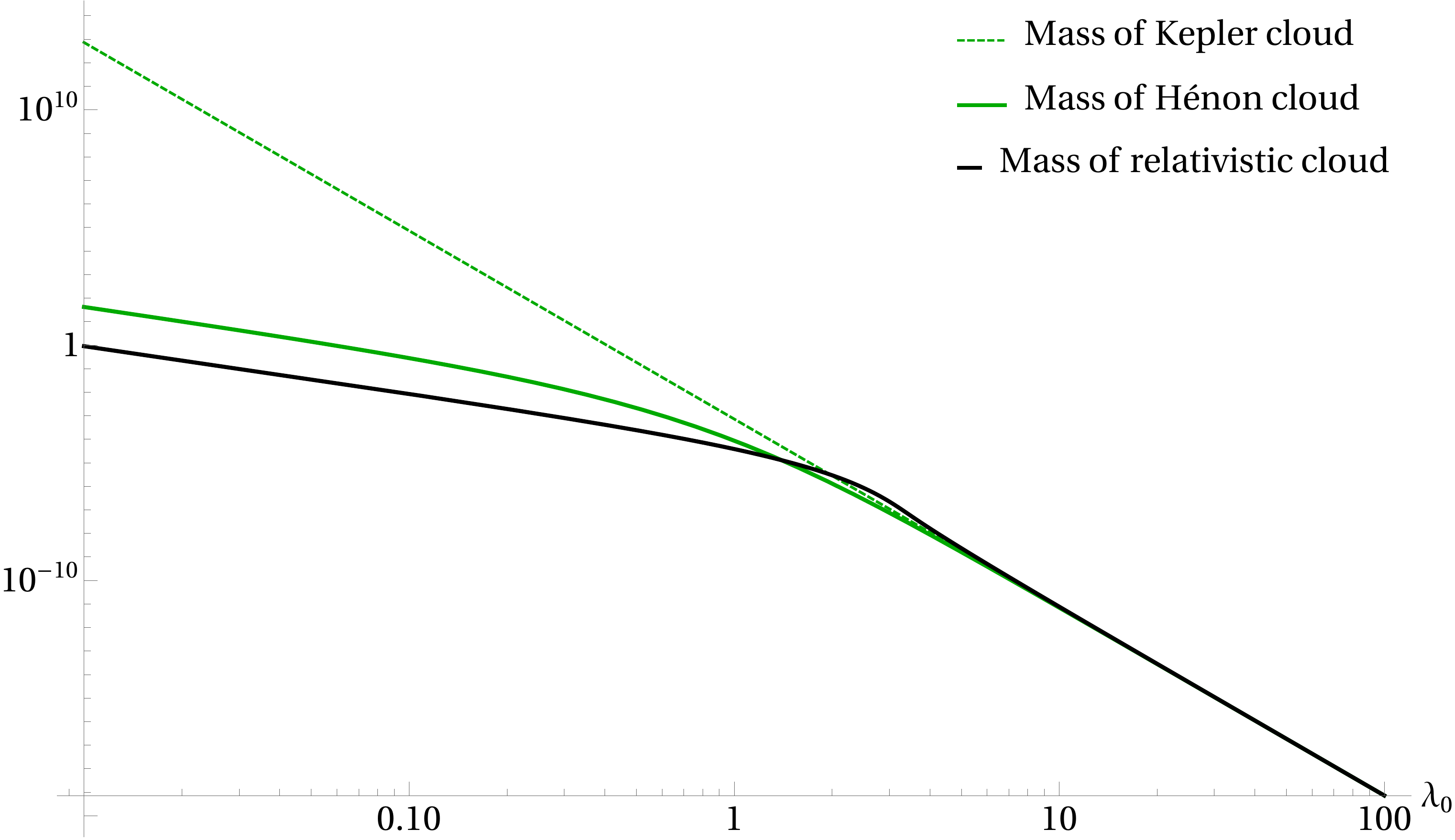}}
\caption{Comparison between the total mass in the Newtonian case with the Kepler ($\kappa=0$) and isochrone ($\kappa=1$) potentials and the total rest mass $m N_{\text{gas}}$ in the relativistic case as a function of $\lambda_0$ for different values of $(k,l)$. In all cases, the total mass is normalized by the factor $\alpha r_g^3 m^4 c^3$. Left panel: $(k,l)=(5,0)$. Right panel: $(k,l)=(7,2)$.}
\label{Fig:ComparisonTotMassTotNumberParticles}
\end{figure}

\begin{figure}[h!]
\centerline{
\includegraphics[scale=0.28]{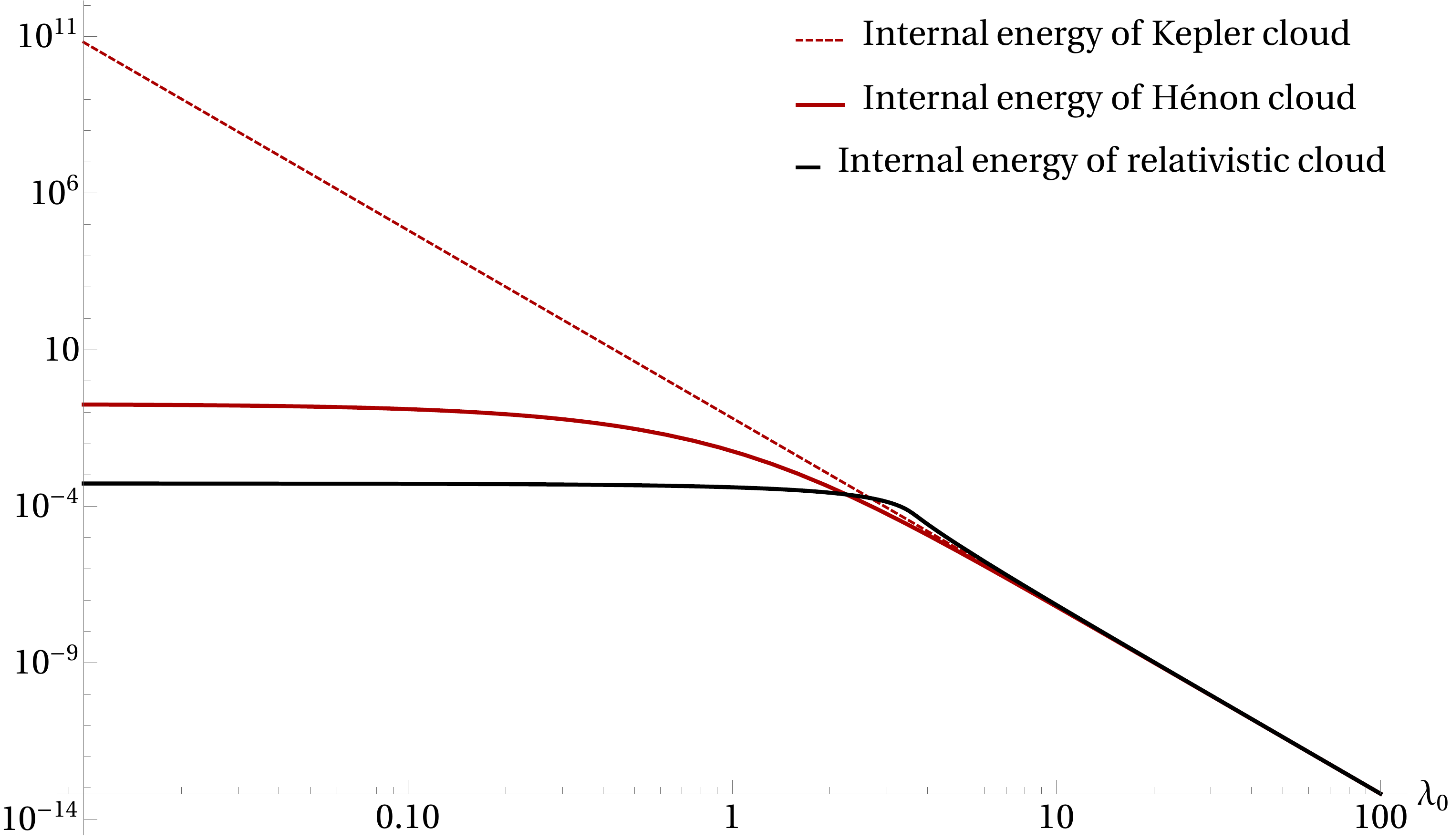}
\includegraphics[scale=0.28]{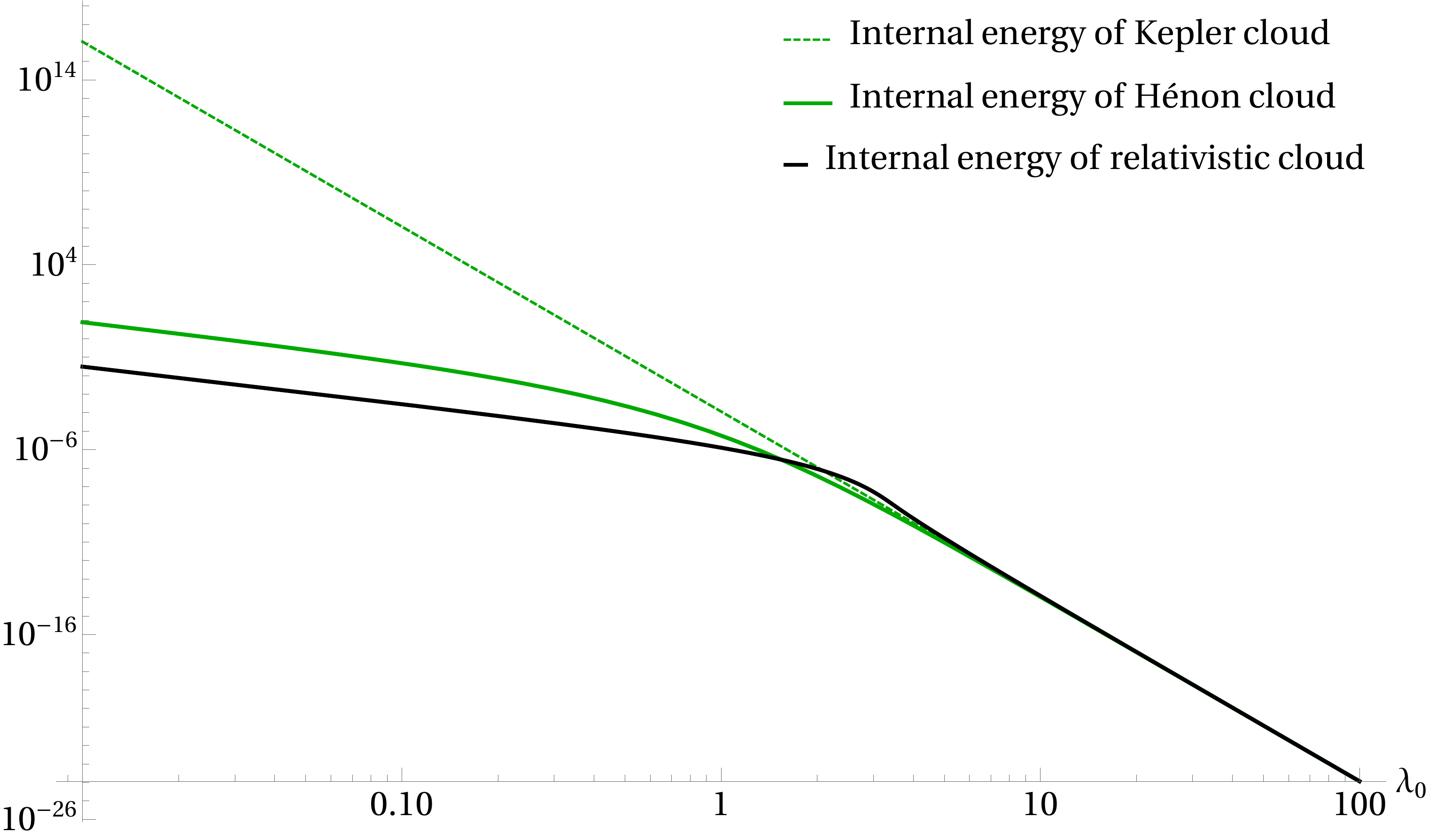}}
\caption{Comparison between the (absolute value of the) internal energy for the same configurations as in the previous figure. In all cases, the energy is normalized by the factor $\alpha r_g^3 m^4 c^5$.}
\label{Fig:ComparisonEnergy}
\end{figure}

In order to analyze the differences between the relativistic and non-relativistic cases, we evaluate numerically the expression in Eq.~(\ref{Eq:NumberParticleseP}) for the total particle number and compare it with the analogous expression from Eq.~(C7) in paper I with $E_0 = m c^2$ for the Newtonian model in the Kepler and isochrone potentials (the latter being regular at the center). This comparison is shown in figure~\ref{Fig:ComparisonTotMassTotNumberParticles} for different values of the parameters $(k,l)$ and the parameter values $\kappa = 0$ (corresponding to the Kepler potential) and $\kappa = 1$ characterizing the isochrone potential. For large values of $\lambda_0$, the three descriptions show the same behavior, as expected. However, the relativistic configurations have a smaller total rest mass than their non-relativistic counterparts for $\lambda_0\lesssim 1$ while their mass is slightly larger for some intermediate values of $\lambda_0$ lying in a small interval between $1$ and $10$.
Similarly, in figure~\ref{Fig:ComparisonEnergy} we provide a comparison between the internal energy $E_{\textrm{gas}} - m N_{\textrm{gas}}$ in the relativistic and Newtonian cases. As in the total mass case, we note that the three descriptions agree with each other when $\lambda_0$ is large, while for small values of $\lambda_0$ the relativistic internal energy has the smallest absolute value.

In the next section we discuss the properties of the spacetime observables associated with the relativistic gas configurations, keeping fixed the total particle number computed from Eq.~(\ref{Eq:NumberParticleseP}).

\section{Properties of the spacetime observables}
\label{Sec:Examples}

In this section we discuss the properties of the spacetime observables derived from the particle current density vector and the energy-momentum-stress tensor. For definiteness, we focus our attention on the particle density,  the kinetic temperature, the pressure anisotropy, and we compare their properties with those of the corresponding Newtonian model in paper I. Recall that these quantities depend on the amplitude $\alpha$ of the DF and the dimensionless parameters $\varepsilon_0$, $\lambda_0$ (see Eq.~(\ref{Eq:DimensionlessQuantities})), $k$ and $l$. As explained previously, we choose the maximum value $\varepsilon_0 = 1$ in all the results shown, which leads to an infinitely extended cloud. The cut-off value $\lambda_0$ for the azimuthal angular momentum is an arbitrary positive parameter, while $l\geq 0$ and $k > 7/2 + l/2$ is necessary in order to guarantee a finite total particle number, energy and angular momentum. In the following, we compare configurations of equal total particle number $N_{\textrm{gas}}$, which fixes the amplitude $\alpha$. At the end of this section we also compare our kinetic configurations with the well-known "polish doughnuts" based on a perfect fluid model.

\subsection{Normalized particle density and morphology of the gas cloud}

\begin{figure}[h!]
\centerline{
\includegraphics[scale=0.275]{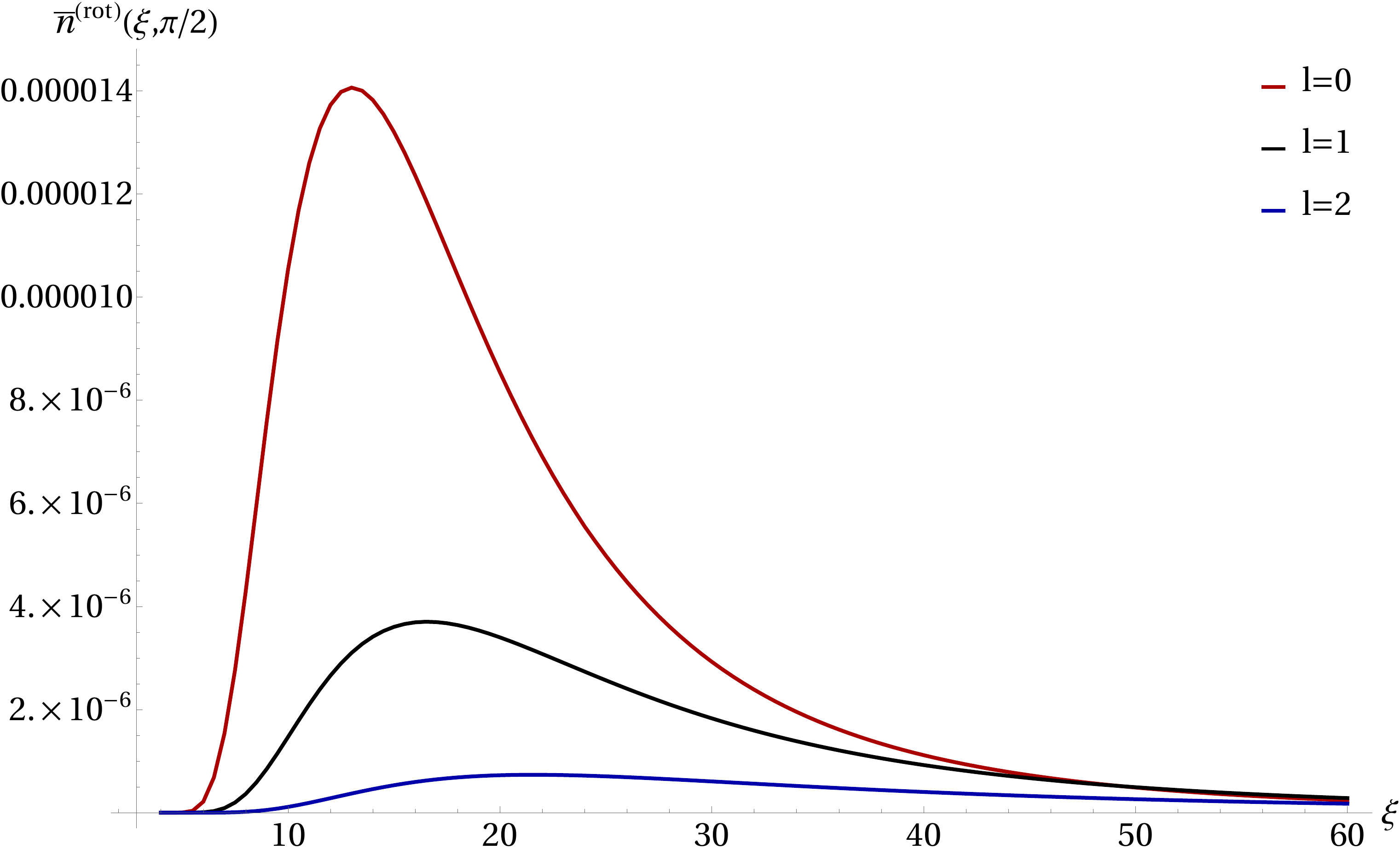}
\includegraphics[scale=0.275]{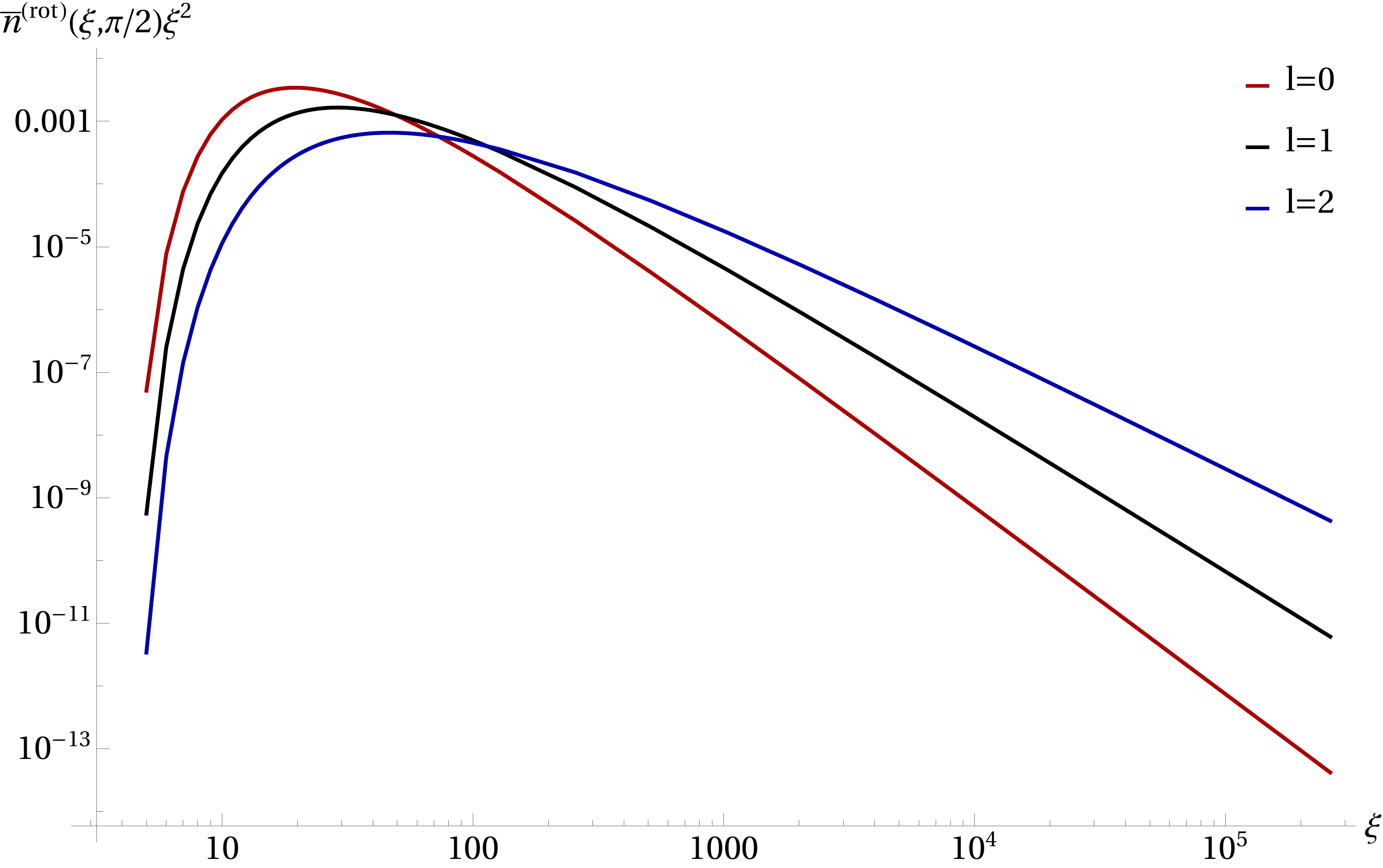}}
\caption{Normalized particle density $\bar{n}^{(\textrm{rot})}$ (left panel) and $\bar{n}^{(\textrm{rot})}$ times the square radius (right panel) as a function of the dimensionless radius $\xi$ in the equatorial plane for $k=5$, $l=0,1,2$, $\lambda_0=4$. As $l$ increases, the location of the maximum moves outwards, as expected. Although configurations with smaller values of $l$ have a higher maximum of $\bar{n}^{(\textrm{rot})}$, we see that they eventually intersect the curves belonging to higher values of $l$ and they decay faster for large $\xi$, as is visible from the plot in the right panel.}
\label{Fig:ParticleDensityk5l012} 
\end{figure}

\begin{figure}[h!]
\centerline{
\includegraphics[scale=0.275]{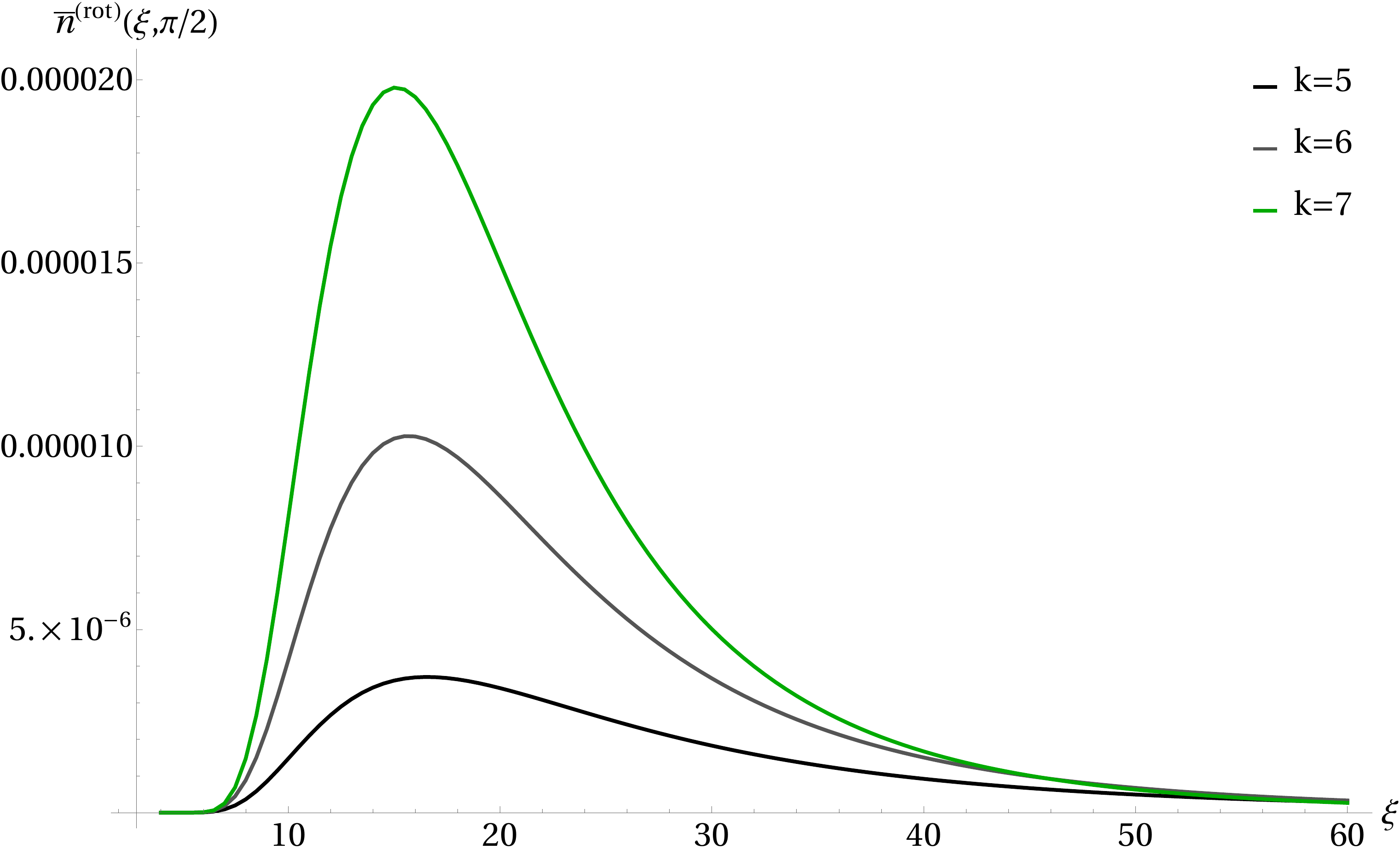}
\includegraphics[scale=0.275]{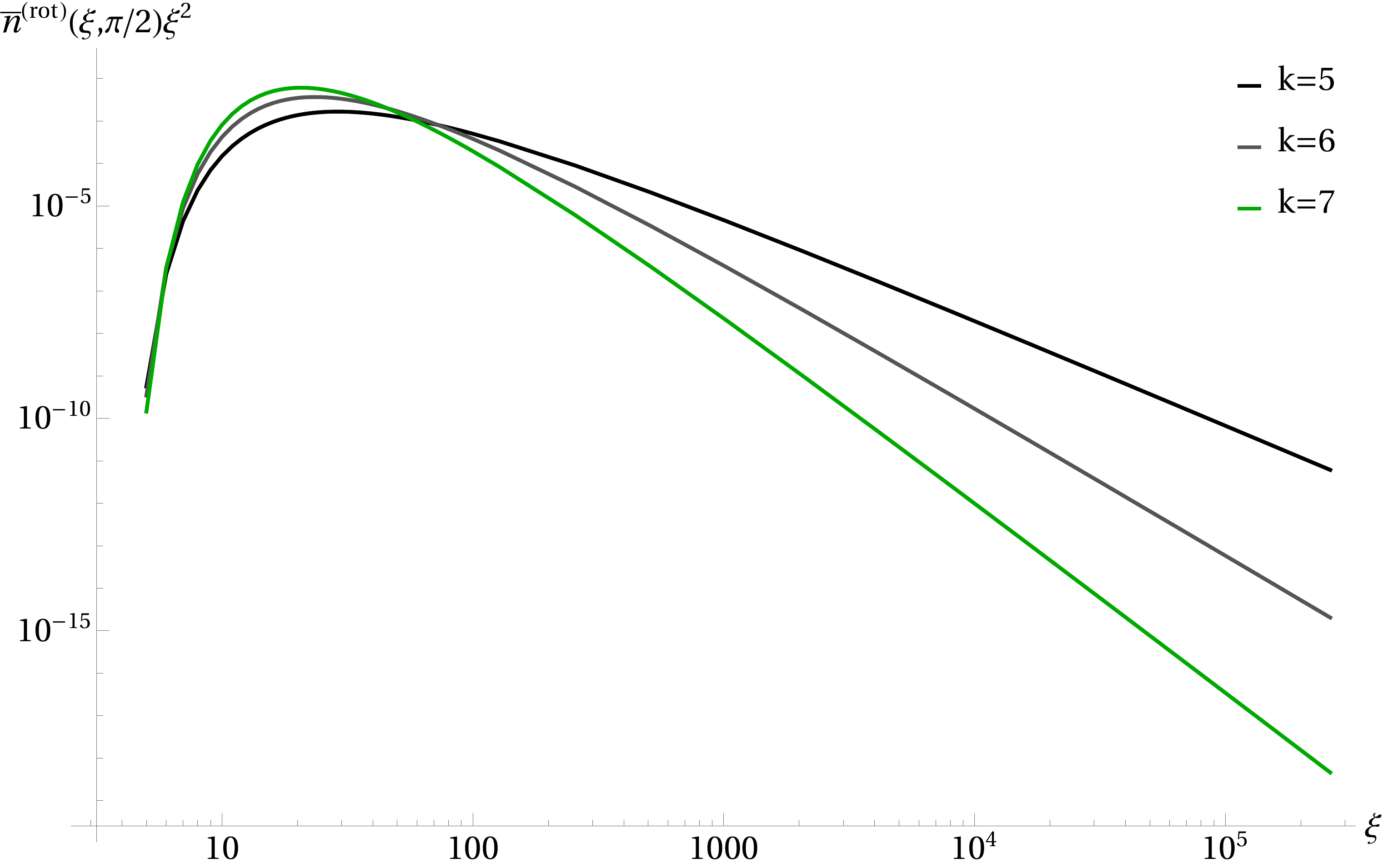}}
\caption{Same as previous plot for the parameter values $l=1$, $k=5,6,7$ and $\lambda_0=4$.}
\label{Fig:ParticleDensityk567l1}
\end{figure}

In this subsection we discuss the properties of the normalized particle density
\begin{equation}
\bar{n}(\xi,\vartheta) := M^3\frac{n(\xi,\vartheta)}{N_{\textrm{gas}}},
\end{equation}
which can be computed from Eqs.~(\ref{Eq:ParticleDensity}) and~(\ref{Eq:NumberParticleseP}) for both the even and rotating models. Although both models have the same total particle number $N_{\textrm{gas}}$, their particle densities differ from each other, $n^{(\textrm{even})} \neq n^{(\textrm{rot})}$, which is due to the fact that $n$ depends on the $\hat{3}$-component of $J$. After fixing $\varepsilon_0 = 1$ the quantity $\bar{n}(\xi,\vartheta)$ depends only on $\lambda_0$, $k$ and $l$, as discussed above. In  figures~\ref{Fig:ParticleDensityk5l012} and~\ref{Fig:ParticleDensityk567l1} we show the profiles of $\bar{n}$ and $\bar{n}\xi^2$ on the equatorial plane ($\vartheta=\pi/2$) for different values of the free parameters for the rotating model. As is visible from these plots, the configurations become more compact as $l$ decreases or $k$ increases. This is the same qualitative behavior as in the Newtonian case, see paper I.

\begin{figure}[h!]
\centerline{
\includegraphics[scale=0.375]{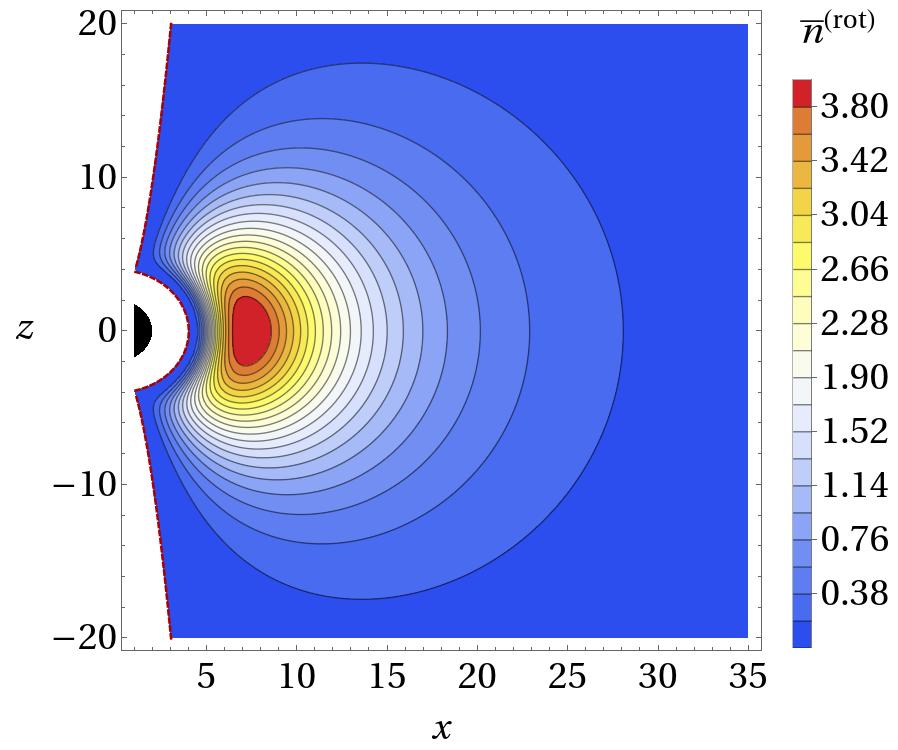}
\includegraphics[scale=0.375]{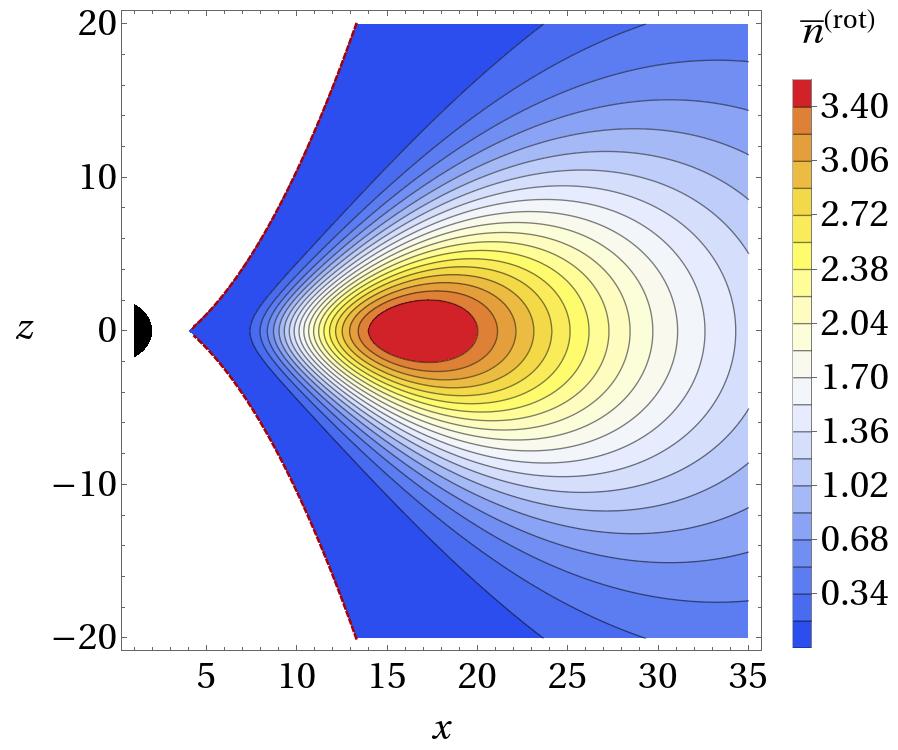}}
\caption{Contour plot of the normalized particle density $\bar{n}^{(\textrm{rot})}$ in the $xz$-plane for $(k,l)=(5,1)$. In the left panel $\lambda_0=1$ and the scale is $1\times 10^{-5}$ while in the right panel $\lambda_0=4$ and the scale is $1\times 10^{-6}$. The dashed red line, computed using Eq.~(\ref{Eq:SupportRel}), denotes the boundary of the corresponding kinetic gas cloud, while the black region represents the black hole.
}
\label{Fig:ContourParticleDensity}
\end{figure}

The morphology of the particle density for configurations with $(k,l)=(5,1)$ and $\lambda_0 = 4$ and $\lambda_0 = 1$ is shown in figure~\ref{Fig:ContourParticleDensity} for the rotating case and reveals the toroidal-like structure of the configuration. As can also be seen from this plot, the particle density is everywhere regular, has a maximum at some circle lying in the equatorial plane, and decays for large radii. Note also the difference  in the shape of the disk near the black hole in the two cases. For $\lambda_0 = 1$ the inner boundary has a spherical part (with radius equal to $4M$, i.e. the radius of the marginally bound orbits) delimited by two cusps, whereas for $\lambda_0 = 4$ the inner part of the disk is wedge-like. This is related to the fact that when $\lambda_0\leq 4$, the ISOs become occupied, which yields a change of behavior in the minimum radius.

\begin{figure}[h!]
\centerline{
\includegraphics[scale=0.3]{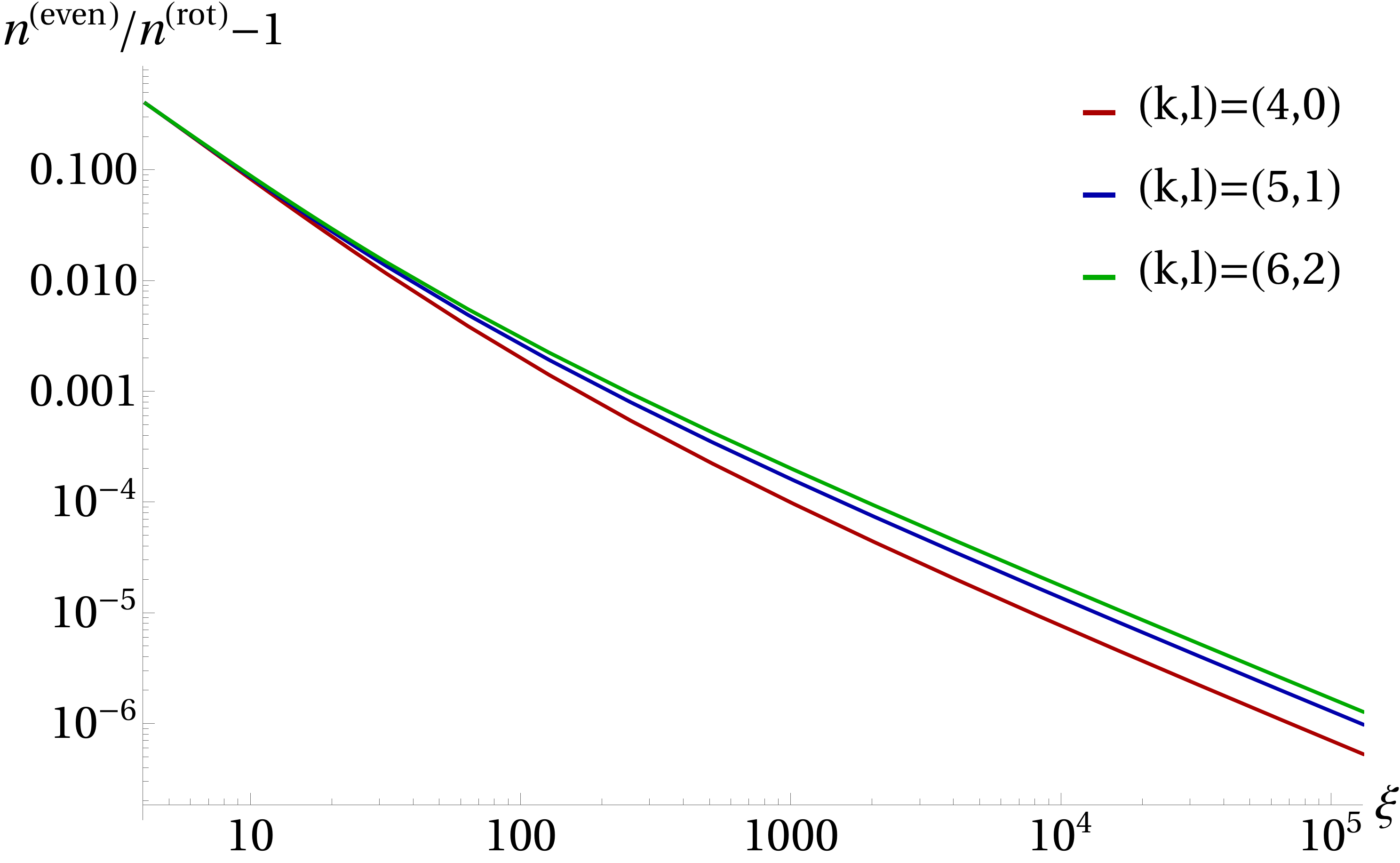}}
\caption{Relative difference $n^{(\textrm{even})}/n^{(\textrm{ro}t)} - 1$ between the even and rotating models for $(k,l) = (4,0)$ (red), $(k,l) = (5,1)$ (blue), $(k,l) = (6,2)$ (green), and $\lambda_0 = 4$.}
\label{Fig:nEvenOdd}
\end{figure}

The differences in the profiles of the particle density between the even and rotating cases is small in all cases we have examined. As an example, we show in figure~\ref{Fig:nEvenOdd} the relative difference $n^{(\textrm{even})}/n^{(\textrm{ro}t)} - 1$ between the two models for different values of $(k,l)$ and $\lambda_0 = 4$. The maximum relative difference is about $0.414$ and occurs at the inner radius of the disk.

\begin{figure}[h!]
\centerline{
\includegraphics[scale=0.3]{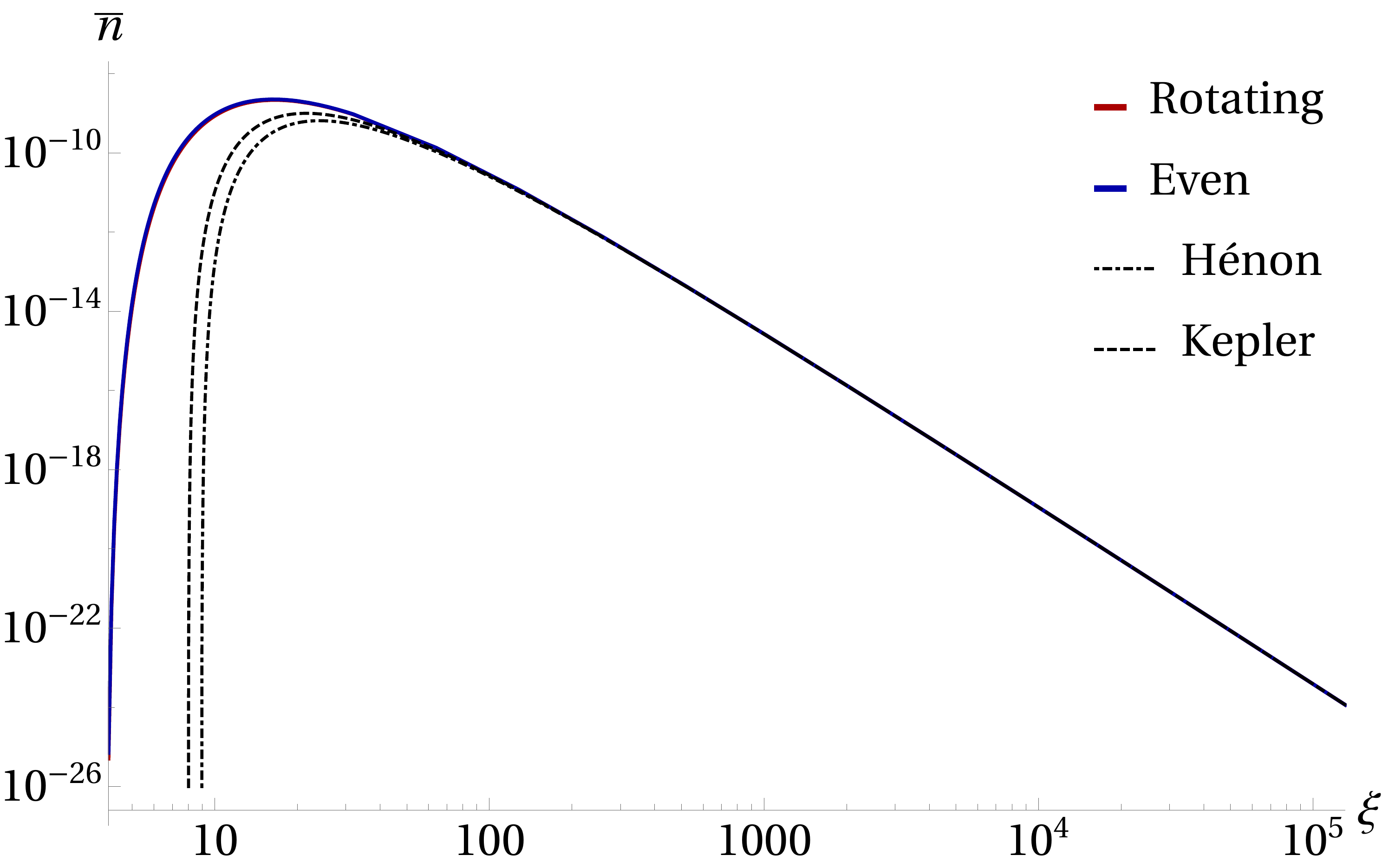}}
\caption{Comparison of the particle density between the even (blue) and rotating (red) models and the corresponding nonrelativistic models for $(k,l) = (5,1)$ and $\lambda_0 = 4$. The difference between the first two models is not visible in this plot. The relative difference is shown in the next plot.
}
\label{Fig:nRelNonRel}
\end{figure}

Finally, in figure~\ref{Fig:nRelNonRel} we compare the particle density profiles and compare them with the corresponding profiles of the nonrelativistic models with the Kepler ($\kappa = 0$) and isochrone potentials ($\kappa=1$). As expected, these profiles agree with each other for large values of $\xi$. However, the nonrelativistic configurations have an inner radius that is larger than their relativistic counterparts, and hence the relativistic and nonrelativistic profiles are considerably different from each other for smaller values of $\xi$.

\subsection{Kinetic temperature}

\begin{figure}[h!]
\centerline{
\includegraphics[scale=0.275]{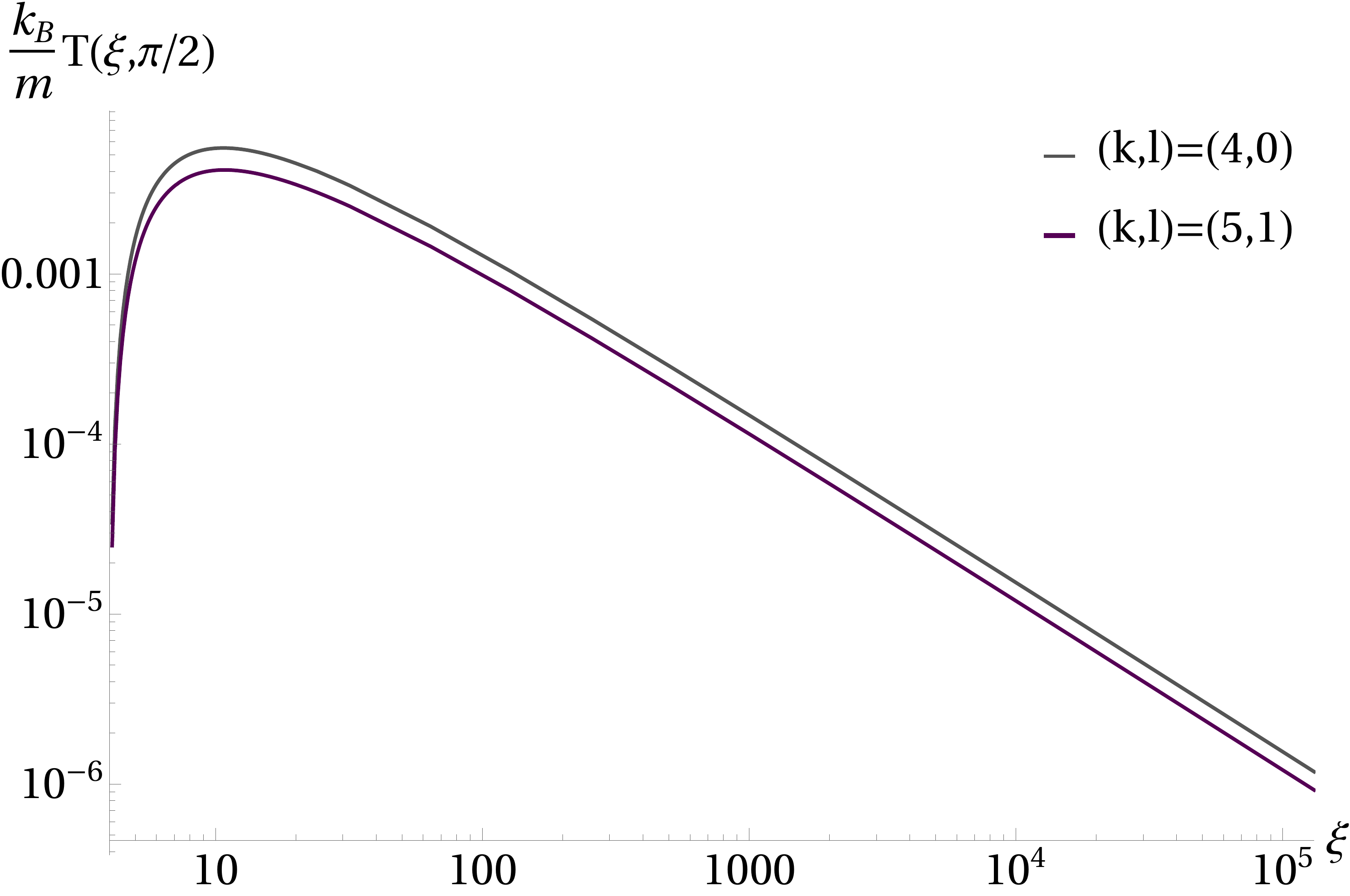}}
\caption{Equatorial kinetic temperature profiles for the model with parameter values $(k,l) = (4,0)$ and $(5,1)$ and $\lambda_0=4$.}
\label{Fig:KineticTemperature}
\end{figure}

As in paper~I, we define the kinetic temperature $T$ through the ideal gas equation $n k_B T = \overline{P} $ with the average pressure $\overline{P} := (P_{\hat{r}} + P_{\hat{\vartheta}} + P_{\hat{\varphi}})/3$ and the particle density $n$. We show the equatorial temperature profile in figure~\ref{Fig:KineticTemperature} for the cases $\lambda_0 = 4$ and $(k,l) = (4,0)$ and $(k,l) = (5,1)$ and the rotating model, along with the corresponding results in the Newtonian cases. The behavior is qualitatively similar to the profile of the particle number density.

\begin{figure}[h!]
\centerline{
\includegraphics[scale=0.3]{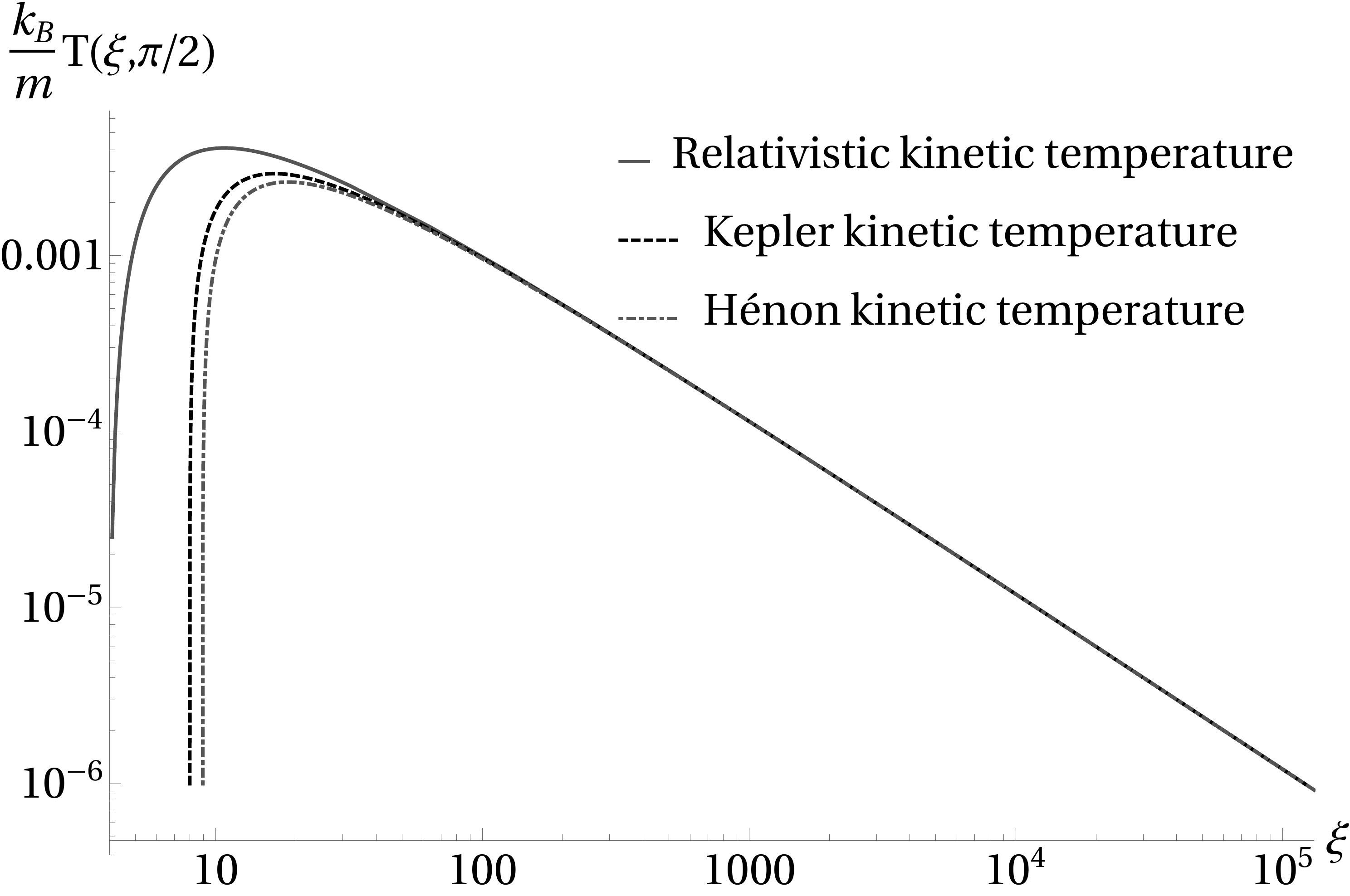}}
\caption{Equatorial kinetic temperature profile for the model with parameter values $k=5$, $l=1$ and $\lambda_0=4$ in the relativistic and Newtonian cases.
}
\label{Fig:KineticTemperature2}
\end{figure}

\subsection{Pressure anisotropy}

\begin{figure}[h!]
\centerline{
\includegraphics[scale=0.28]{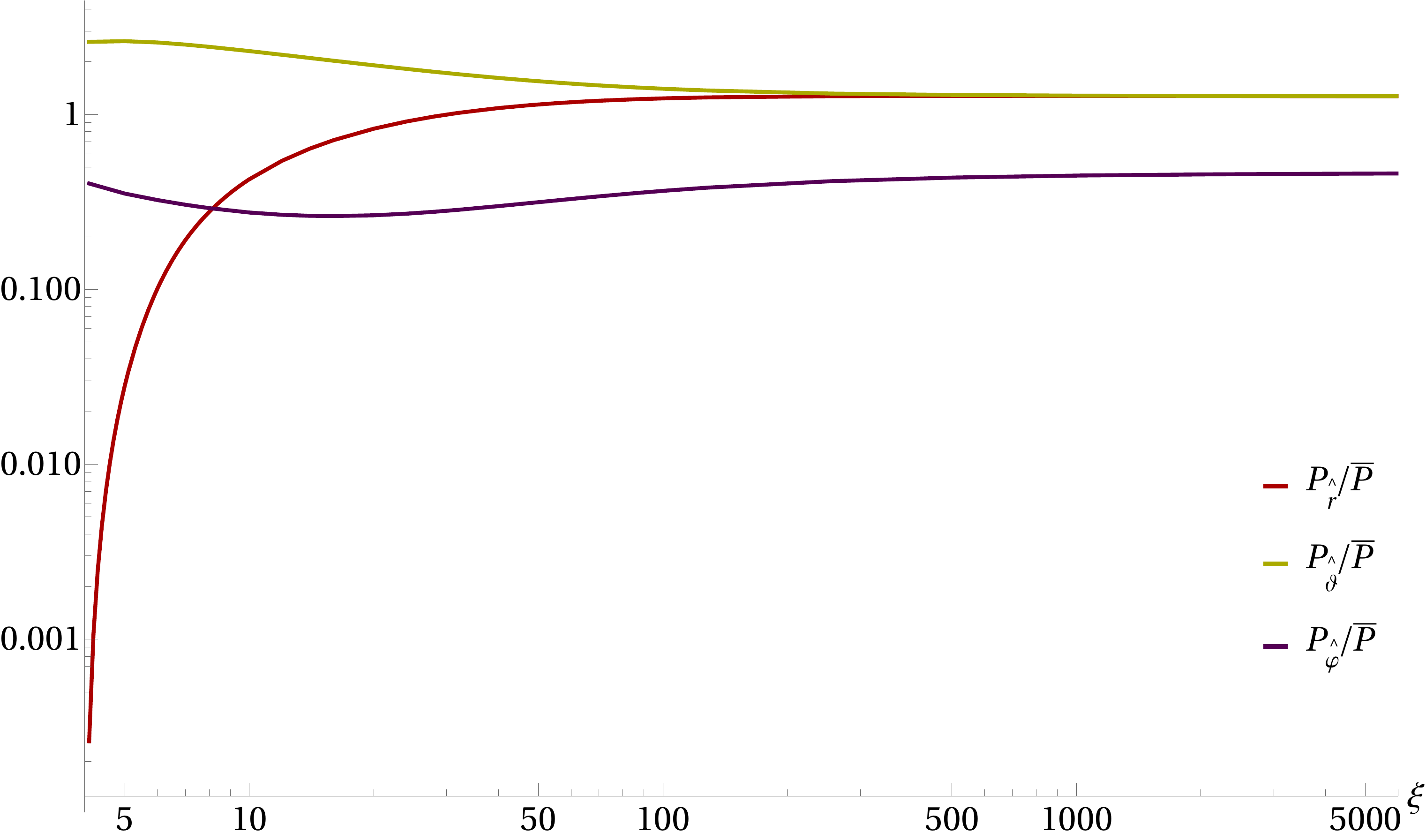}
\includegraphics[scale=0.28]{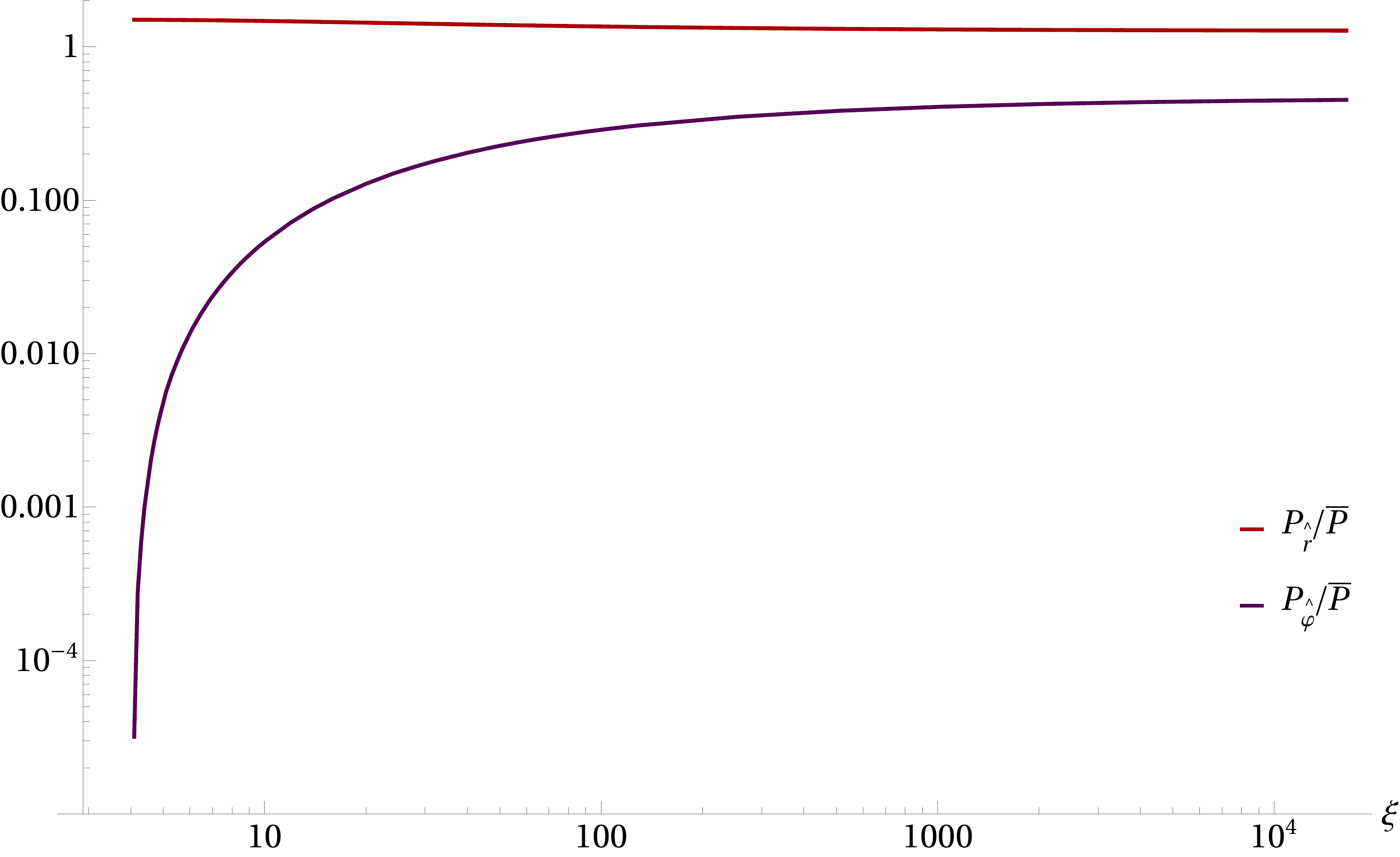}}
\caption{Log-log plot showing the behavior of the principal pressures (normalized by the average pressure $\overline{P}$) as a function of the areal radius $r$ in the equatorial plane  for the parameter values $k=5$, $l=1$, $\varepsilon_0 = 1$ and the rotating model. Left panel: $\lambda_0 = 1$. Note that in this case the three pressures are different from each other, with $P_{\hat{r}}$ converging to $P_{\hat{\vartheta}}$ for large $r$. Right panel: $\lambda_0 = 4$. In this case, $P_{\hat{r}} = P_{\hat{\vartheta}}$ everywhere.
}
\label{Fig:PrincipalPressures1}
\end{figure}

\begin{figure}[h!]
\centerline{
\includegraphics[scale=0.275]{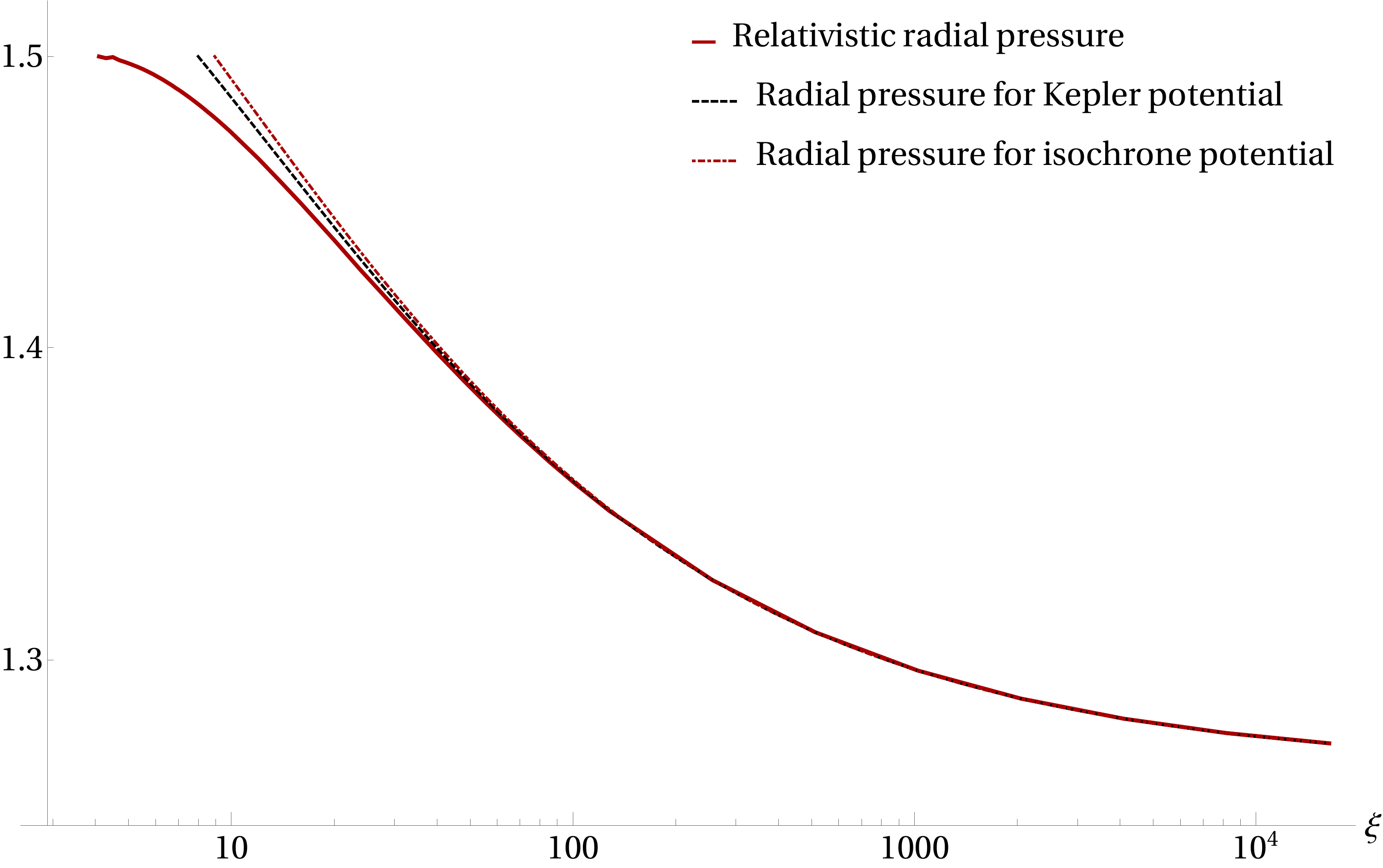}
\includegraphics[scale=0.275]{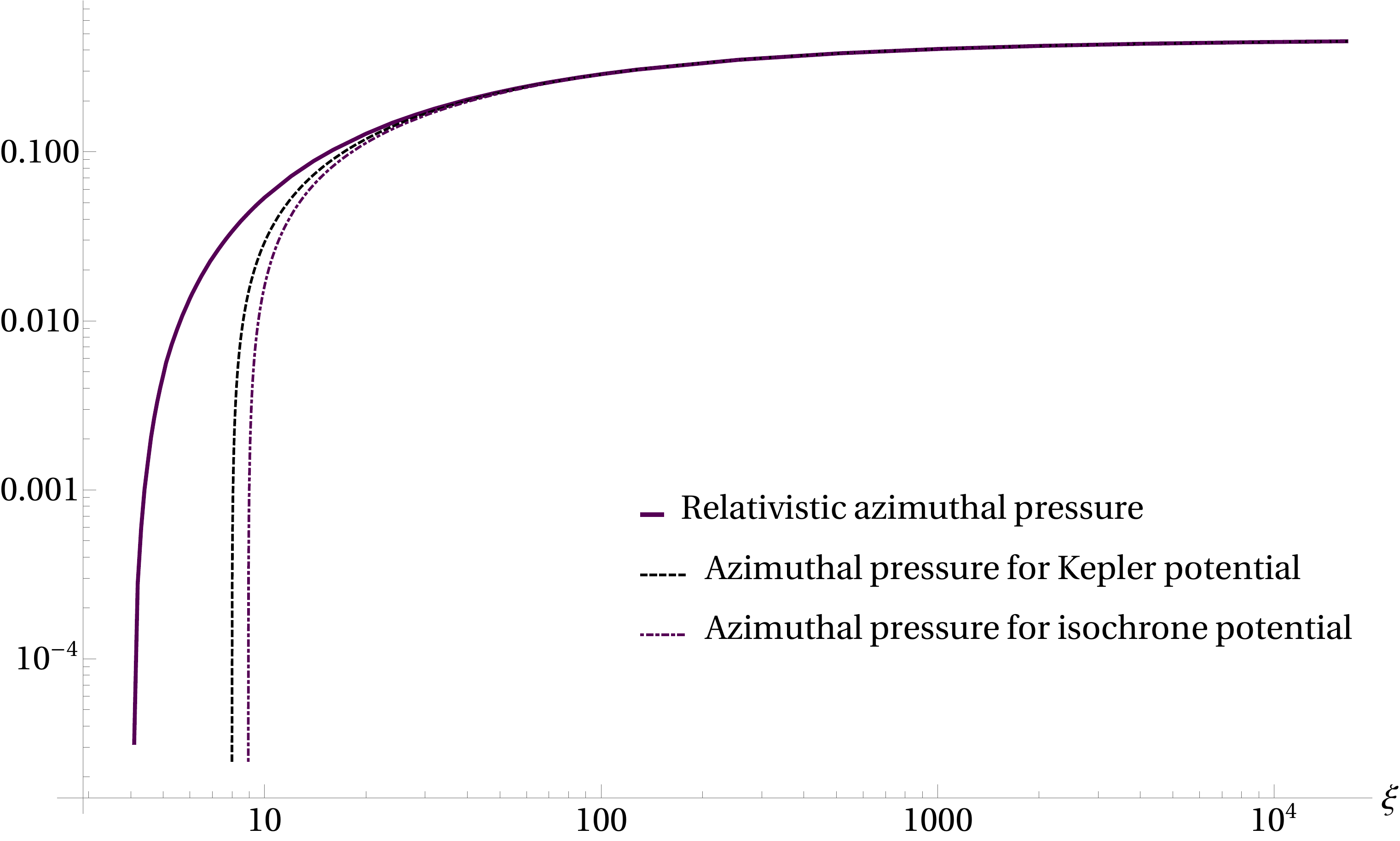}}
\caption{Behavior of the principal pressures $P_{\hat{r}}$ and $P_{\hat{\varphi}}$ (normalized by the average pressure $\overline{P}$) as a function of the areal radius $r$ in the equatorial plane (note that we use a logarithmic scale in $r$) for the parameter values $k=5$, $l=1$, $\varepsilon_0 = 1$, $\lambda_0 = 4$ and the rotating model. For comparison, we also show the corresponding results from the Newtonian case with the Kepler ($\kappa = 0$) and the isochrone $(\kappa = 1$) potentials. Left panel: The principal pressure $P_{\hat{r}}$ corresponding to the radial direction. Right panel: Principal pressure $P_{\hat{\varphi}}$ corresponding to the azimuthal direction. Note that the inner radius of the Newtonian configurations is located at $\xi=8$ (for the Kepler case) and $\xi = 4\sqrt{5}\simeq 8.94$ (for the isochrone potential with $\kappa=1$) while the relativistic configurations have their inner edge located at $\xi = 4$. As this inner radius is approached, $P_{\hat{\varphi}}/\overline{P}$ converges to zero in all cases. As expected, for large values of $r$, the relativistic and Newtonian profiles agree with each other.}
\label{Fig:PrincipalPressures2}
\end{figure}

Figures~\ref{Fig:PrincipalPressures1} and~\ref{Fig:PrincipalPressures2} show the principal pressures as a function of the dimensionless areal coordinate $\xi = r/M$ on the equatorial plane for the rotating model. Note that for $\lambda_0 = 1$ these pressures are different from each other, while for $\lambda_0\geq 4$ the principle pressures $P_{\hat{r}}$ and $P_{\hat{\vartheta}}$ corresponding to the radial and polar directions are always equal to each other. This can be understood by realizing that the expressions~(\ref{Eq:T11},\ref{Eq:T22}) differ from each other only in the sign of the last term in the integrand. However, this term is proportional to $\tilde{\textbf{K}}_l(a,0)$ which vanishes if $\lambda_0\geq 4$. As expected, for large values of $\xi$, the results agree with the corresponding results from the Newtonian models discussed in paper I. As seen from the plots in figure~\ref{Fig:PrincipalPressures2}, as $r$ increases from its value at the inner edge of the configuration to large values, the radial pressure decays while the azimuthal pressure increases monotonically.

\subsection{Comparison with fluid model}

We end this section by comparing our kinetic configurations to the well-known "polish doughnuts" hydrodynamics configurations whose construction is briefly reviewed in Appendix~\ref{App:PolishDoughnuts}. We start with the comparison of the boundary surface delimiting the support of the gas. In the kinetic case, this boundary is determined by Eq.~(\ref{Eq:SupportRel}). Written in terms of the dimensionless quantities introduced in Eq.~(\ref{Eq:DimensionlessQuantities}) this gives
\begin{equation}
\frac{2}{\xi-2} - \frac{\lambda_0^2}{\xi^2 \sin^2\vartheta} \geq 0,\qquad
\xi \geq 4,
\label{Eq:DimensionlessBoundary1}
\end{equation}
which is equivalent to the condition $\xi\geq \xi_{\textrm{min}}(\vartheta)$ with
\begin{equation}
\xi_{\textrm{min}}(\vartheta) := \left\{ \begin{array}{cl}
\displaystyle \frac{\lambda_0^2}{4\sin^2\vartheta} \left[ 1 + \sqrt{1 - \frac{16\sin^2\vartheta}{\lambda_0^2}} \right] & \hbox{if}\quad 4\sin\vartheta\leq \lambda_0, \\
4 & \hbox{if} \quad 4\sin\vartheta > \lambda_0.
\end{array} \right.
\label{Eq:DimensionlessBoundary2}
\end{equation}
In the fluid case, restricting ourselves to the region $r > 4M$, the boundary surface is determined by the level one set of the enthalpy function $h$. According to Eq.~(\ref{Eq:W}), this yields precisely the same condition $\xi\geq \xi_{\textrm{min}}(\vartheta)$ where now $\lambda_0$ stands for the (constant) azimuthal angular momentum per energy of the fluid elements divided by $M$. As explained in paper I, this coincidence is due to the fact that as one approaches the boundary surface from the inside of the fluid configuration, the pressure's gradient converges to zero implying that the fluid elements follow timelike geodesics with specific azimuthal angular momentum $M\lambda_0$, as in the kinetic case.

Although the boundary surface in the kinetic and fluid models coincide exactly with each other, the structure of the interior cloud cannot be identical in both models. Indeed, in the fluid model, the pressure is enforced to be isotropic and each fluid element has constant azimuthal angular momentum per energy along the stream lines, whereas in the kinetic model, the pressure is anisotropic and the gas particles' azimuthal angular momenta obey the distributions $I^{(\text{even})}$ or $I^{(\text{rot})}$, see Eqs.~(\ref{Eq:PolytropeLzEven},\ref{Eq:PolytropeLzRot}).

\begin{figure}[h!]
\centerline{
\includegraphics[scale=0.275]{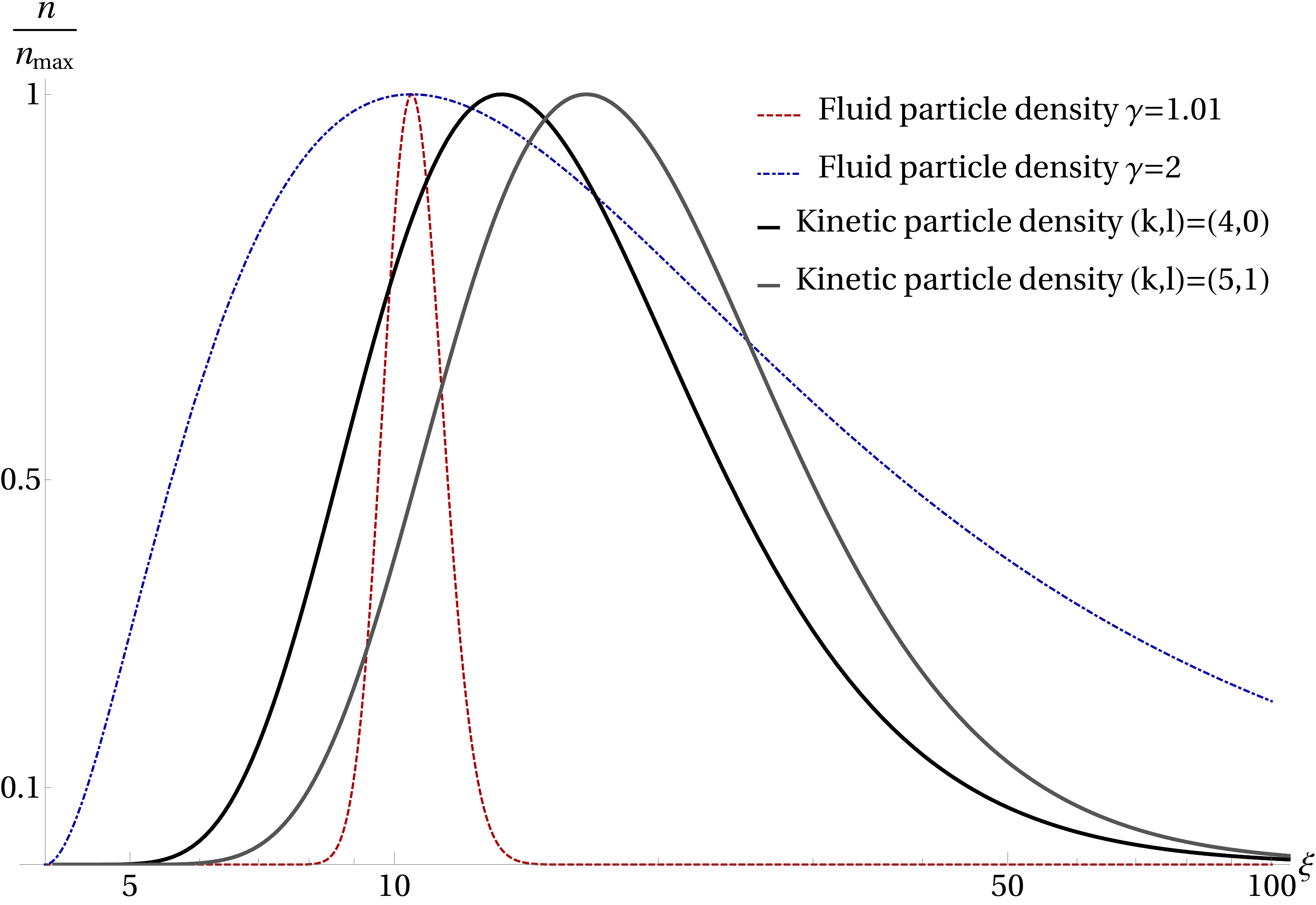}\quad
\includegraphics[scale=0.275]{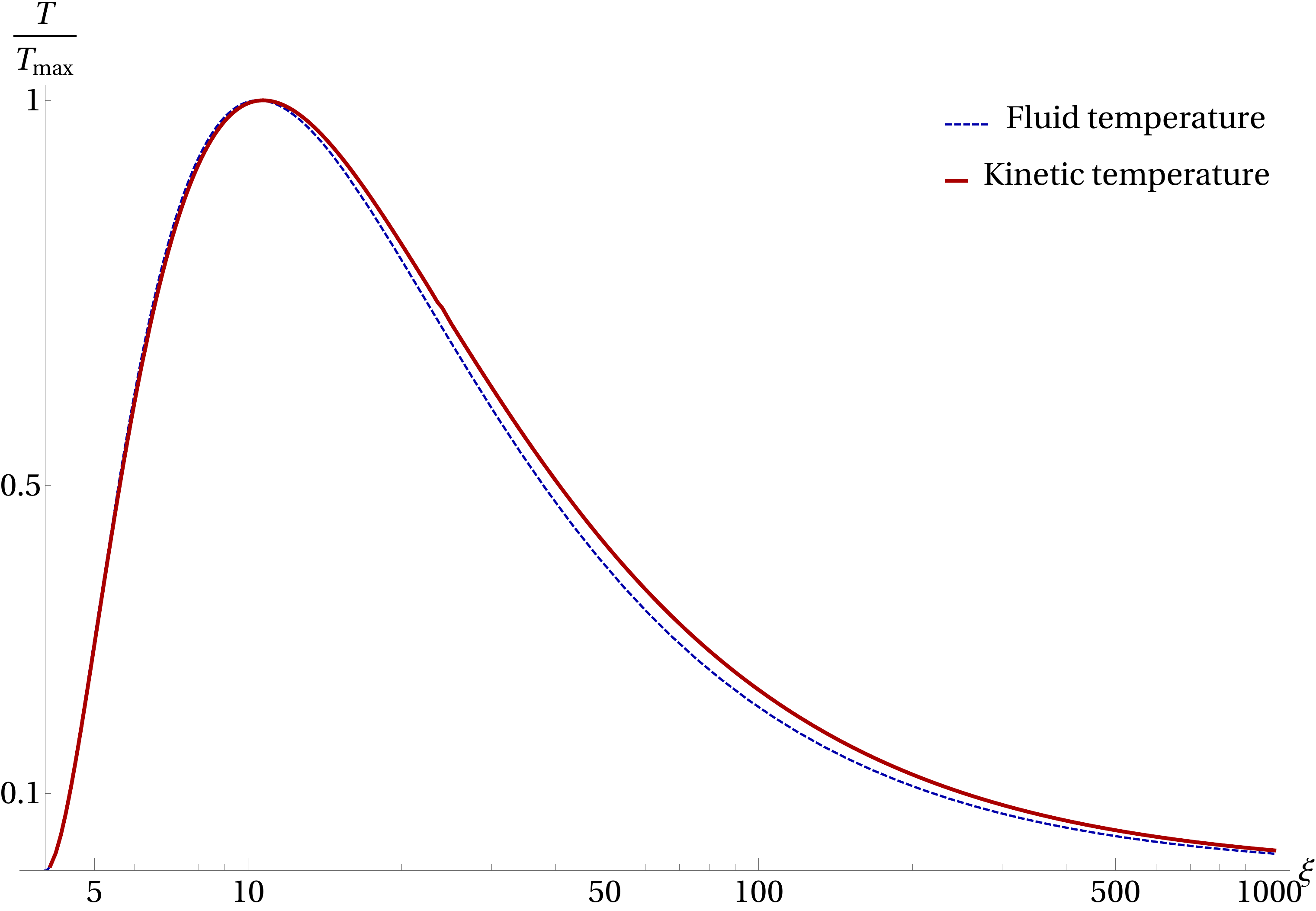}}
\caption{Normalized profiles between the kinetic and fluid descriptions. The left panel shows the profile of the particle density for the fluid model in the extreme cases with adiabatic indices $\gamma=1.01$ (dashed red) and $\gamma=2$ (dotted dashed blue) and the corresponding particle density for the rotating kinetic models with $(k,l)=(4,0)$ (black) and $(k,l)=(5,1)$ (gray). The right panel shows the normalized profile for the fluid (dashed blue) and kinetic temperature with $(k,l)=(4,0)$ (red). The temperature profile for the kinetic model with $(k,l) = (5,1)$ lies almost on the top of the profile for the $(k,l)=(4,0)$ model, and thus it is not plotted. In both panels the normalization is chosen such that the maximum is one.
}
\label{Fig:FluidvsKinetic}
\end{figure}

In order to compare both model's morphology with each other, for the following we use the particle density $n$ and the temperature $T$. In the fluid case, we assume a polytropic equation of state $P = K n^\gamma$ with $K$ a constant and $\gamma$ the adiabatic index subject to $1 < \gamma \leq 2$, and obeying the ideal gas equation $P = n k_B T$. Integrating the first law one obtains the following relation between the specific enthalpy $h$, $n$ and $T$ (see Appendix~\ref{App:PolishDoughnuts} for details):
\begin{equation}
h - 1 
= \frac{\gamma}{\gamma-1}\frac{K}{\bar{m}} n^{\gamma-1}
= \frac{\gamma}{\gamma-1}\frac{k_B}{\bar{m}} T,
\end{equation}
with $\bar{m}$ the averaged rest mass per particle. Figure~\ref{Fig:FluidvsKinetic} shows the normalized (with respect to its maximum value) density and temperature profiles for the fluid and the kinetic models with different values of $(k,l)$ and $\lambda_0 = 4$. As in the Newtonian case, the fluid configurations are more compact and slightly hotter than the kinetic ones. Nevertheless, remarkably, the equatorial temperature profile of the fluid configuration (which is independent of the adiabatic index $\gamma$) agrees very well with the corresponding profile of the kinetic model with $(k,l) = (4,0)$ or $(k,l) = (5,1)$. A quantitative comparison between the locations and values of the maximum temperature in both models are given in table~\ref{Table:RatioTemperatures}. These values should be compared with the corresponding values reported in Table~III of paper I.

\begin{table}
\centering
\caption{Top row: Ratio between the radii corresponding to the maxima of the fluid and kinetic temperature profiles. Bottom row: Ratio between the corresponding maxima of the temperature. Here we choose $(k,l) = (5,1)$, and in the fluid case the adiabatic index is determined as in Paper I, i.e. $\gamma = \gamma^{(\text{kinetic})}$ which fits the asymptotic decay of the kinetic configuration. Three significant figures are shown.}
\label{Table:RatioTemperatures}
\begin{tabular}{|c|c|c|c|}
\hline
 & $\lambda_0 = 4$ & $\lambda_0 = 7$ & $\lambda_0 = 10$ \\ 
\hline
& & & \\ 
$\displaystyle \frac{r_\text{max}^{(\text{fluid})}}{r_\text{max}^{(\text{kinetic})}}$ 
& 0.970 & 0.970 & 0.968 \\ 
& & & \\ 
\hline
 & & & \\
$\displaystyle\frac{m}{\bar{m}}\frac{T_\text{max}^{(\text{fluid})}}{T_\text{max}^{(\text{kinetic})}}$ & 1.31 & 1.30 & 1.30 \\
& & & \\ 
\hline
\end{tabular}
\end{table}

\section{Conclusions}
\label{Sec:Conclusions}

In this work we derived analytic solutions describing relativistic stationary and axisymmetric collisionless kinetic gas clouds consisting of identical massive neutral particles surrounding a non-rotating black hole. Based on the assumption that the gravitational field is dominated by the black hole, the self-gravity of the gas is neglected, such that the gas particles follow bound geodesic orbits in the Schwarzschild exterior spacetime. As we have shown, our gas configurations agree with their non-relativistic counterparts constructed in paper I~\cite{cGoS2022b} in the limit in which the angular momentum cut-off parameter $L_0$ tends to infinity. The one-particle DF in our models depends on the energy $E$ and the azimuthal component of the angular momentum $L_z$ of the particles  through the generalized polytropic ans\"atze~(\ref{Eq:DistributionFunctionProduct},\ref{Eq:PolytropeEnergy},\ref{Eq:PolytropeLzEven},\ref{Eq:PolytropeLzRot}) describing both rotating and nonrotating stationary and axisymmetric configurations. In the rotating case all particles have positive values of $L_z$ larger than some cut-off value $L_0$, giving rise to a net angular momentum while in the nonrotating case the DF is an even function of $L_z$. We have derived explicit expressions for the spacetime observables in terms of a single integral over a function depending solely on the energy. This has been achieved by rewriting the fibre integrals over the particles' momenta as integrals over the constants of motion $E$, $L$ and $L_z$. A challenging problem in this change of variables is the determination of the correct range of integration over which $E$, $L$ and $L_z$ vary, as this range depends on the observer's radius $r$. The result which was derived in section~\ref{SubSec:ExplicitObservables} and Appendix~\ref{App:LimitsIntegration} is summarized in Eqs.~(\ref{Eq:RangeIntegration},~\ref{Eq:MinimumEnergy2},\ref{Eq:AngularMomentumC},\ref{Eq:MaximumAngularMomentum}), and it should have applications extending beyond the ones given in the current work.

By choosing the polytropic index $k$ sufficiently large, it follows from the asymptotic analysis in paper I that our configurations have finite total particle number, energy and angular momentum.  By introducing action-angle variables we have been able to reduce the expressions for these total quantities to a triple integrals which involve the period function $T_r(E,L)$ describing the period of the radial motion for an orbit with energy $E$ and total angular momentum $L$. For our models, the integral over $L_z$ can be performed analytically, which leaves a double integral over $E$ and $L$. To perform this integral, we used the fact that for the case of a Schwarzschild spacetime, $T_r(E,L)$ can be expressed in terms of Legendre's elliptic integrals whose arguments depend on the eccentricity and the ``semi-latus rectum", and one ends up with a numerical integral over these new variables. The behavior of the total quantities was compared with those of the corresponding non-relativistic quantities, and we have verified that they agree with each other in the limit $L_0\to \infty$, see figures~\ref{Fig:ComparisonTotMassTotNumberParticles} and~\ref{Fig:ComparisonEnergy}.

In addition to the aforementioned total quantities, we have computed and analyzed the behavior of the particle density, the principal pressures, and the kinetic temperature as a function of the free parameters $k,l,E_0,L_0$ in our models (the first two corresponding to the polytropic indices and the last two describing the cut-off parameters associated with the energy and angular momentum). We have shown that these quantities agree with their non-relativistic counterparts computed in paper I in the limit in which $L_0$ tends to infinity. However, when $L_0/(M m)$ is of the order one or smaller, several relativistic effects become visible. The most important difference consists in the morphology of the inner part of the torus. In the Newtonian case the boundary surface is completely smooth and its minimum radius shrinks continuously to zero as $L_0\to 0$. However, in the relativistic case, there are no bound orbits at all in the region $r < 4M$. This is due to the presence of the maximum of the effective potential $V_{m,L}$ which drops below the asymptotic value $m^2$ of $V_{m,L}$ when $L_0/(M m) < 4$. Hence, in this case, the minimum inner radius is given by the location of the turning point of the orbit with maximum energy $E = m$ for orbits with $L > 4M m$, while for $L < 4M m$ this inner radius is determined by the ISOs. This transition leads to the two cusps which are visible in the left panel of figure~\ref{Fig:ContourParticleDensity}. Another relativistic effect that appears when $L_0$ drops below $4 M m$ can be seen from the pressure anisotropies: in this case the three principal pressures are different from each other (see figures~\ref{Fig:PrincipalPressures1}) while for $L_0 \geq 4 M m$ the radial and polar principal pressures are always equal to each other, like in the Newtonian case. As explained in the article, this is again related to the presence of the ISOs. Finally, though less dramatic, a further relativistic effect is related to the difference between the particle density in the even and rotating models. In the Newtonian case the density is exactly equal in both models. In contrast, in the relativistic case the, this density is slightly larger in the even model as shown in the figure~\ref{Fig:nEvenOdd}. This is due to the fact that the relativistic invariant particle density $n$ depends on all components of the current density four-vector, see Eq.~(\ref{Eq:ParticleDensity}). Therefore, even though the orthonormal component $J_{\hat{0}}$ in the time direction coincides exactly in both models, the presence of the azimuthal component in the rotating case diminishes the invariant particle density $n$. 

Finally, we have compared our kinetic configurations with their hydrodynamic analogues, namely the well-known ``polish doughnuts" configurations. This comparison revealed the following properties: (i) by matching the value of the cut-off parameter $L_0/(M m)$ with the (constant) fluid angular momentum $\ell$, we have shown that both configurations are delimited by exactly the same boundary surface. (ii) The normalized equatorial radial profile of the particle density has its maximum lying closer to the black hole in the fluid case, meaning that the fluid configurations are generally more compact than the kinetic ones. (iii) Surprisingly however, the normalized equatorial radial profile of the temperature, which is independent of the adiabatic index in the fluid case and largely insensitive to the polytropic indices $(k,l)$ in the kinetic case, is very similar in both cases (see the right panel of figure~\ref{Fig:FluidvsKinetic}). This correspondence was also found in the nonrelativistic configurations of paper I. Like in the Newtonian case, choosing the adiabatic index such that it fits the asymptotic temperature decay of the kinetic configuration, it was found that the fluid configurations are slightly hotter than the kinetic ones as can be inferred from table~\ref{Table:RatioTemperatures}.

We close this article by emphasizing that, although they have finite total rest mass, energy and angular momentum, the configurations we have considered in this article  extend all the way to infinity. A related model in which the function $I(L_z)$ is replaced by a function of $L_z/L$ in Eq.~(\ref{Eq:DistributionFunctionProduct}) was studied in~\cite{cGoS2022a}. Solutions with finite support could also have been considered by choosing $E_0 < m$ instead of $E_0 = m$ in the ansatz~(\ref{Eq:PolytropeEnergy}). The recent work by Jabiri~\cite{fJ2021,fJ2022} suggests that the effects from the self-gravity can be included for such finite configurations, provided the amplitude of the DF is sufficiently small. It should also be interesting to generalize the models studied here to a rotating (Kerr) black hole or to include effects from binary collisions or an electromagnetic field. We leave these generalizations to future work.

\acknowledgments

We thank Francisco Astorga, Ana Laura Garc\'ia, Ulises Nucamendi, Emilio Tejeda, and Thomas Zannias for fruitful comments and discussions throughout this work. C.G. was supported by a PhD CONACyT fellowship. O.S. was partially supported by a CIC Grant to Universidad Michoacana. We also acknowledge support from the CONACyT Network Project No. 376127 ``Sombras, lentes y ondas gravitatorias generadas por objetos compactos astrof\'isicos".

\appendix
%
%
\section{Parametrization of bound trajectories in terms of the variables $(p,e)$}
\label{App:Parametrization}

In this appendix we recollect some useful formulae that can be used to parametrize the spatially bound timelike geodesics in the Schwarzschild exterior. In the body of the article these orbits have mainly been parametrized in terms of the conserved quantities $(E,L)$, corresponding to the energy and total angular momentum of the particle. Here we show that these orbits can also be parametrized in terms of their turning points $(r_1,r_2)$ or the associated dimensionless ``semi-latus rectum" $p$ and eccentricity $e$ defined by $r_1 = M p/(1+e)$, $r_2 = M p/(1-e)$. For more details and generalizations to the Kerr spacetime, see Refs.~\cite{wS02,jBmGtH15,pRoS18}.

\begin{figure}[h!]
\includegraphics[scale=0.3]{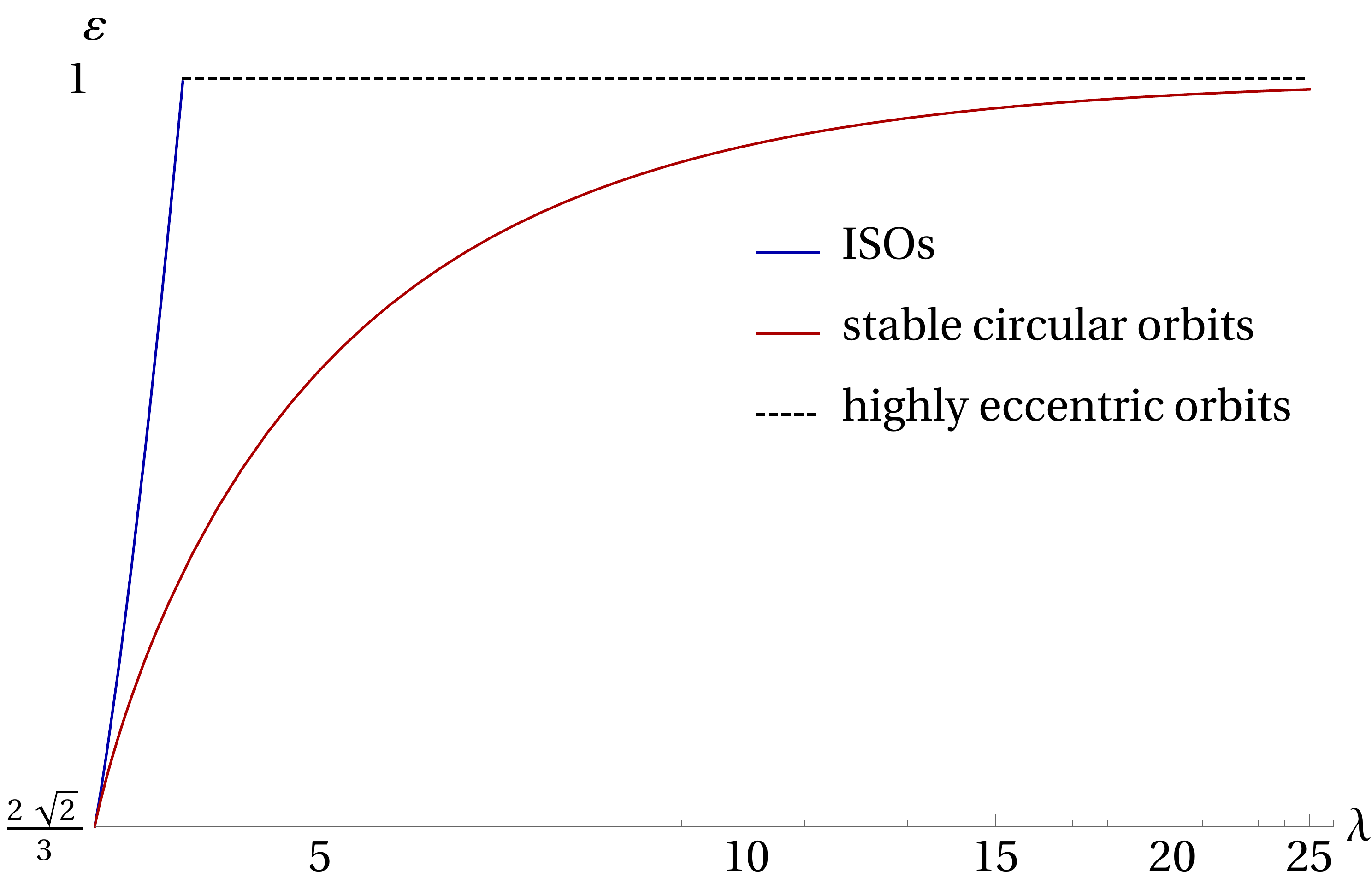}
\caption{Image of the domain $p > 6 + 2e$, $0 < e < 1$ under the transformation~(\ref{Eq:EnergyAngularMomentum}). Its boundary consists of the dashed line, corresponding to highly eccentric orbits $(e=1)$, the blue line representing the ISOs and the purple line corresponding to the stable circular orbits.}
\label{Fig:ELspace}
\end{figure}

The turning points are determined by the equation $V_{m,L}(r) = E^2$, which, in terms of the dimensionless variables introduced in Eq.~(\ref{Eq:DimensionlessQuantities}) yields the  following quartic equation for $\xi = r/M$:
\begin{equation}
R(\xi) := \xi \left[ \left(\varepsilon^2 - 1\right)\xi^3 + 2 \xi^2 - \lambda^2 \xi + 2 \lambda^2\right]
 = 0.
\label{Eq:R1}
\end{equation}
For bound orbits, there are four real roots $0$, $\xi_0$, $\xi_1$, and $\xi_2$ of $R(\xi)$ satisfying $0 < \xi_0 < \xi_1 < \xi_2$, the largest two ones corresponding to the turning points. Comparing Eq.~(\ref{Eq:R1}) with $R(\xi) = (\varepsilon^2 - 1)\xi(\xi - \xi_0)(\xi - \xi_1)(\xi - \xi_2)$ one  obtains the following relations between the conserved quantities $(\varepsilon,\lambda)$ and the roots $\xi_0$, $\xi_1$ and $\xi_2$:
\begin{equation}
\varepsilon^2 = 1-\frac{2}{\xi_0 + \xi_1 + \xi_2}, \qquad
\lambda^2 = \frac{\xi_0 \xi_1 \xi_2}{\xi_0 + \xi_1 + \xi_2}, \qquad
2 = \frac{\xi_0 \xi_1 \xi_2}{\xi_0 \xi_1 + \xi_0 \xi_2 + \xi_1 \xi_2}.
\label{Eq:Relations}
\end{equation}
The last relation allows one to express the roots explicitly in terms of $(p,e)$:
\begin{equation}
\xi_0 = \frac{2p}{p-4},\qquad
\xi_1 := \frac{p}{1+e}, \qquad \xi_2 := \frac{p}{1-e},
\label{Eq:ParametrizationeP}
\end{equation}
where $0 < e < 1$ and the condition $\xi_1 > \xi_0$ leads to the restriction $p > p_{ISO}(e) := 6 + 2e$. Note that the limits $e=0$ and $p = p_{ISO}(e)$ correspond to stable circular and ISOs, respectively. Inserting this into the first two relations in Eq.~(\ref{Eq:Relations}), one obtains
\begin{equation}
\varepsilon^2 = \frac{(p-2)^2 - 4e^2 }{p\left( p - e^2 - 3\right)}, \qquad
\lambda^2 = \frac{p^2}{p - e^2 - 3}.
\label{Eq:EnergyAngularMomentum}
\end{equation}
This yields the following relation between the elements $\lambda d\lambda d\varepsilon$ and $de dp$ (cf.~Eq.~(14) in Ref.~\cite{pRoS18}):
\begin{equation}
\lambda d\lambda d\varepsilon = \frac{e \sqrt{p} \left[ (p - 6)^2 - 4 e^2 \right]}{2\sqrt{\left(p - e^2 - 3 \right)^5 \left[(p - 2)^2 - 4 e^2 \right]}}de dp.
\label{Eq:Elements}
\end{equation}
For illustrative purpose, we show in figure~\ref{Fig:ELspace} the image of the domain $p > 6 + 2e$, $0 < e < 1$ under the transformation~(\ref{Eq:EnergyAngularMomentum}).

\section{Limits of integration for the fiber integrals defining the spacetime observables}
\label{App:LimitsIntegration}

In this appendix we provide an alternative derivation for the integration limits for $E$ and $L$ in Eq.~(\ref{Eq:RangeIntegration}), given an observer located at a certain radius $r_{\textrm{obs}}$. As in the previous appendix, for simplicity we work with the dimensionless variables introduced in Eq.~(\ref{Eq:DimensionlessQuantities}).

\begin{figure}[h!]
\centerline{
\subfigure{\includegraphics[scale=0.275]{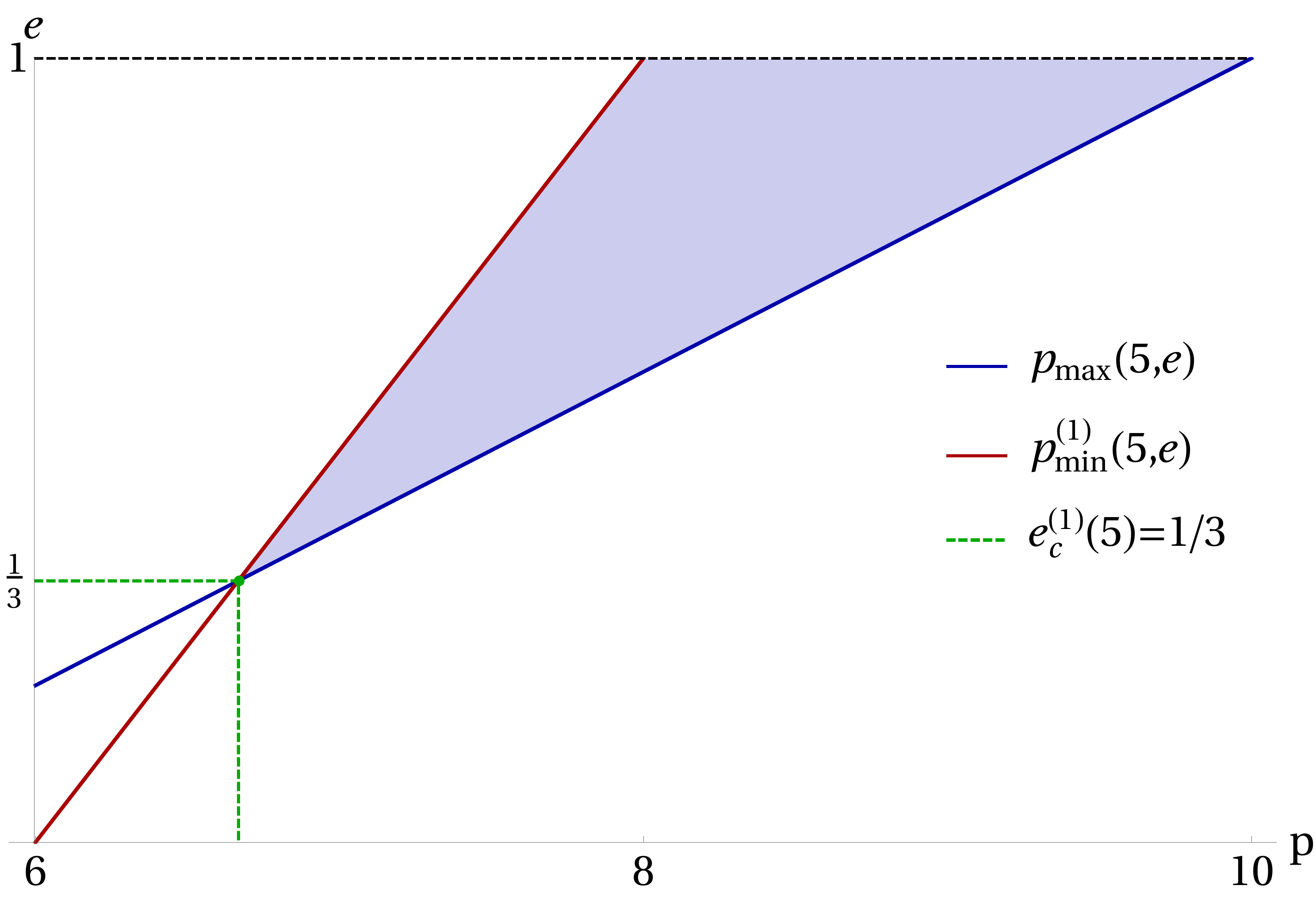}}
\subfigure{\includegraphics[scale=0.275]{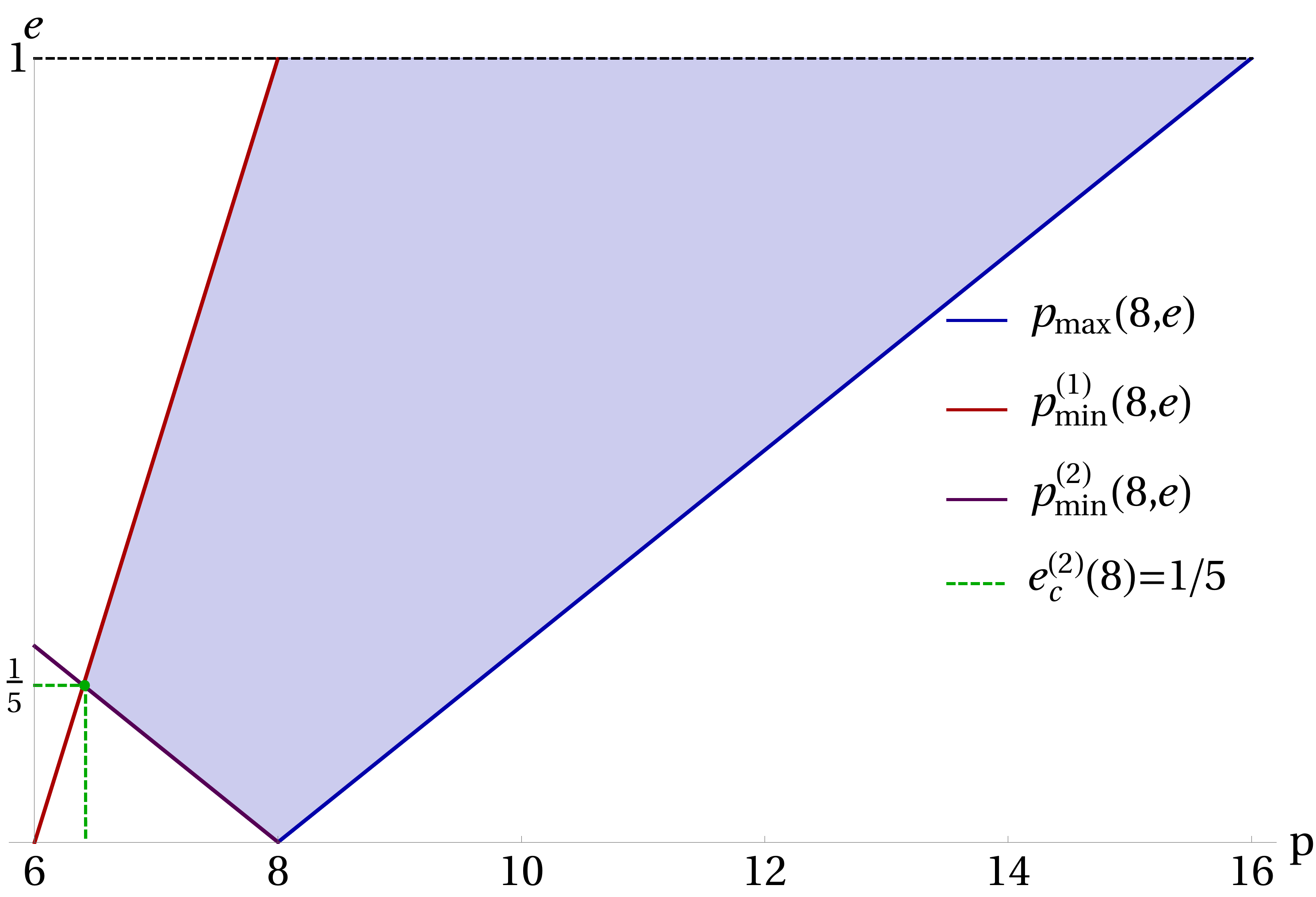}}}
\caption{Illustration for the allowed region of integration in the $(p,e)$-plane. Left panel: the observer is located at $\xi_{\text{obs}} = 5$ and the domain is delimited by a triangle. Right panel: the observer is located at $\xi_{\textrm{obs}} = 8$. In this case the domain is delimited by a quadrilateral.}
\label{Fig:ParametrizationeP}
\end{figure}

The first key observation is that the observer must lie between the turning points of the orbits, in order to perceive it; that is one must have $\xi_1 < \xi_{\textrm{obs}} < \xi_2$. Using Eq.~(\ref{Eq:ParametrizationeP}) and the restriction $p > 6 + 2e$ this yields the following restrictions for $p$:
\begin{equation}
p^{(1)}_{\textrm{min}}(\xi_{\textrm{obs}}, e) , p^{(2)}_{\textrm{min}}(\xi_{\textrm{obs}}, e) 
 < p < p_{\textrm{max}}(\xi_{\textrm{obs}}, e),
\end{equation}
where
\begin{eqnarray}
\label{Eq:Pmax}
p_{\textrm{max}}(\xi_{\textrm{obs}}, e) &:=& (1 + e) \xi_{\textrm{obs}}, \\
\label{Eq:Pmin12}
p^{(1)}_{\textrm{min}}(\xi_{\textrm{obs}}, e) &:=& p_{ISO}(e) = 6 + 2e,\qquad
p^{(2)}_{\textrm{min}}(\xi_{\textrm{obs}}, e) := (1 - e) \xi_{\textrm{obs}}.
\end{eqnarray}
To determine this range more explicitly, we first note that the equations $p_{\text{min}}^{(1)}(\xi_{\text{obs}}, e) = p_{\text{max}}(\xi_{\text{obs}}, e)$ and  $p_{\text{min}}^{(1)}(\xi_{\text{obs}}, e) = p_{\text{min}}^{(2)}(\xi_{\text{obs}}, e)$ are satisfied if and only if $e = e^{\text{(1)}}_{\text{c}}(\xi_{\text{obs}})$ and $e = e^{\text{(2)}}_{\text{c}}(\xi_{\text{obs}})$, respectively, with
\begin{equation}
e^{\text{(1)}}_{\text{c}}(\xi_{\text{obs}}) = \frac{\xi_{\text{obs}} - 6}{2 - \xi_{\text{obs}}},\qquad
e^{\text{(2)}}_{\text{c}}(\xi_{\text{obs}}) = \frac{\xi_{\text{obs}} - 6}{2 + \xi_{\text{obs}}},\label{Eq:ec}
\end{equation}
where the first function lies in the required range $0\leq e < 1$ if $4 < \xi_{\text{obs}} \leq 6$ and the second one if $\xi_{\text{obs}} \geq 6$. These observations allow one to conclude that the permitted values of $(p,e)$ are given by:
\begin{eqnarray}
4 < \xi_{\text{obs}}\leq 6 &:& p_{\text{min}}^{(1)}(\xi_{\text{obs}}, e) < p < p_{\text{max}}(\xi_{\text{obs}}, e)  \quad\hbox{and}\quad e_{\text{c}}^{(1)}(\xi_{\text{obs}}) < e < 1. \\
\xi_{\text{obs}} > 6 &:& p_{\text{min}}^{(2)}(\xi_{\text{obs}}, e) < p < p_{\text{max}}(\xi_{\text{obs}}, e)  \quad\hbox{for} \quad 0 < e < e_{\text{c}}^{(2)}(\xi_{\text{obs}}), \nonumber\\ 
 && p_{\text{min}}^{(1)}(\xi_{\text{obs}}, e) < p < p_{\text{max}}(\xi_{\text{obs}}, e)
  \quad \hbox{for} \quad e_{\text{c}}^{(2)}(\xi_{\text{obs}}) < e < 1. 
\end{eqnarray}
We illustrate two representative examples of the resulting domain in the $(p,e)$-plane in figure~\ref{Fig:ParametrizationeP}. Note that $e$ cannot approach zero when $\xi_{\text{obs}} < 6$, which is compatible with the fact that there are no stable circular orbits in this region.

\begin{figure}[h!]
\centerline{
\subfigure{\includegraphics[scale=0.275]{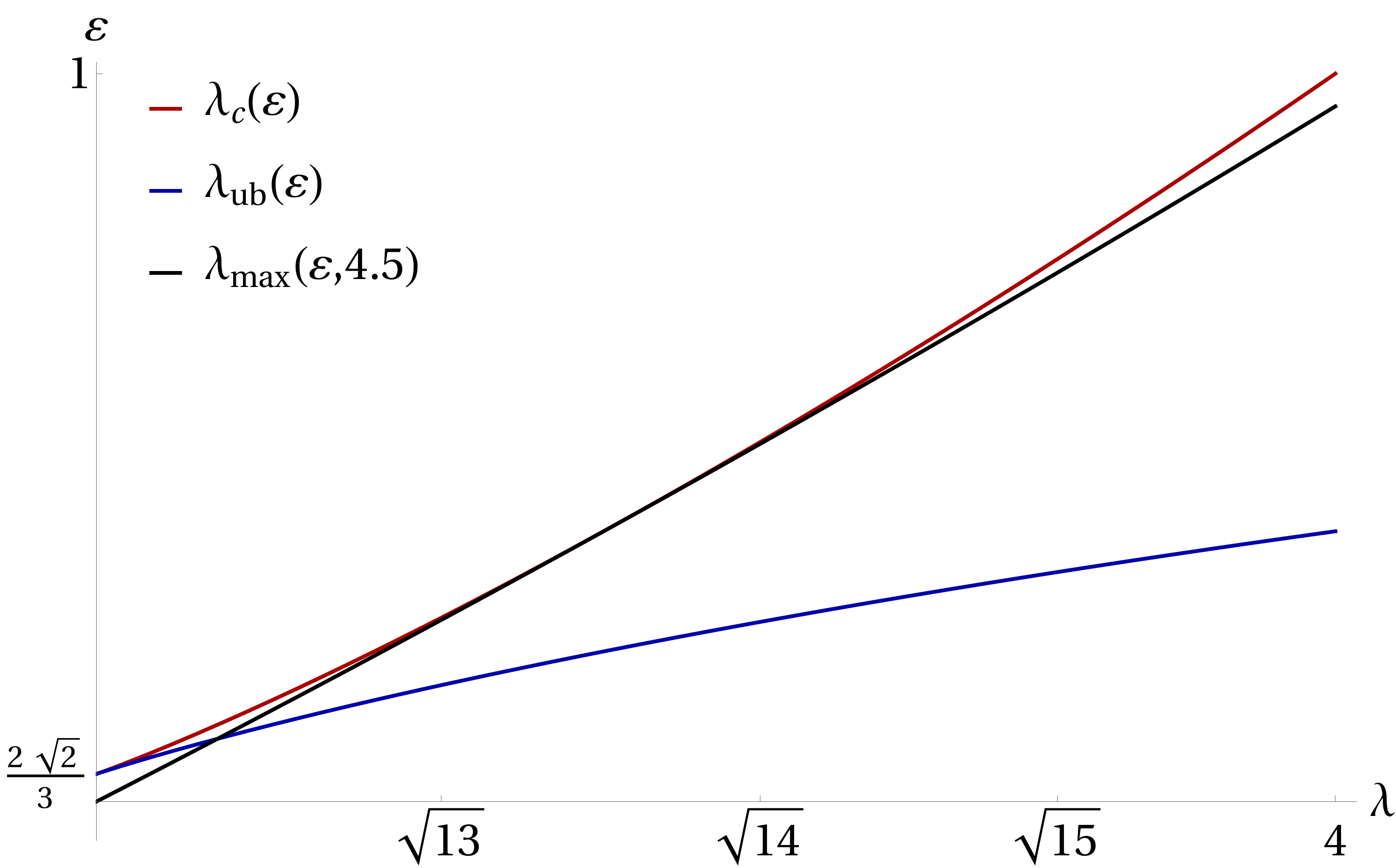}} 
\subfigure{\includegraphics[scale=0.275]{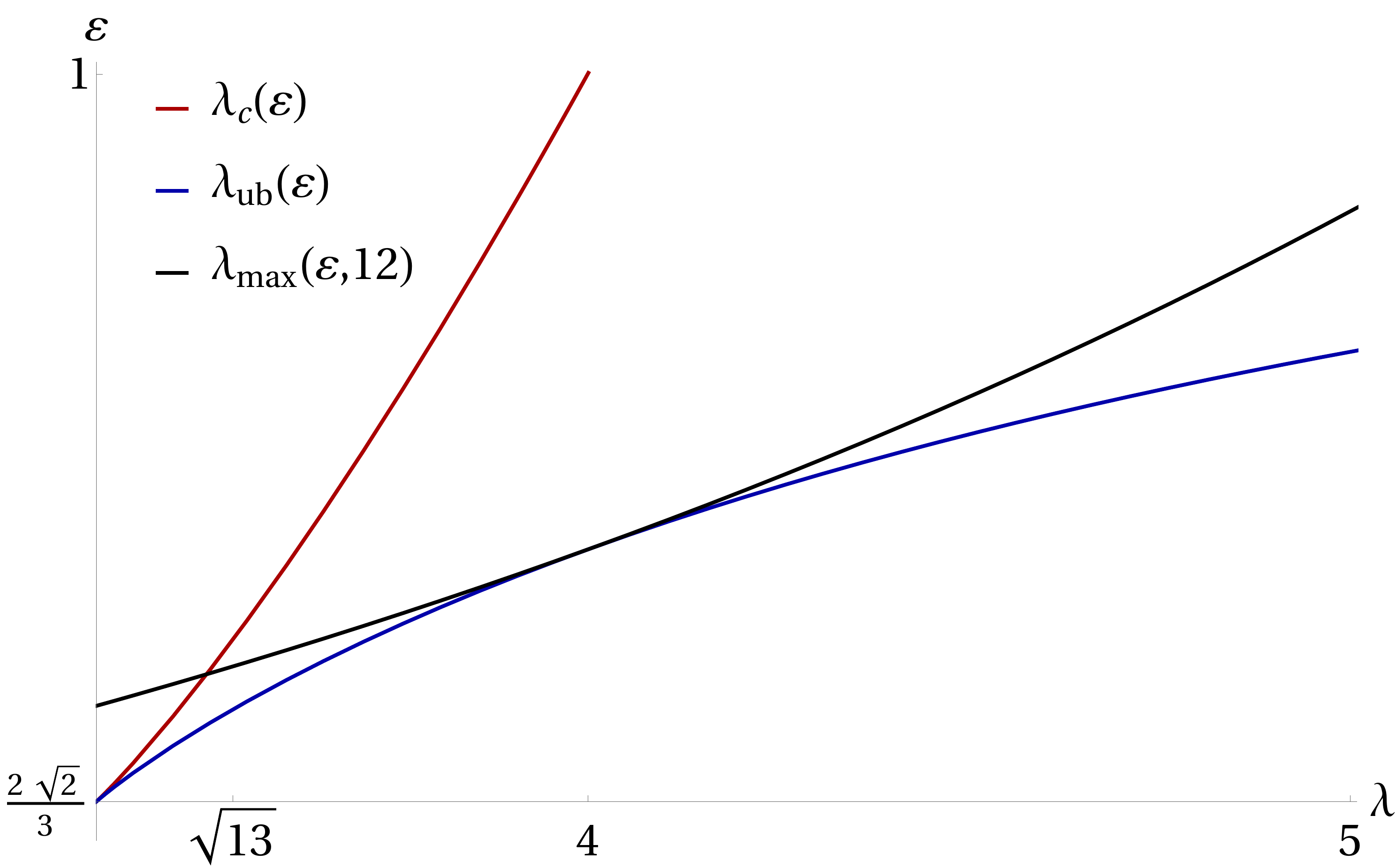}}}
\caption{The curves $\lambda = \lambda_{\textrm{c}}(\varepsilon)$, $\lambda = \lambda_{\textrm{up}}(\varepsilon)$ and $\lambda = \lambda_{\textrm{max}}(\varepsilon,\xi)$ which are relevant for computing the integration domain corresponding to a spacetime observable at $\xi$. Left panel: $\xi = 4.5$. Note that the curves $\lambda = \lambda_{\textrm{c}}(\varepsilon)$ and $\lambda = \lambda_{\textrm{max}}(\varepsilon,\xi)$ ``touch each other'' at $m\varepsilon = E_c(r)$. Right panel: $\xi = 12$. In this case, it is the curves $\lambda = \lambda_{\textrm{up}}(\varepsilon)$ and $\lambda = \lambda_{\textrm{max}}(\varepsilon,\xi)$ that touch each other. In both cases the relevant domain is delimited by the curves  $\lambda = \lambda_{\textrm{c}}(\varepsilon)$ and $\lambda = \lambda_{\textrm{max}}(\varepsilon,\xi)$.}
\label{Fig:ParametrizationEL}
\end{figure}

Now that we have determined the correct domain of integration in terms of the variables $(p,e)$, we translate the results to the variables $(E,L)$ by means of the transformation $(p,e)\mapsto (\varepsilon,\lambda)$ defined in Eq.~(\ref{Eq:EnergyAngularMomentum}). To understand the resulting region, it is sufficient to map the boundary of the allowed region in the $(p,e)$ plane, which consists of straight line segments. First, note that the line segment with $e=1$ is mapped to $\varepsilon = 1$. Next, consider the line $p = p_{\text{min}}^{(1)}(\xi_{\text{obs}}, e)$ corresponding to the ISOs. Its image under the the transformation $(p,e)\mapsto (\varepsilon,\lambda)$ yields
\begin{equation}
\varepsilon^2_{\text{ISO}} = \frac{8}{9-e^2},\qquad
\lambda^2_{\text{ISO}} = \frac{4(3+e)^2}{(1+e)(3-e)},\qquad 0 < e < 1.
\label{Eq:EpsilonLambdaISO}
\end{equation}
Eliminating $e$ from these expression one obtains $\lambda = \lambda_{\textrm{c}}(\varepsilon)$ with
\begin{equation}
\lambda_{\textrm{c}}(\varepsilon) := \frac{4\sqrt{2}}{\sqrt{36 \varepsilon^2 - 8 - 27\varepsilon^4 + \varepsilon \sqrt{\left[ 9\varepsilon^2 - 8\right]^3}}},\qquad
\frac{8}{9} < \varepsilon^2 < 1,
\label{Eq:Lambdac}
\end{equation}
which is precisely the critical angular momentum $L_{\textrm{c}}/(M m)$ introduced below Eq.~(\ref{Eq:SchEffectivePotential}). Finally, by noticing that $p = p_{\text{min}}^{(2)}(\xi_{\text{obs}}, e)$ and $p = p_{\text{max}}(\xi_{\text{obs}}, e)$ correspond to the turning points, where the effective potential $V_{m,L}(r)$ is equal to $E^2$, one obtains $\lambda = \lambda_{\textrm{max}}(\varepsilon,\xi)$ with
\begin{equation}
\lambda_{\textrm{max}}(\varepsilon,\xi) := \xi \sqrt{\frac{\varepsilon^2}{1 - \frac{2}{\xi}} - 1}.
\label{Eq:MaximalAngularMomentum}
\end{equation}
In fact, this expression can also be obtained by substituting either $p = (1+e)\xi_{\textrm{obs}}$ or $p = (1-e)\xi_{\textrm{obs}}$ into Eq.~(\ref{Eq:EnergyAngularMomentum}) and eliminating $e$. At first sight, it seems curious that the wedge-like portion of the boundary consisting of the two line segments $p = (1\pm e)\xi_{\textrm{obs}}$ is transformed into the smooth curve $\lambda = \lambda_{\textrm{max}}(\varepsilon,\xi)$. The reason for this relies in the factor $e$ in Eq.~(\ref{Eq:Elements}), which implies that the transformation $(p,e)\mapsto (\varepsilon,\lambda)$ is singular at $e=0$, which is the location where the two line segments cross each other. The point corresponding to the minimum value of $p$ corresponds to the point of minimal energy $\varepsilon$. It can be obtained by substituting Eq.~(\ref{Eq:ec}) into the expression for the energy in Eq.~(\ref{Eq:EpsilonLambdaISO}), which yields
\begin{equation}
\varepsilon_{\text{c}}(\xi) := \left\{
\begin{array}{lcl}
\displaystyle \frac{\xi - 2}{\sqrt{\xi(\xi - 3)}} & \hbox{for} & 4 \leq \xi \leq 6, \\
 & & \\
\displaystyle \frac{\xi + 2}{\sqrt{\xi\left(\xi + 6\right)}} & \hbox{for} & \xi \geq 6. 
\end{array}
\right.
\label{Eq:MinimumEnergy}
\end{equation}
We conclude that the allowed region for bound trajectories which intersect an observer located at $\xi = \xi_{\textrm{obs}}$ is given by
\begin{equation}
\lambda_{\textrm{c}}(\varepsilon) < \lambda < \lambda_{\textrm{max}}(\varepsilon,\xi) \quad \hbox{and} \quad \varepsilon_{\text{c}}(\xi) < \varepsilon < 1.
\end{equation}
This coincides with the result in Eq.~(\ref{Eq:RangeIntegration}) which is derived in section~\ref{SubSec:ExplicitObservables} by different means. Figure~\ref{Fig:ParametrizationEL} shows the behavior of the curves $\lambda = \lambda_{\textrm{c}}(\varepsilon)$, $\lambda = \lambda_{\textrm{ub}}(\varepsilon)$ and $\lambda = \lambda_{\textrm{max}}(\varepsilon,\xi)$ for two representative values of $\xi$.

\section{Integrals over angular momentum}
\label{App:Integrals}

In this appendix we list the relevant integrals that are used in subsection~\ref{SubSec:Observables} in order to compute the spacetime observables. For the integrals over the total angular momentum $L$, we first introduce the notation $L_1 := |L_z|/\sin\vartheta$, $L_2 := L_{\text{max}}(E,r)$, and $L_3 := \max\{ L_{\text{c}}(E), |L_z|/\sin\vartheta \}$. By virtue of Eq.~(\ref{Eq:RangeIntegration}) these quantities satisfy the inequality $L_1\leq L_3\leq L_2$. By means of the variable substitution
\begin{equation}
\sin^2\phi = \frac{L^2 - L_1^2}{L_2^2 - L_1^2},\qquad 0\leq \phi\leq \frac{\pi}{2},
\end{equation}
one finds
\begin{eqnarray}
\int\limits_{L_3}^{L_2}  \frac{L dL}{\sqrt{L^2 - L_1^2}\sqrt{L_2^2 - L^2}} 
&=& \frac{\pi}{2} - \arcsin\sqrt{\frac{L_3^2 - L_1^2}{L_2^2 - L_1^2}}
 = \frac{\pi}{2} - \arctan\sqrt{\frac{L_3^2 - L_1^2}{L_2^2 - L_3^2}},
 \\
\int\limits_{L_3}^{L_2} \frac{\sqrt{L_2^2 - L^2}}{\sqrt{L^2 - L_1^2}} L dL 
&=& \frac{1}{2}\left(L_2^2 - L_1^2\right)\left(\frac{\pi}{2} - \arctan\sqrt{\frac{L_3^2 - L_1^2}{L_2^2 - L_3^2}}\right) - \frac{1}{2}\sqrt{L_3^2 - L_1^2}\sqrt{L_2^2 - L_3^2},
\\
\int\limits_{L_3}^{L_2}  \frac{\sqrt{L^2 - L_1^2}}{\sqrt{L_2^2 - L^2}} L dL
&=& \frac{1}{2}\left(L_2^2 - L_1^2\right)\left(\frac{\pi}{2} - \arctan\sqrt{\frac{L_3^2 - L_1^2}{L_2^2 - L_3^2}}\right) + \frac{1}{2}\sqrt{L_3^2 - L_1^2}\sqrt{L_2^2 - L_3^2}.
\end{eqnarray}
Note that the contributions from the $\arctan$ on the right-hand side of these equations vanish when $L_3 = L_1$, that is, when $|L_z|/\sin\vartheta\geq L_c(E)$.

In order to treat the integrals over the azimuthal angular momentum for the models~(\ref{Eq:PolytropeLzEven},\ref{Eq:PolytropeLzRot}) which appear in equations~(\ref{Eq:Jhat03},~\ref{Eq:T00T03T33},~\ref{Eq:A},~\ref{Eq:B}) we use the following equations for $a > b > 0$ which can be obtained using integration by parts,
\begin{eqnarray}
a \int\limits_{1/a}^{1} d\lambda_z \sqrt{1 - \lambda_z^2}\left(a\lambda_z-1\right)^l_+ &=& \frac{1}{l+1} \int\limits_{1/a}^{1}\frac{d\lambda_z \lambda_z}{\sqrt{1-\lambda_z^2}}\left(a\lambda_z -1\right)^{l+1}_+ ,
\\
a \int\limits_{1/a}^{1} d\lambda_z \left(a\lambda_z-1\right)^l_+ \arctan\sqrt{\frac{1-\lambda_z^2}{(1/b^2) - 1}} &=& \frac{b\sqrt{1-b^2}}{l+1}\int\limits_{1/a}^1 \frac{\lambda_z d\lambda_z}{\sqrt{1-\lambda_z^2}}\frac{(a\lambda_z-1)^{l+1}_+}{(1-b^2\lambda_z^2)} = \frac{\pi}{2}\frac{b}{l+1}\tilde{\text{K}}_l(a,b),
\\
a^2 \int\limits_{1/a}^{1} d\lambda_z \lambda_z \left(a\lambda_z-1\right)^l_+ \arctan\sqrt{\frac{1-\lambda_z^2}{(1/b^2) - 1}} &=& \int\limits_{1/a}^1 \frac{\lambda_z d\lambda_z}{\sqrt{1-\lambda_z^2}}\frac{b\sqrt{1-b^2}}{(1-b^2\lambda_z^2)} \left[\frac{(a\lambda_z-1)^{l+1}_+}{l+1} + \frac{(a\lambda_z-1)^{l+2}_+}{l+2}\right] \nonumber\\
&=& \frac{\pi}{2} b \left[ \frac{1}{l+1} \tilde{\text{K}}_l(a,b) + \frac{1}{l+2}\tilde{\text{K}}_{l+1}(a,b) \right],
\\
a^3 \int\limits_{1/a}^{1} d\lambda_z \lambda_z^2 \left(a\lambda_z-1\right)^l_+ \arctan\sqrt{\frac{1-\lambda_z^2}{(1/b^2) - 1}} &=& \frac{\pi}{2} b \left[ \frac{1}{l+1}\tilde{\text{K}}_l(a,b) + \frac{2}{l+2}\tilde{\text{K}}_{l+1}(a,b) + \frac{1}{l+3}\tilde{\text{K}}_{l+2}(a,b) \right],
\end{eqnarray}
where we recall the definition of the function $\tilde{\text{K}}_l(a,b)$ defined in Eq.~(\ref{Eq:KtildeDef}).

\section{Properties of the integral kernels $\tilde{\text{K}}_l(a,b)$}
\label{App:IntegralKernel}

In this appendix we summarize some elementary properties of the integral
\begin{equation}
\tilde{\text{K}}_l(a,b) := \frac{2}{\pi} \sqrt{1 - b^2} \int\limits_{1/a}^1 \frac{d\lambda \lambda}{\sqrt{1-\lambda^2}} \frac{(a\lambda - 1)^{l+1}}{1 - b^2\lambda^2},\qquad
a > 1, \quad 0 < b < 1,
\label{Eq:Kltilde}
\end{equation}
which was defined in Eq.~(\ref{Eq:KtildeDef}). First, observe that $\tilde{\text{K}}_l(a,b)$ is continuous in $(a,b)$ and that
\begin{equation}
\lim\limits_{a \rightarrow 1} \tilde{\text{K}}_l(a,b) = 0.
\end{equation}
In order to analyze the behavior for large $a$ and the limits $b\to 0$ and $b\to 1$ it is convenient to perform the variable substitution $\mu := \sqrt{1-\lambda^2}/\sqrt{1-b^2}$, which transforms the integral into
\begin{equation}
\tilde{\text{K}}_l(a,b) := \frac{2}{\pi}\int\limits_{0}^{\frac{1}{a}\sqrt{ \frac{a^2-1}{1-b^2}}}
\frac{d\mu}{1 + b^2\mu^2} \left[ a\sqrt{1 - (1-b^2)\mu^2} - 1 \right]^{l+1}.
\label{Eq:KltildeNew}
\end{equation}
From this, we see that for all $a > 1$,
\begin{equation}
\lim\limits_{b\rightarrow 1} \tilde{\text{K}}_l(a,b) = \frac{2}{\pi}\int\limits_0^\infty \frac{d\mu}{1+\mu^2} (a-1)^{l+1} = (a-1)^{l+1}.
\end{equation}
Furthermore, using the fact that $\sqrt{a^2 - 1}/a \leq 1$, one obtains the estimate
\begin{equation}
0\leq \tilde{\text{K}}_l(a,b) \leq \frac{2}{\pi b}\arctan\left( \frac{b}{\sqrt{1- b^2}} \right) (a-1)^{l+1}
\leq (a-1)^{l+1},\qquad
a > 1,\quad 0 < b < 1.
\end{equation}
From Eq.~(\ref{Eq:Kltilde}) and the variable substitution $\lambda = (1 - 1/a) t + 1/a$ one also obtains the limit
\begin{equation}
\tilde{\text{K}}_l(a,0) := \lim\limits_{b\rightarrow 0} \tilde{\text{K}}_l(a,b) 
 = \frac{1}{\sqrt{\pi}}\frac{\Gamma(l+2)}{\Gamma\left(l + \frac{5}{2}\right)}
 \sqrt{1 - \frac{1}{a^2}} (a-1)^{l+1}
 {}_{2}F_{1}\left( -\frac{1}{2}, l + 1, l + \frac{5}{2},-\frac{a-1}{a+1} \right),
\end{equation}
with ${}_{2}F_1(a,b;c; z)$ the Gauss hypergeometric function as defined, for instance, in Eq.~(9.111) of~\cite{iGiR2007}.

\begin{figure}[h]
\centerline{
\subfigure[\; $l=0$.]{\includegraphics[scale=0.29]{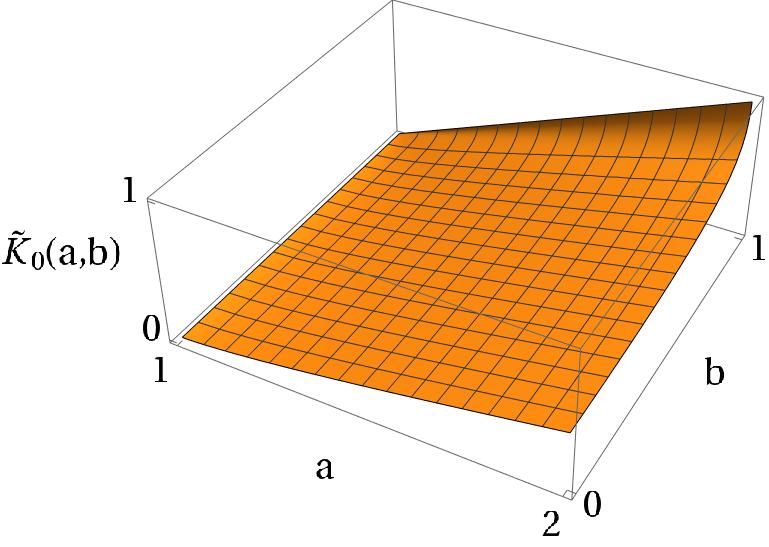}}
\subfigure[\;$l=1$.]{\includegraphics[scale=0.29]{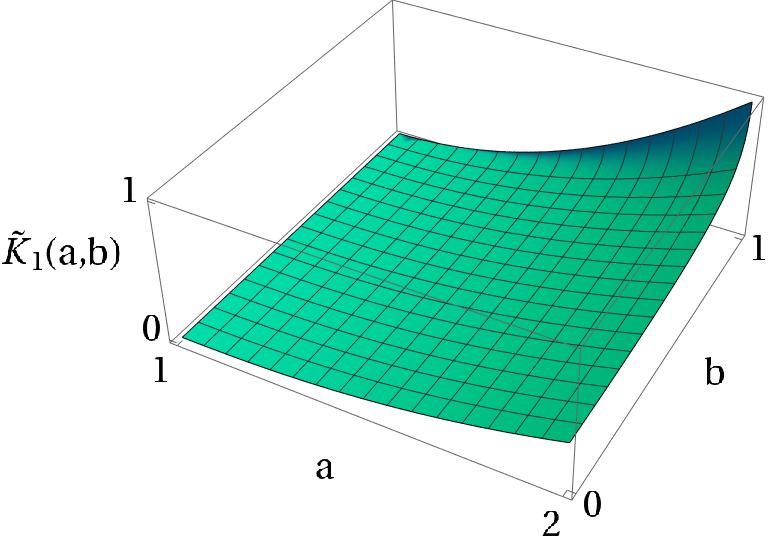}}
\subfigure[\;$l=2$.]{\includegraphics[scale=0.29]{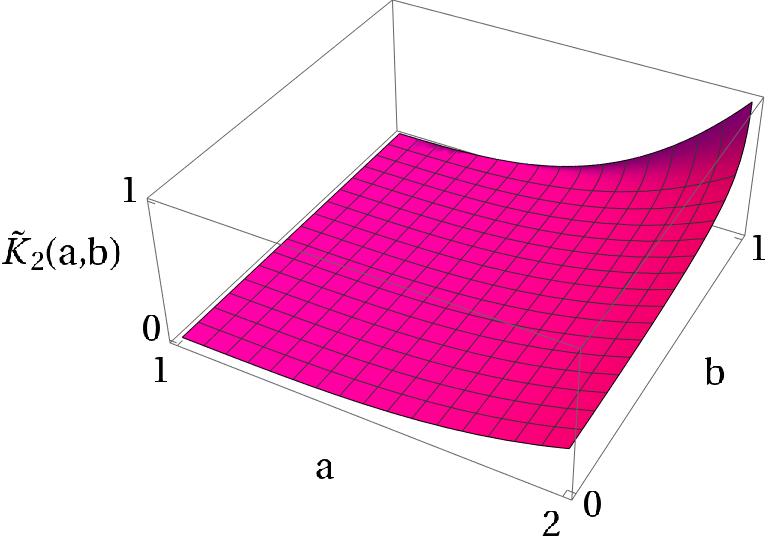}}}
\caption{Behavior of the integrals $\tilde{\text{K}}_l(a,b)$ for $l = 0,1,2$.}
\label{Fig:Kltilde}
\end{figure}

Although we have not found an explicit expression for $\tilde{\text{K}}_l(a,b)$ which is valid for \emph{generic} values of $l$, the integral can be computed explicitly for the specific values of $l=0,1,2$:
\begin{eqnarray}
\tilde{\text{K}}_0(a,b) &=& \frac{a}{b^2}\left[ 1 - \sqrt{1 - b^2}\left(1 - \frac{2}{\pi}\arcsin\frac{1}{a} \right)\right] \nonumber\\
&& +\; \frac{a-b}{\pi b^2}\arctan\frac{ab-1}{\sqrt{(a^2 -1)(1-b^2)}} - \frac{a+b}{\pi b^2}\arctan\frac{ab+1}{\sqrt{(a^2 -1)(1-b^2)}},
\\ 
\tilde{\text{K}}_1(a,b) &=& -\frac{2a}{b^2}\left[\frac{1}{\pi}\sqrt{(a^2 -1)(1-b^2)} + 1 - \sqrt{1 - b^2}\left(1 - \frac{2}{\pi}\arcsin\frac{1}{a} \right)\right] \nonumber\\
& & +\; \frac{(a-b)^2}{\pi b^3}\arctan\frac{ab-1}{\sqrt{(a^2 -1)(1-b^2)}} + \frac{(a+b)^2}{\pi b^3}\arctan\frac{ab+1}{\sqrt{(a^2 -1)(1-b^2)}},
\\
\tilde{\text{K}}_2(a,b) &=& \frac{a^3}{b^4}\left[1-\sqrt{1-b^2}\left(1+\frac{b^2}{2}\right)\left(1-\frac{2}{\pi}\arcsin\frac{1}{a}\right) \right] + \frac{3 a}{b^2}\left[\frac{5}{3\pi}\sqrt{(a^2-1)(1-b^2)}+1-\sqrt{1-b^2}\left(1-\frac{2}{\pi}\arcsin\frac{1}{a} \right)\right] \nonumber\\
& & +\; \frac{(a-b)^3}{\pi b^4}\arctan\frac{ab-1}{\sqrt{(a^2 -1)(1-b^2)}} - \frac{(a+b)^3}{\pi b^4}\arctan\frac{ab+1}{\sqrt{(a^2 -1)(1-b^2)}}.
\end{eqnarray}
We show corresponding plots of these functions in figure~\ref{Fig:Kltilde}.

\section{Short review of polish doughnuts configurations}
\label{App:PolishDoughnuts}

In this appendix we briefly review the ``polish doughnuts" configurations describing stationary and axisymmetric hydrodynamic thick disks surrounding black holes (for more details see Refs.~\cite{lFvM1976,mAmJmS1978,mKmJmA1978,jFfD2002,jFfD2004,oSetal2012,Straumann-Book,Rezzolla-Book}). To describe these configurations, we assume that the spacetime is stationary, axisymmetric and circular (the Kerr spacetime satisfies these properties; however for the moment we may assume a metric more general than Kerr). It can be shown (see for instance~\cite{Heusler-Book}) that such a spacetime admits local coordinates $(t,r,\vartheta,\varphi)$ in which the metric coefficients depend neither on the time coordinate $t$ (stationarity) nor the azimuthal coordinate $\varphi$ (axisymmetry), and such that $g_{tr} = g_{t\vartheta} = g_{\varphi r} = g_{\varphi\vartheta} = 0$ (circularity), that is,
\begin{equation}
g = g_{tt}dt^2 + 2g_{t\varphi} dtd\varphi + g_{rr}dr^2 + g_{\vartheta\vartheta}d\vartheta^2 + g_{\varphi\varphi}d\varphi^2.
\end{equation}
The Killing vector fields associated with this spacetime are $k = \partial_t$ and $v = \partial_\varphi$.

Next, consider a perfect fluid flow in circular motion, which means that its four-velocity $u$ is a linear combination of $k$ and $v$, such that
\begin{equation}
u = A\left( k + \Omega v \right),
\label{Eq:Circular}
\end{equation}
for some normalization constant $A > 0$ and a function $\Omega$ describing the angular velocity of the flow. Denoting by $\rho = n m$, $h$ and $P$ the rest mass density, specific enthalpy and pressure, the relativistic Euler equations can be written as
\begin{equation}
a_\nu = -\frac{1}{\rho h}(\nabla_\nu P + u_\nu\nabla_u P),
\label{Eq:Euler}
\end{equation}
with $a_\nu = (\nabla_u u)_\nu = u^\mu\nabla_\mu u_\nu$ the acceleration of the fluid elements.

For the following, we impose the Killing symmetries on the fluid quantities, such that $\rho$, $h$, $p$ and $u$ are invariant with respect to the flows generated by $k$ and $v$. Assuming an adiabatic fluid in local thermodynamic equilibrium, such that $dh = dp/\rho$, one can show that the Bernoulli-type quantities
\begin{equation}
\mathcal{B} := -h u_\mu k^\mu = - h u_t, \qquad
\mathcal{L} := h u_\mu v^\mu = h u_\varphi,
\end{equation}
are constant along the flow lines. For the following, it is convenient to introduce the ``fluid angular momentum" (the angular momentum per energy of a fluid element), defined as
\begin{equation}
\ell := \frac{\mathcal{L}}{\mathcal{B}} = -\frac{u_\varphi}{u_t}.
\end{equation}
Combining this definition with the equation $\Omega = u^\varphi/u^t$ and the normalization condition $u_\mu u^\mu = -1$ yields
\begin{equation}
u_t u^t = -\frac{1}{1 - \Omega\ell},
\label{Eq:utut}
\end{equation}
which allows one to express the components $u^t$, $u^\varphi = \Omega u^t$, $u_\varphi = -\ell u_t$ in terms of $u_t$, $\Omega$ and $\ell$. Computing the normalization constant $A = u^t$ and using Eq.~(\ref{Eq:utut}) one also finds
\begin{equation}
u_t = -\frac{1}{1-\Omega\ell}\sqrt{-g_{tt} - 2\Omega g_{t\varphi} - \Omega^2 g_{\varphi\varphi}}.
\label{Eq:ut}
\end{equation}
Using Eq.~(\ref{Eq:utut}), the formula $a_\nu = u^\mu(\partial_\mu u_\nu - \partial_\nu u_\mu)$, and taking into account the Killing symmetries, one obtains\footnote{An elegant way of deriving Eq.~(\ref{Eq:FluidAcceleration}) makes use of the calculus for differential forms. Denoting by $\underline{u} = u_\mu dx^\mu$ the one-form associated with the four-velocity, one obtains from Eq.~(\ref{Eq:Circular})
\begin{equation}
a = i_u d\underline{u} = A(i_k d\underline{u} + \Omega i_v d\underline{u}) 
= -A(d i_k\underline{u} + \Omega d i_v\underline{u})
= -A\left[ d u_t - \Omega d(\ell u_t) \right]\nonumber
\end{equation}
where we have used the Cartan formula $\pounds_k\underline{u} = i_k d\underline{u} + di_k\underline{u}$ in the third step. Taking into account that $A = u^t$ and using Eq.~(\ref{Eq:utut}) yields the desired result.
}
\begin{equation}
a_\nu = \partial_\nu \log|u_t| - \left(\frac{\Omega}{1-\Omega\ell} \right)\partial_\nu \ell.
\label{Eq:FluidAcceleration}
\end{equation}
Euler's equation~(\ref{Eq:Euler}) together with $\nabla_u P = 0$ and the first law lead to
\begin{equation}
-d\log[h] = d\log|u_t| - \left( \frac{\Omega}{1-\Omega\ell}\right) d \ell.
\label{Eq:FluidAccelerationh}
\end{equation}
Taking the exterior derivative on both sides yields $d\Omega\wedge d\ell = 0$, implying that $d\Omega$ is proportional to $d\ell$. This in turn implies that the angular velocity $\Omega$ only depends on the specific angular momentum (or vice-versa)~\cite{mA1971}. On the other hand, using $u_\mu = g_{\mu\nu} u^\nu$ one finds the following relation between $\ell$ and $\Omega$:
\begin{equation}
g_{tt}\ell +g_{t\varphi}\left(1+\Omega\ell \right) + g_{\varphi\varphi}\Omega = 0.
\label{Eq:vonZeipelCylinder}
\end{equation}
The surfaces of constant $\ell$ and $\Omega$ describe the ``von Zeipel cylinders".

After these remarks, stationary, axisymmetric perfect fluid configuration in circular motion can be described as follows. One possibility is to specify a function $\Omega_0(\ell)$ of $\ell$, such that Eq.~(\ref{Eq:FluidAccelerationh}) can be integrated to yield
\begin{equation}
\log(h) = -\log|u_t|
 + \int\limits^\ell \frac{\Omega_0(\ell') d\ell'}{1 - \Omega_0(\ell')\ell'}.
\label{Eq:logh}
\end{equation}
Here, $u_t$ is given by Eq.~(\ref{Eq:ut}) with $\Omega$ replaced by $\Omega_0(\ell)$ and $\ell$ is the function of $(r,\vartheta)$ which is determined from Eq.~(\ref{Eq:vonZeipelCylinder}) by setting $\Omega = \Omega_0(\ell)$. The support of the fluid configuration is determined by the region for which the right-hand side of Eq.~(\ref{Eq:logh}) is positive, such that $h > 1$. Note that the location of the boundary surface $h=1$ depends on a free integration constant. By construction, the quantity $h |u_t|$ is constant on the von Zeipel cylinders.

Alternatively, one can fix $\ell = \ell_0$ to be constant throughout the fluid configuration, such that
\begin{equation}
\log(h) = -\log|u_t| + const,
\label{Eq:loghBis}
\end{equation}
with $u_t$ given again by Eq.~(\ref{Eq:ut}) where now $\Omega$ is the function of $(r,\vartheta)$ determined by Eq.~(\ref{Eq:vonZeipelCylinder}), which yields
\begin{equation}
\Omega(r,\vartheta) = -\frac{g_{t\varphi} + g_{tt}\ell_0}{g_{\varphi\varphi} + g_{t\varphi}\ell_0}.\label{Eq:Omega}
\end{equation}
Like in the previous case, the support of the fluid configuration corresponds to the region for which the right-hand side of Eq.~(\ref{Eq:loghBis}) is positive. For a Schwarzschild black hole, these expressions simplify considerably and one obtains
\begin{equation}
\log h(r,\vartheta) = W_{\textrm{in}} -W_{\ell_0}(r,\vartheta)
\end{equation}
with an integration constant $W_{\textrm{in}}$ and the ``effective potential"
\begin{equation}
W_{\ell_0}(r,\vartheta) := \log|u_t| = -\frac{1}{2}\log\left( \frac{r}{r - 2M} - \frac{\ell^2_0}{r^2\sin^2\vartheta} \right),\qquad
r > 2M,\quad \sin^2\vartheta > \ell_0^2\frac{r-2M}{r^3}.
\label{Eq:W}
\end{equation}
As can easily be verified, the critical points of $W_{\ell_0}(r,\vartheta)$ are located on the equatorial plane $\vartheta = \pi/2$ at radii determined by the equation
\begin{equation}
\ell_0^2(r-2M)^2 = M r^3,
\label{Eq:ellcritical}
\end{equation}
where $W_{\ell_0}(r,\pi/2) = -\log[\sqrt{r(r-3M)}/(r-2M)]$. There are no critical points when $\ell_0 < \ell_{\textrm{c}} = \sqrt{27/2} M \approx 3.67M$. For $\ell_0 > \ell_{\textrm{c}}$, there are two critical points with radii $2M < r_{\textrm{saddle}} < r_{\textrm{min}}$, the first one describing a saddle and the second one a minimum. As $\ell_0$ increases from $\ell_{\textrm{c}}$ to $\sqrt{27} M$, $r_{\textrm{saddle}}$ decreases monotonously from $6M$ to $3M$ and $W_{\ell_0}(r_{\textrm{saddle}},\pi/2)$ increases from $\log\sqrt{8/9}$ to $\infty$ while $r_{\textrm{min}}$ increases monotonously from $6M$ to $6(2+\sqrt{3})M$ and $W_{\ell_0}(r_{\textrm{min}},\pi/2)$ from $\log\sqrt{8/9}$ to $-\log[3\sqrt{2\sqrt{3}-3}/2]$.

\begin{figure}[h!]
\centerline{
\includegraphics[scale=0.25]{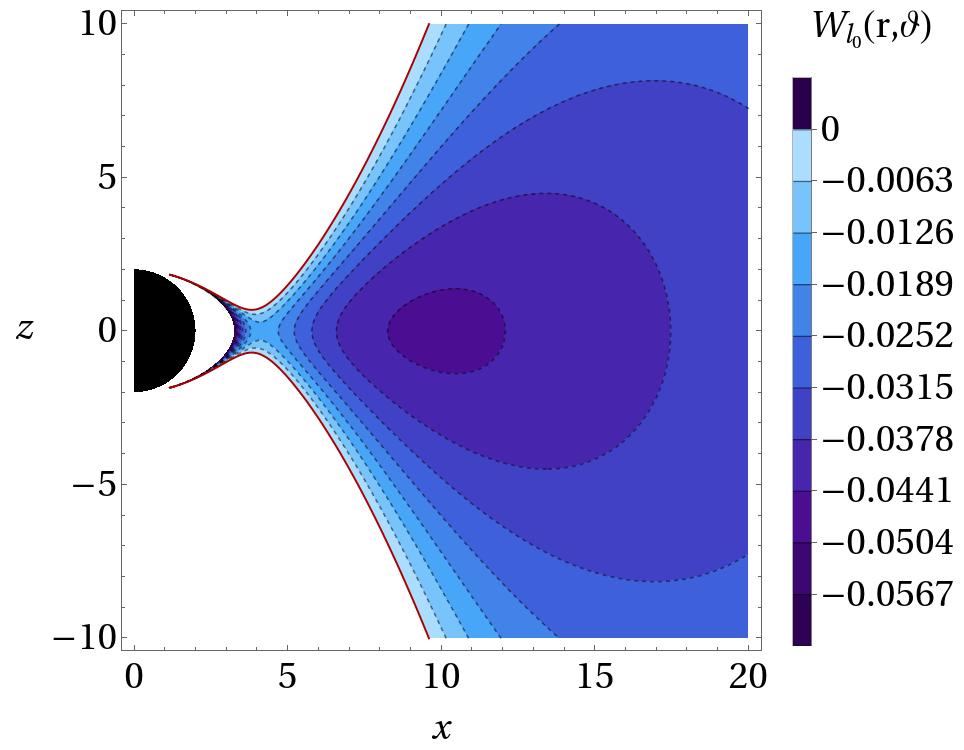}
\includegraphics[scale=0.25]{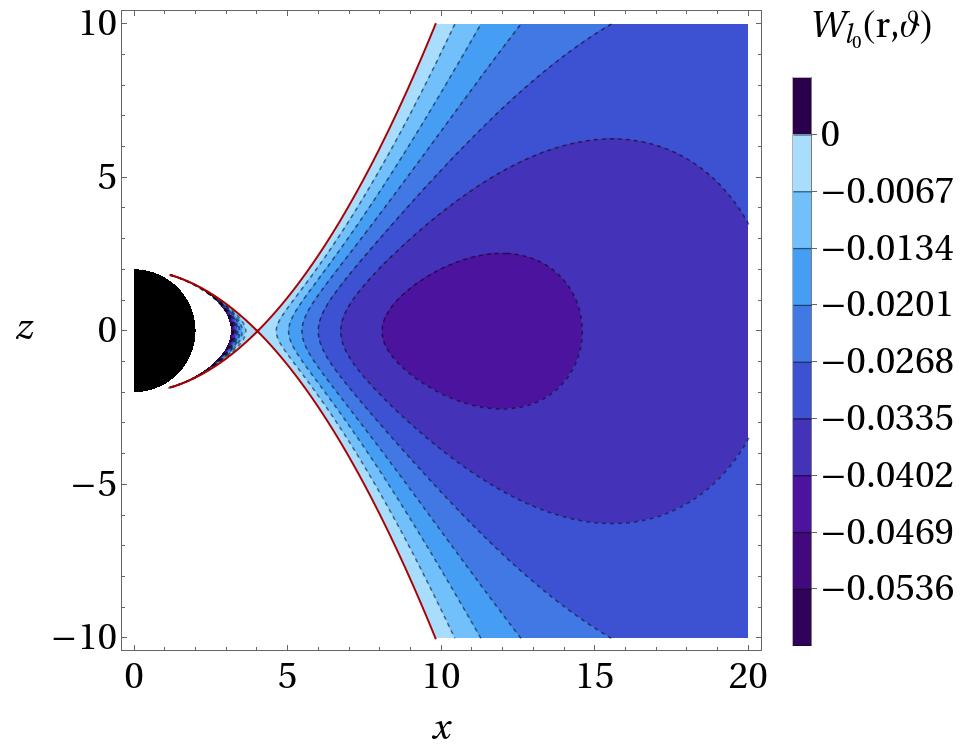}
\includegraphics[scale=0.25]{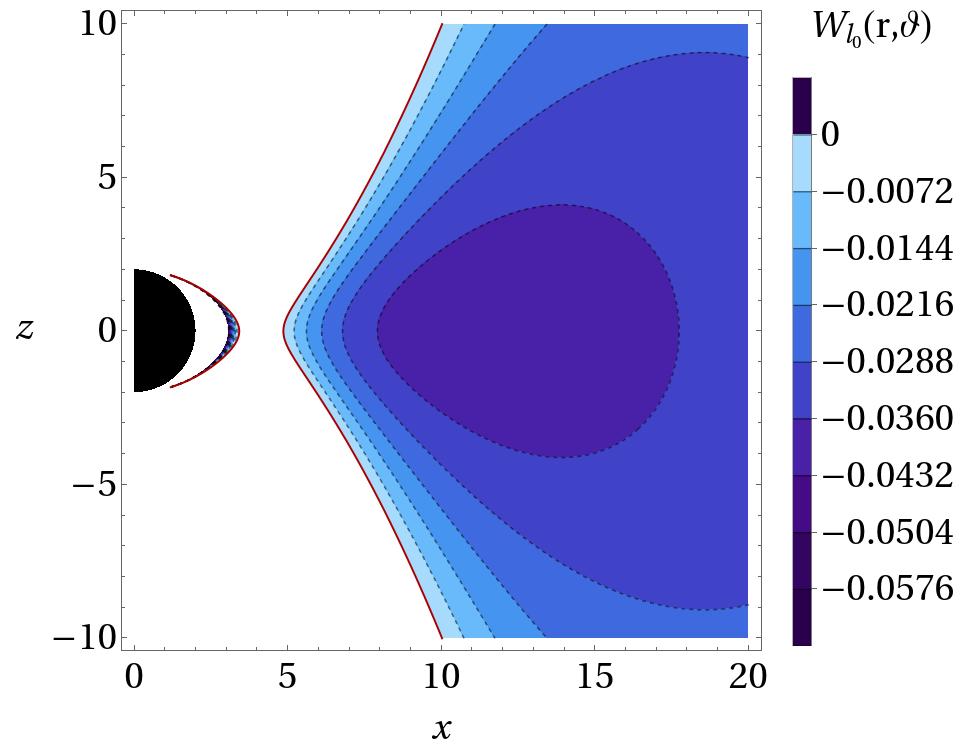}}
\caption{Contour plots for the effective potential defined in Eq.~(\ref{Eq:W}) in the $xz$-plane for the values for the fluid angular momentum given by $\ell_0 = \sqrt{31/2}M,\sqrt{32/2}M,\sqrt{33/2}M$ as shown in the left, center and right panels, respectively. In all plots the continuous red line represents the boundary surface where $W_{\ell_0} = 0$ and the black region represents the Schwarzschild black hole region.}
\label{Fig:PolishDoughnuts}
\end{figure}

From these observations, we see that as long as $\ell_0 > \ell_{\textrm{c}}$, there are bound fluid configurations whose maximum specific enthalpy is located at $r_{\textrm{min}}$ and is constant on the closed level sets surrounding $r_{\textrm{min}}$ (see figure~\ref{Fig:PolishDoughnuts}). The surface of the configuration can be chosen as any of these closed level sets, setting $W_{\textrm{in}}$ equal to the value of $W$ at this level set, such that $h = 1$ at the surface and $h > 1$ inside it. The most extended surface that can be chosen is the one that approaches the separatrix through the saddle point. 

The maximum of $W_{\ell_0}$ lies below its asymptotic value $0$ if $\ell_{\textrm{c}} < \ell_0 < \ell_{\textrm{mb}} = 4M$. The upper limit $\ell_0 = \ell_{\textrm{mb}}$ corresponds to  $r_{\textrm{saddle}} = 4M$ which is the radius of the marginally bound  circular trajectory. For $\ell_0 > \ell_{\textrm{mb}}$ the saddle lies above $0$ and hence bound fluid configurations cannot spill over the saddle and fall into the black hole. Therefore, the range $\ell_{\textrm{c}} < \ell_0 < \ell_{\textrm{mb}} = 4M$ is the one that is relevant for accretion disk models. Note also that at the minimum of $W_{\ell_0}$ the fluid elements follow circular geodesics due to the fact that the gradient of the pressure vanishes there. Recalling that $\ell$ represents the angular momentum per energy of the particles and making use of the formulae $L(r) = m M^{1/2}r^{3/2}/\sqrt{r(r-3M)}$ and $E(r) = m(r - 2M)/\sqrt{r(r-3M)}$ for the angular momentum and energy of the circular timelike geodesics (see Eq.~(\ref{Eq:EnergyAngularMomentum}) with $e=0$), one obtains $\ell_0 = L(r)/E(r) = M^{1/2}r^{3/2}/(r-2M)$ which satisfies equation~(\ref{Eq:ellcritical}).

\bibliographystyle{unsrt}
\bibliography{refs_kinetic}

\begin{thebibliography}{10}

\bibitem{cGoS2022b}
C.~Gabarrete and O.~Sarbach.
\newblock Axisymmetric, stationary collisionless gas clouds trapped in a
  {N}ewtonian potential. \textit{In preparation}.
\newblock 2022.

\bibitem{EHTC}
Event {H}orizon {T}elescope {C}ollaboration.
\newblock \url{http://www.eventhorizontelescope.org}.

\bibitem{EHTCI}
K.~Akiyama and et~al.
\newblock First {M}87 {E}vent {H}orizon {T}elescope {R}esults. {I}. {T}he
  {S}hadow of the {S}upermassive {B}lack {H}ole.
\newblock {\em Astrophys. J. Letters}, 875:17, 2019.

\bibitem{EHTCXII}
The EHT~Collaboration et~al.
\newblock First {S}agittarius {A}* {E}vent {H}orizon {T}elescope {R}esults.
  {I}. {T}he {S}hadow of the {S}upermassive {B}lack {H}ole in the {C}enter of
  the {M}ilky {W}ay.
\newblock {\em Astrophys. J. Letters}, 930:21, 2022.

\bibitem{mKjSeQ2016}
M.~Kunz, J.~Stone, and E.~Quataert.
\newblock Magnetorotational turbulence and dynamo in a collisionless plasma.
\newblock {\em Phys. Rev. Lett.}, 117:235101, Dec 2016.

\bibitem{Heusler-Book}
M.~Heusler.
\newblock {\em Black Hole Uniqueness Theorems}.
\newblock Cambridge University Press, Cambridge, England, 1996.

\bibitem{lrr-2012-7}
P.T. Chru{\'{s}}ciel, J.~Lopes Costa, and M.~Heusler.
\newblock Stationary black holes: Uniqueness and beyond.
\newblock {\em Liv. Rev. Relativ.}, 15(7), 2012.

\bibitem{pRoS2020}
P.~Rioseco and O.~Sarbach.
\newblock Phase space mixing in an external gravitational central potential.
\newblock {\em Class. Quantum Grav.}, 37(19):195027, 2020.

\bibitem{pRoS18}
P.~Rioseco and O.~Sarbach.
\newblock Phase space mixing in the equatorial plane of a {K}err black hole.
\newblock {\em Phys. Rev. D}, 98(12):124024, 2018.

\bibitem{pRoS17a}
P.~Rioseco and O.~Sarbach.
\newblock Accretion of a relativistic, collisionless kinetic gas into a
  {S}chwarzschild black hole.
\newblock {\em Class. Quantum Grav.}, 34(9):095007, 2017.

\bibitem{sSsT85a}
S.L. Shapiro and S.A. Teukolsky.
\newblock {Relativistic stellar dynamics on the computer. I - Motivation and
  numerical method}.
\newblock {\em Astrophys. J.}, 298:34--79, 1985.

\bibitem{sSsT85b}
S.L. Shapiro and S.A. Teukolsky.
\newblock {Relativistic stellar dynamics on the computer. II - Physical
  Applications}.
\newblock {\em Astrophys. J.}, 298:58--79, 1985.

\bibitem{sSsT1993a}
S.L. Shapiro and S.A. Teukolsky.
\newblock {Relativistic Stellar Systems with Spindle Singularities}.
\newblock {\em Astrophysical Journal}, 419:622, December 1993.

\bibitem{sSsT1993b}
S.L. Shapiro and S.A. Teukolsky.
\newblock {Relativistic Stellar Systems with Rotation}.
\newblock {\em Astrophysical Journal}, 419:636, December 1993.

\bibitem{hAmKgR11}
H.~Andr{\'{e}}asson, M.~Kunze, and G.~Rein.
\newblock Existence of axially symmetric static solutions of the
  {E}instein-{V}lasov system.
\newblock {\em Commun. Math. Phys.}, 308:23--47, 2011.

\bibitem{hAmKgR14}
H.~Andr{\'{e}}asson, M.~Kunze, and G.~Rein.
\newblock Rotating, stationary, axially symmetric spacetimes with collisionless
  matter.
\newblock {\em Commun. Math. Phys.}, 329:787--808, 2014.

\bibitem{eAhAaL16}
E.~Ames, H.~Andr{\'{e}}asson, and A.~Logg.
\newblock On axisymmetric and stationary solutions of the self-gravitating
  {V}lasov system.
\newblock {\em Class. Quantum Grav.}, 32:155008, 2016.

\bibitem{eAhAaL19}
E.~Ames, H.~Andr{\'{e}}asson, and A.~Logg.
\newblock Cosmic string and black hole limits of toroidal {V}lasov bodies in
  general relativity.
\newblock {\em Phys. Rev. D}, 99(2):024012, 2019.

\bibitem{fJ2022}
F.E. Jabiri.
\newblock Stationary axisymmetric {E}instein-{V}lasov bifurcations of the
  {K}err spacetime.
\newblock 2 2022.
\newblock arXiv:2202.10245 [math.AP].

\bibitem{fJ2021}
F.E. Jabiri.
\newblock Static spherically symmetric {E}instein-{V}lasov bifurcations of the
  {S}chwarzschild spacetime.
\newblock {\em Annales Henri Poincar{\'e}}, 22(7):2355--2406, 2021.

\bibitem{pDeJmAeMdN17}
P.~Dom\'inguez, E.~Jim\'enez, M.~Alcubierre, E.~Montoya, and D.~N\'u{\~n}ez.
\newblock Description of the evolution of inhomogeneities on a dark matter halo
  with the {V}lasov equation.
\newblock {\em Gen.Rel.Grav.}, 49:123, 2017.

\bibitem{pRoS17b}
P.~Rioseco and O.~Sarbach.
\newblock Spherical steady-state accretion of a relativistic collisionless gas
  into a {S}chwarzschild black hole.
\newblock {\em J. Phys. Conf. Ser.}, 831(1):012009, 2017.

\bibitem{aCpM2020}
A.~Cie\ifmmode \acute{s}\else \'{s}\fi{}lik and P.~Mach.
\newblock Accretion of the {V}lasov gas on {R}eissner-{N}ordstr\"om black
  holes.
\newblock {\em Phys. Rev. D}, 102:024032, Jul 2020.

\bibitem{pMoA2021a}
P.~Mach and A.~Odrzywo\l{}ek.
\newblock Accretion of the relativistic {V}lasov gas onto a moving
  {S}chwarzschild black hole: Exact solutions.
\newblock {\em Phys. Rev. D}, 103:024044, Jan 2021.

\bibitem{aGetal2021}
A.~Gamboa, C.~Gabarrete, P.~Dom\'{\i}nguez-Fern\'andez, D.~N\'u{\~n}ez, and
  O.~Sarbach.
\newblock Accretion of a {V}lasov gas onto a black hole from a sphere of finite
  radius and the role of angular momentum.
\newblock {\em Phys. Rev. D}, 104:083001, Oct 2021.

\bibitem{pMoA2021b}
P.~Mach and A.~Odrzywo\l{}ek.
\newblock Accretion of dark matter onto a moving {S}chwarzschild black hole: An
  exact solution.
\newblock {\em Phys. Rev. Lett.}, 126:101104, Mar 2021.

\bibitem{pMaO2022}
P.~Mach and A.~Odrzywo\l{}ek.
\newblock Accretion of the relativistic {V}lasov gas onto a moving
  {S}chwarzschild black hole: low-temperature limit and numerical aspects.
\newblock {\em Acta Phys. Pol. B Proc. Suppl.}, 15(1-A7), January 2022.
\newblock Presented at the 7th conference of the {P}olish {S}ociety on
  {R}elativity, \L{}\'od\'z, {P}oland, 20-23 september 2021.

\bibitem{aCpMaO22}
A.~Cie\'slik, P.~Mach, and A.~Odrzywolek.
\newblock {Accretion of the relativistic Vlasov gas in the equatorial plane of
  the Kerr black hole}.
\newblock {\em arXiv e-prints}, page arXiv:2203.12401, 2022.

\bibitem{hA11}
H.~Andr{\'{e}}asson.
\newblock The {E}instein-{V}lasov system/kinetic theory.
\newblock {\em Living Reviews in Relativity}, 14(4), 2011.

\bibitem{oStZ13}
O.~Sarbach and T.~Zannias.
\newblock Relativistic kinetic theory: An introduction.
\newblock {\em AIP Conf. Proc.}, 1548:134--155, 2013.

\bibitem{oStZ14b}
O.~Sarbach and T.~Zannias.
\newblock The geometry of the tangent bundle and the relativistic kinetic
  theory of gases.
\newblock {\em Class. Quantum Grav.}, 31:085013, 2014.

\bibitem{rAcGoS2022}
R.~Acu{\~n}a{-}C{\'a}rdenas, C.~Gabarrete, and O.~Sarbach.
\newblock An introduction to the relativistic kinetic theory on curved
  spacetimes.
\newblock {\em General Relativity and Gravitation}, 54(23), February 2022.

\bibitem{CercignaniKremer-Book}
C.~Cercignani and G.M. Kremer.
\newblock {\em The Relativistic Boltzmann Equation: Theory and Applications}.
\newblock {Birkh\"auser}, Basel, 2002.

\bibitem{cGoS2022a}
C.~Gabarrete and O.~Sarbach.
\newblock Kinetic gas disks surrounding {S}chwarzschild black holes.
\newblock {\em Acta Phys. Pol. B Proc. Suppl.}, 15(1-A10), January 2022.
\newblock Presented at the 7th conference of the {P}olish {S}ociety on
  {R}elativity, \L{}\'od\'z, {P}oland, 20-23 september 2021.

\bibitem{Synge2-Book}
J.L. Synge.
\newblock {\em Relativity: The Special Theory}.
\newblock Elsevier Science, Amsterdam, 1956.

\bibitem{DLMF}
Digital library of mathematical functions.
\newblock \url{http://dlmf.nist.gov/}.

\bibitem{wS02}
W.~Schmidt.
\newblock Celestial mechanics in {K}err space-time.
\newblock {\em Class. Quantum Grav.}, 19:2743--2764, 2002.

\bibitem{jBmGtH15}
J.~Brink, M.~Geyer, and T.~Hinderer.
\newblock Astrophysics of resonant orbits in the {K}err metric.
\newblock {\em Phys. Rev. D}, 91(8):083001, 2015.

\bibitem{iGiR2007}
I.~Gradshteyn and I.~Ryzhik.
\newblock {\em Table of {I}ntegrals, {S}eries, and {P}roducts}.
\newblock Academic Press, INC., San Diego, CA, 2007.

\bibitem{lFvM1976}
L.G. Fishbone and V.~Moncrief.
\newblock Relativistic fluid disks in orbit around {K}err black holes.
\newblock {\em The Astrophysical Journal}, 207:962--976, 1976.

\bibitem{mAmJmS1978}
M.~Abramowicz, M.~Jaroszy\'nski, and M.~Sikora.
\newblock Relativistic, accreting disks.
\newblock {\em Astronomy and Astrophysics}, 63:221--224, 1978.

\bibitem{mKmJmA1978}
M.~Kozlowski, M.~Jaroszynski, and M.~A. Abramowicz.
\newblock {The analytic theory of fluid disks orbiting the Kerr black hole.}
\newblock {\em Astronomy and Astrophysics}, 63(1-2):209--220, February 1978.

\bibitem{jFfD2002}
J.~Font and F.~Daigne.
\newblock The runaway instability of thick discs around black holes-{I}. {T}he
  constant angular momentum case.
\newblock {\em Monthly Notices of the Royal Astronomical Society},
  334:383--400, 2002.

\bibitem{jFfD2004}
J.~Font and F.~Daigne.
\newblock The runaway instability of thick disks around black holes-{II}.
  {N}on-constant angular momentum discs.
\newblock {\em Monthly Notices of the Royal Astronomical Society},
  349(3):841--868, 2004.

\bibitem{oSetal2012}
O.~Straub et. al.
\newblock {Modelling the black hole silhouette in {S}gr {A}* with ion tori}.
\newblock {\em Astron. Astrophys.}, 543:A83, 2012.

\bibitem{Straumann-Book}
N.~Straumann.
\newblock {\em General Relativity}.
\newblock Springer-Verlag, Berlin, 2013.

\bibitem{Rezzolla-Book}
L.~Rezzolla and O.~Zanotti.
\newblock {\em Relativistic Hydrodynamics}.
\newblock Oxford University Press, Great Clarendon Street, Oxford, OX2 6DP,
  United Kingdom, 2013.

\bibitem{mA1971}
M.~A. Abramowicz.
\newblock The {R}elativistic von {Z}eipel's {T}heorem.
\newblock {\em Acta Astronomica}, 21(1):81--85, 1971.

\end{thebibliography}
%
\end{document}